\documentclass[10pt,a4paper]{article}
 \pdfoutput=1
\usepackage{amsmath,fourier,amssymb,amsthm,graphicx,color,epstopdf}
\usepackage{chngcntr}
\counterwithout{figure}{section}
\usepackage[T1]{fontenc}
\allowdisplaybreaks
\numberwithin{equation}{section}

 \newtheorem{innercustomthm}{Theorem}
 \newenvironment{customthm}[1]
   {\renewcommand\theinnercustomthm{#1}\innercustomthm}
   {\endinnercustomthm}

 \theoremstyle{plain}
 \newtheorem {hypo}{\bf\hspace{-\parindent}Hypothesis}[section]

 \newtheorem {ass}[hypo]{Assumption}
 \newtheorem {prop}[hypo]{Proposition}
 \newtheorem {lemma}[hypo]{Lemma}
 \newtheorem {theo}[hypo]{Theorem}
 \newtheorem {defin}[hypo]{Definition}
 \newtheorem {cor}[hypo]{Corollary}
 
 \theoremstyle{remark}
 \newtheorem {rmk}[hypo]{Remark}
  
 \newcommand{\pf}{\begin{bpf}}

 \newcommand{\pfms}{\begin{bpfms}}
 \newcommand{\epf}{\end{bpf}\hfill$\square$\vspace{0.1cm}}
 \newcommand{\epfms}{\end{bpfms}\hfill$\square$\\ }
 \newcommand\ben{\begin{equation*}}
 \newcommand\ebn{\end{equation*}}
 \newcommand\beq{\begin{equation}}
 \newcommand\eeq{\end{equation}}
 \newcommand\ds{\displaystyle}
  \newcommand\lb{\left(}
  \newcommand\rb{\right)} 
  
   \newcommand\Cb{\mathbb{C}} 
   \newcommand\Zb{\mathbb{Z}}
   \newcommand\Pb{\mathbb{P}}

\begin{document}
\LARGE
\noindent
\textbf{Monodromy dependence and connection formulae \vspace{0.1cm}\\
for isomonodromic tau functions}
\normalsize
 \vspace{1cm}\\
 \textit{
 A. R. Its$\,^{a,b,}$\footnote{aits@iupui.edu}, 
 O. Lisovyy$\,^{c,}$\footnote{lisovyi@lmpt.univ-tours.fr}, 
 A. Prokhorov$\,^{a,b,}$\footnote{aprokhor@iupui.edu}}
 \vspace{0.2cm}\\
 $^a$ Department of Mathematical Sciences,
 Indiana University-Purdue University,
 402 N. Bla\-ckford St.,
 Indianapolis, IN 46202-3267,
 USA 
 \vspace{0.1cm}\\
  $^b$ St. Petersburg State University,
  Universitetskaya emb., 7/9, St. Petersburg, Russia
 \vspace{0.1cm}\\
 $^c$ Laboratoire de Math\'ematiques et Physique Th\'eorique CNRS/UMR 7350,  Universit\'e de Tours, Parc de Grandmont,
  37200 Tours, France

\begin{abstract}
We discuss an extension of the Jimbo-Miwa-Ueno differential 1-form  to a form closed  on the
full space of extended monodromy data of systems of linear ordinary  differential equations 
with rational coefficients. This extension is based on the results of M. Bertola generalizing a previous construction by B. Malgrange. We show how this 1-form can be used to solve a long-standing problem of evaluation
of the connection formulae for the isomonodromic tau functions which would include an explicit computation
of the relevant constant factors. We explain how this scheme works for Fuchsian systems and, in particular,
calculate the connection constant for generic Painlev\'e VI tau function. The result proves the conjectural formula
for this constant proposed in \cite{ILT13}. We also apply the method to non-Fuchsian systems and 
evaluate constant factors in the asymptotics of Painlev\'e II tau function.
\end{abstract}
\section{Introduction and statement of results}

\subsection{Linear systems}
Consider a system of linear ordinary differential equations with rational coefficients,
  \beq\label{gensys0} 
 \frac{d\Phi}{dz}= A\lb z\rb\Phi,
 \eeq
 where $ A\lb z\rb$ is an $N\times N$, $N>1$ matrix-valued rational function. We are concerned with its isomonodromic deformations.
 More precisely, the object of our study is the  global asymptotic analysis of the associated Jimbo-Miwa-Ueno   tau function. Let us remind, following
 \cite{JMU}, the general set-up associated with this notion.

 Denote the poles of matrix function $ A\lb z\rb$ on $\mathbb P\equiv
 \mathbb P^1\lb\Cb\rb$ by $a_1,\ldots ,a_n,\infty$  and  write the leading terms of the  Laurent expansions of $ A\lb z\rb$ at  these points  in the form
 \ben
 A\lb z\rb=\begin{cases}\ds
 \frac{A_{\nu}}{(z-a_{\nu})^{r_{\nu} +1}}+O\lb\lb z-a_{\nu}\rb^{-r_{\nu}}\rb\qquad & \text{as }z\to a_{\nu},\vspace{0.1cm} \\
 -z^{r_{\infty} -1}A_{\infty}+O\lb z^{r_{\infty} -2}\rb\qquad &
 \text{as }z\to \infty,
 \end{cases}
 \ebn
 where $r_{1},\ldots,r_n,r_{\infty}\in\mathbb{Z}_{\ge 0}$.
  We are going to make the standard assumption that all $A_{\nu}$ with $\nu=1,\ldots,n,\infty$ are diagonalizable and have pairwise distinct eigenvalues. We further require these eigenvalues to be non-resonant, i.e. distinct modulo $\Zb$ whenever $r_{\nu}=0$. Fix the diagonalizations
 \ben
A_{\nu}=G_{\nu}\Theta_{\nu,-r_\nu}G_{\nu}^{-1},\qquad \Theta_{\nu,-r_{\nu}}=\operatorname{diag}\left\{\theta_{\nu,1},
\ldots, \theta_{\nu,N}\right\}.
\ebn
At each singular point, the system (\ref{gensys0}) admits a {\it formal} solution
\beq\label{formalsol}
 \Phi_{\text{form}}^{(\nu)}\left(z\right) =  G^{\lb\nu\rb}\left(z\right)e^{\Theta_{\nu}\lb
 z\rb}, \qquad \nu = 1, \ldots , n,\,\infty,
\eeq
where $G^{\lb\nu\rb}\left(z\right)$ is a formal series,
$$
G^{\lb\nu\rb}\left(z\right) =
G_{\nu}\hat{\Phi}^{\lb\nu\rb}\lb z\rb,\qquad 
\hat{\Phi}^{\lb\nu\rb}\lb z\rb=
\begin{cases} \mathbf{1}+\sum_{k=1}^{\infty}g_{\nu,k}\left(z-a_{\nu}\right)^k,
\qquad &\nu = 1, \ldots, n,\\
\mathbf{1}+\sum_{k=1}^{\infty}g_{\infty,k}z^{-k},\qquad & \nu=\infty,
\end{cases}
$$
and  $\Theta_{\nu}(z)$ are diagonal matrix-valued functions,
\ben
\Theta_{\nu}(z)=\sum_{k=-r_\nu}^{-1}
\frac{\Theta_{\nu,k}}{k}\lb z-a_\nu\rb^{k}+\Theta_{\nu,0}\ln \lb z-a_\nu\rb,\qquad
\Theta_{\infty}\lb z\rb =-\sum_{k=1}^{r_{\infty}}\frac{\Theta_{\infty,-k}}{k}z^{k}-\Theta_{\infty,0}\ln z.
\ebn
For every $\nu\in\{1, \ldots, n,\infty\}$,  the matrix coefficients $g_{\nu,k}$ and $\Theta_{\nu,k}$ can be uniquely and  explicitly computed  in terms 
of the coefficients of the matrix-valued rational function $G^{-1}_{\nu} A\lb z \rb G_{\nu}$, see \cite{JMU}. If the point is {\it Fuchsian}
($r_{\nu} = 0$), then the series (\ref{formalsol})  converges and represents a genuine solution of (\ref{gensys0}). If $r_{\nu} > 0$, i.e.
$a_{\nu}$ is an  irregular point, the
series (\ref{formalsol}) usually diverges. In this case there are  $2r_{\nu} + 1$ genuine  {\it canonical solutions}  of (\ref{gensys0}); each   canonical 
solution has  (\ref{formalsol}) as its asymptotic series in the corresponding Stokes sector.  These fundamental concepts will be reviewed in the main text (Sections \ref{sec_FS} and \ref{secIR}).

 The non-formal global properties of solutions of the equation (\ref{gensys0}) are described by its
 {\it monodromy data} $M$ which include: i) formal monodromy exponents $\Theta_{\nu,0}$, ii) appropriate connection
 matrices between canonical solutions at different singular points, and iii) relevant Stokes matrices at irregular singular
 points.  All these objects will be also described  in more
detail in the main body of the paper.  
Let us denote the space of monodromy data of the system (\ref{gensys0}) by $ \mathcal{M} $.

Assume that irregular singular points (i.e. the points with $r_{\nu} > 0$) are $\infty$ and the first $m \leq n$  among the
singular points $a_{1},\ldots,a_n$.
Introduce the set $\mathcal{T}$  of {\it isomonodromic times}
\begin{equation}\label{istime}
a_{1},\ldots, a_{n}, \qquad  \lb \Theta_{\nu,k}\rb_{ll}, \qquad k=-r_{\nu}, \ldots, -1,\quad \nu = 1,\ldots, m, \infty,
\quad l = 1, \ldots, N.
\end{equation}
Let us also denote by $\mathcal{A}$ the variety of all rational matrix-valued 
functions $ A \lb z\rb$ with a fixed number of poles  of fixed orders.
The so-called Riemann-Hilbert correspondence states that, up to  submanifolds where 
the inverse monodromy problem for (\ref{gensys0}) is not solvable, the space $ \mathcal{A}$ can be identified with
the product $\widetilde{\mathcal T}\times \mathcal{M}$, where $\widetilde{\mathcal T}$ denotes the universal covering of $\mathcal{T}$. We shall loosely write,
$$
\mathcal{A} \simeq \widetilde{\mathcal T}\times \mathcal{M}.
$$

 \subsection{Tau function}
The {\it Jimbo-Miwa-Ueno 1-form}  is defined as the following differential form on $\mathcal{A}$:
\begin{equation}\label{jmu}
\omega_{\mathrm{JMU}}=-\sum_{\nu=1,\ldots, n, \infty} \mathop{\mathrm{res}}\nolimits_{z=a_\nu} \operatorname{Tr}\left(\hat\Phi^{(\nu)}
\lb z\rb^{-1}\partial_z\hat\Phi^{(\nu)}\lb z\rb\, d_{\mathcal T}\Theta_{\nu}\lb z\rb\right),
\end{equation}
where we put $a_{\infty} \equiv \infty$. The notation $d_{\mathcal T}\Theta_{\nu}\lb z\rb$ stands for
$$
d_{\mathcal T}\Theta_{\nu}\lb z\rb = \sum_{k=1}^{L}\frac{\partial\Theta_{\nu}\lb z\rb}{\partial t_{k}}\,dt_k,\qquad 
L = n + N\left(\sum_{\nu=1}^{m}r_{\nu} + r_{\infty}\right),
$$
where $t_1, \ldots,t_L$ are parameters from (\ref{istime}). While the series $\hat\Phi^{(\nu)}(z)$ is a formal object, the right side of (\ref{jmu}) is well-defined. Indeed, the residue of the product of the formal series appearing in each term
of the sum is defined as the coefficient of the term $\lb z-a_{\nu}\rb^{-1}$,  and as such it involves only
finitely many coefficients. 

Let us fix a point $M \in \mathcal{M} $ and consider the 
{\it isomonodromic} family in the space $\mathcal{A}$,
$$
A\lb z\rb \equiv  A\lb z; \vec{t};  M\rb,\qquad \vec{t} = \lb t_1,\ldots, t_L\rb,
$$
that is, the family of  systems (\ref{gensys0}) that are characterized the same set $M$ of monodromy data.
The significance of the form $\omega_{\mathrm{JMU}}$ is that, being restricted to the isomonodromic family
$ A\lb z; \vec{t}; M\rb$,
it  becomes closed with respect to times $\mathcal T$, i.e.
\begin{equation}\label{jmu2}
d_{\mathcal T}\left(\omega_{\mathrm{JMU}}\bigl|_{ A\lb z; \vec{t}; M\rb}\right) = 0.
\end{equation}
It should be  also noticed  that in the non-resonant case (the situation we are exclusively interested in here)  $\mathcal{T}$ gives a complete set of independent isomonodromic  deformation parameters so that the equation $A\lb z\rb \equiv  A\lb z; \vec{t};  M\rb$ describes all possible monodromy preserving deformations of 
(\ref{gensys0}). This is a nontrivial  fact, and it has been proved in \cite{JMU}.

The closedness of the 1-form $\omega_{\mathrm{JMU}}$ with respect to $\mathcal T$ in turn implies that locally there is a function $\tau \equiv \tau\lb\vec{t};M\rb$ on ${\mathcal{T}}
\times \mathcal M$ such that
\begin{equation}\label{jmu3}
d_{\mathcal T}\ln \tau = \omega_{\mathrm{JMU}}.
\end{equation}
A remarkable property of this {\it tau function}  $\tau\lb\vec{t}; M\rb$, which was established in   \cite{Malgrange} and \cite{Miwa}, is that it 
admits analytic continuation  as an entire function to the whole universal covering $\widetilde{\mathcal T}$ of the parameter space ${\mathcal T}$.
Furthermore, zeros of $\tau\lb\vec{t}; M\rb$ correspond to the points in ${\mathcal T}$ where the inverse monodromy problem
for (\ref{gensys0}) is not solvable for a given set $M$ of monodromy data  (or, equivalently, where a certain holomorphic vector
bundle over $\Pb$ determined by $ M$ becomes nontrivial). Hence the tau function plays a central role in the monodromy theory
of systems of linear differential equations. 

The  tau function has several other striking properties. Among them we shall single out the  hamiltonian aspect. A key fact of the monodromy  theory of linear systems  is that 
the isomonodromic family $ A\lb z; \vec{t}; M\rb$ can be described in terms of
solutions of an integrable (in the sense of Frobenius)  {\it monodromy preserving deformation equation},
\begin{equation}\label{isomeq0}
{d_{\mathcal{T}} A}=\partial_z U+[U, A].
\end{equation}
Here $U$  is a matrix-valued differential form,  $U \equiv \sum_{k=1}^{L}U_k(z)dt_k$,  whose coefficients $U_k\lb z\rb$ are rational matrix-valued functions of $z$ uniquely determined by the coefficients of the system (\ref{gensys0}). The space  $\mathcal{A}$ can be equipped with a canonical
symplectic structure so that the isomonodromy equation (\ref{isomeq0})   induces $L$ commuting Hamiltonian
flows on $\mathcal{A}$.  It turns out that in many cases the logarithm of the tau function serves as the {\it generating function} of their Hamiltonians
$H_k$:
\begin{equation}\label{ham00}
\frac{\partial \ln\tau\lb\vec{t}; M\rb}{\partial t_k} = {H_k}\bigl|_{ A\lb z; \vec{t}; M\rb}.
\end{equation}

Isomonodromy equation (\ref{isomeq0}) is of great interest on  its own. Indeed, it includes as special cases practically
all known integrable differential equations. The first nontrivial cases of (\ref{isomeq0}), where the set of isomonodromic times 
effectively reduces to a single variable $t$,  cover all six classical Painlev\'e equations. Solutions of the latter are dubbed as {\it nonlinear special functions},
and they indeed play this role in many areas  of modern nonlinear science. Besides the canonical applications of Painlev\'e transcendents such as integrable systems \cite{WMTB,JMMS,AS}, two-dimensional quantum gravity \cite{BK,DS,GrM} and random matrix theory \cite{TW1,TW2}, we would like to mention a few very recent examples concerned with black hole scattering \cite{NC}, Rabi model \cite{CCR} and Fermi gas spectral determinants arising in supersymmetric Yang-Mills theory \cite{BGT}.

\subsection{Connection problem}
The principal  analytic issue concerning the tau function, in particular from the point of view of applications, is its  behavior near the critical hyperplanes,
where either $a_{\mu} = a_{\nu}$ for some $\mu \neq \nu$, or $\theta_{\nu,\alpha} = \theta_{\nu,\beta}$
for some~$\nu$ and some $\alpha \neq \beta$.  This is the question we are addressing in this paper. We are going to study two nontrivial examples corresponding to the sixth and the second Painlev\'e equations. The critical hyperplanes reduce
to three branching points $t = 0, 1, \infty$  in the case of Painlev\'e VI, and to one essential singularity
$t = \infty$ in the case of Painlev\'e II. Our goal is to express the parameters of the  asymptotic behavior
of the corresponding  tau functions at these critical points explicitly in terms of  monodromy data of the associated linear
systems~(\ref{gensys0}).

 A convenient tool of the  global asymptotic analysis of Painlev\'e transcendents as
 well as solutions of an arbitrary monodromy preserving deformation equation (\ref{isomeq0}) is
 provided by the Riemann-Hilbert method. It is based on the
 {\it Riemann-Hilbert  representation}  of solutions, i.e. on the
 representation of the coefficients of matrix $A\lb z\rb$ in terms of the 
 inverse monodromy map, 
 \begin{equation}\label{mmap}
 \mathcal{RH}^{-1}:
 \mathcal{M} \to \mathcal{A}.
 \end{equation}
 Analytically, this map is realized as  a matrix Riemann-Hilbert problem. It has been proven to be extremely efficient
 in the asymptotic analysis of Painlev\'e equations; the reader is referred to the monograph \cite{FIKN}
 for detailed exposition and history of the  subject. 
 The Riemann-Hilbert technique, however, addresses directly the coefficients of the  matrix $ A\lb z\rb$,
 i.e. in the $2\times 2$ case, it deals with conventional Painlev\'e functions and not the associated tau functions.
In order to obtain a complete asymptotic information about the latter, 
one has to evaluate, according to (\ref{ham00}), integrals of  certain combinations  of  Painlev\'e 
transcendents and their derivatives.  This would mean the evaluation of the   tau function asymptotics 
including {\it constant factors}. More precisely, since the tau function is itself defined up to a multiplicative 
constant, we are actually  talking about the evaluation, in terms of monodromy data,  of the ratios of constant factors corresponding
to different critical points (Painlev\'e~VI) or to different critical directions (Painlev\'e II).

For a long time, the ``constant  problem''  has been successfully  handled only for rather special 
 solutions of  Painlev\'e equations whose tau functions 
 admit additional representations in terms of certain Fredholm or Toeplitz or Hankel  determinants. The aim of the present paper is to develop a technique which would be applicable to general two-parameter families of Painlev\'e tau functions and which would not rely on determinant formulae.
 The key idea of our approach is to find an  extension of the Jimbo-Miwa-Ueno differential form $\omega_{\mathrm{JMU}}$
 to a closed  1-form on the whole space ${\mathcal A}\simeq \widetilde{\mathcal T}\times  \mathcal{M}$. This means 
 the construction of  a differential  1-form $\hat{\omega}\equiv \hat{\omega}\lb A\rb
 \equiv \hat{\omega}\lb \vec{t}; M\rb$ such that
$$
d \hat{\omega} \equiv  d_{\mathcal T}\hat{\omega} + d_{\mathcal M}\hat{\omega} = 0,
$$
and such that the compatibility condition
 $$
 \hat{\omega}\lb\partial_{t_k}\rb = \omega_{\mathrm{JMU}}\lb \partial_{t_k}\rb
 $$
is satisfied for all isomonodromic times $t_1,\ldots, t_L\in \mathcal T$.
Having such a 1-form expressed in terms of the fundamental matrix solution of (\ref{gensys0}),  we will be able to define the tau function  by
the formula
\begin{equation}\label{exttau0}
\ln\tau = \int \hat{\omega}.
 \end{equation}
 Equation (\ref{exttau0})  allows one to use the asymptotic behavior  of $\Phi\lb z\rb$ to evaluate the 
 asymptotics of the associated tau function  up to a numerical (i.e. independent of monodromy data) constant. The  latter can be calculated  by
 applying the final formulae to special solutions with known constant factors in the relative asymptotics.

 The program outlined above has been first realized in \cite{IP} for Painlev\'e III equation of the most degenerate type $D_8$, where it allowed to give a proof of the connection formula for PIII ($D_8$) tau function earlier conjectured in \cite{ILT14}.  
 In the present paper, we use 
 it  to solve the ``constant problem'' for the sixth and second Painlev\'e equations. The key ingredient of our approach
 is yet another 1-form which  we shall denote $\omega$:
 \beq\label{formomega}
 \omega=\sum_{\nu=1,\ldots,n,\infty}\operatorname{res}_{z=a_{\nu}}
 \operatorname{Tr}\lb 
 {G^{\lb \nu\rb}\lb z\rb}^{-1}
 A\lb z\rb \,dG^{\lb \nu\rb}\lb z\rb \rb,\qquad
 \qquad
 d=d_{\mathcal T}+d_{\mathcal M}.
 \eeq
 This expression is inspired by the works of Malgrange \cite{Malgrange}
 and Bertola \cite{Bertola}. It extends the  Jimbo-Miwa-Ueno form $\omega_{\mathrm{JMU}}$ and can be defined for
 an arbitrary system (\ref{isomeq0}). However, $\omega$ is {\it not} closed. Instead, its exterior  differential
 turns out to be a 2-form on $\mathcal M$ only and it furthermore turns out to be independent of isomonodromic times ${\mathcal T}$. This fact in conjunction with computable asymptotics of $\Phi\lb z\rb$  determines what should be added to the form $\omega$ to make it closed, i.e. to transform it into the form $\hat{\omega}$.

 Though we do not pursue the analysis of the form  $\hat{\omega}$ for the general system (\ref{gensys0})  in this paper, we strongly believe that $\hat{\omega}$ can be identified with the extension suggested by Bertola in \cite{Bertola} and \cite{Bertola1}. This, in view of general results of
 \cite{Malgrange,Miwa,Palmer,AB}, would mean that the extended tau function defined by (\ref{exttau0})
is entire  on the extended  phase space $\widetilde{\mathcal T}\times \mathcal{M}$ and it  vanishes  when the extended inverse monodromy problem, 
 $\widetilde{\mathcal T}\times \mathcal{M} \to \mathcal A$ is not solvable. We give more  details on  the Malgrange-Bertola extension of the Jimbo-Miwa-Ueno form
and its relation to $\hat{\omega}$ in Section \ref{secIR}, see Remark \ref{bertola_malgrange}.

 \subsection{Summary of results\label{subsec_results}}
 
 Let us now present the main technical results of this work --- a complete solution of the connection problems for the Painlev\'e VI and Painlev\'e  II tau functions.
 These are representative examples associated to two basic classes of linear systems (\ref{gensys0}): a  purely Fuchsian  system (PVI) and a system with irregular singularities (PII). The extension of the method developed in \cite{IP} for PIII ($D_8$) to equations studied here has required, especially  in the  Painlev\'e VI case, the development of new conceptual and technical features of the scheme itself.  We shall start with the PVI tau function.

The Painlev\'e VI equation describes  the first nontrivial case of the Fuchsian  isomonodromic deformations, corresponding to
the $2\times2$ linear system with four regular singularities, i.e. to $n=3$. Using affine transformations, one can always fix two of three finite singular points to be at $0$ and $1$. Denote the only remaining isomonodromic time  $a_3$  by $t$.
That is, we are dealing with the system
\begin{equation}\label{PVIlinsys0}
\frac{d\Phi}{d z}=A\left(z\right)\Phi,\qquad A\left(z\right) = \frac{A_0}{z} + \frac{A_t}{z-t} + \frac{A_1}{z-1},
\end{equation}
where $A_{0,t,1}$ are traceless $2\times2$ matrices. Denote by $\pm \theta_{\nu}$ the eigenvalues of $A_{\nu}$, including those of ${A_{\infty}=-A_0 - A_t - A_1} $. The latter matrix may be assumed diagonal so that $A_{\infty}=\operatorname{diag}\left\{\theta_{\infty}, -\theta_{\infty}\right\}$. The non-resonance assumption takes the form
\begin{equation}\label{nonresp6}
2\theta_\nu \neq \Zb, \qquad \nu = 0,t,1,\infty.
\end{equation}
In this setting, 
$$
\mathcal{A} = \{ (A_0, A_t,  A_1,  t)\}, \qquad \mathcal{T} =\left \{ t\in \Cb: t\neq 0, 1\right\}.
$$
We shall  also assume that $t \in (0,1)$. The general isomonodromy equation (\ref{isomeq0}) in the case of the system (\ref{PVIlinsys0}) reduces to a system of matrix ODEs,
\begin{equation}\label{p60}
 \frac{dA_0}{dt}=\frac{\left[A_t,A_0\right]}{t},\qquad 
 \frac{dA_1}{dt}=\frac{\left[A_t,A_1\right]}{t-1},
\end{equation}
while the general Jimbo-Miwa-Ueno definition (\ref{jmu}) produces the following formula for the tau function:
$$
d\ln \tau = \omega_{\mathrm{JMU}} = \lb
\frac{\operatorname{Tr}A_t A_0}{t}+
  \frac{\operatorname{Tr}A_t A_1}{t-1}\rb\,dt.
$$

The system (\ref{p60}) in turn yields the sixth Painlev\'e equation for the function $u(t)$, defined as the unique zero of $A_{12}(z)$
(note that this matrix entry is a linear function of $z$ due to the diagonal form of $A_{\infty}$),
\begin{subequations}\label{pviforms}
\begin{equation}\label{PVI0}
u_{tt}= \left( \frac{1}{u}+
\frac{1}{u-1} + \frac{1}{u-t} \right)
\frac{u_t^{\,2}}{2} - \left( \frac{1}{t} +\frac{1}{t-1}
+\frac{1}{u-t} \right) u_t
+\frac{u\lb u-1\rb \lb u-t\rb}{t^2 \lb t-1\rb^2} \left( \alpha +
\frac{\beta t}{u^2}+
\frac{\gamma \lb t-1\rb}{\lb u-1\rb^2} +
\frac{\delta t\lb t-1\rb}{\lb u-t\rb^2} \right ).
\end{equation}
 The parameters $\alpha$, $\beta$, $\gamma$, and $\delta$ are determined by the eigenvalues $\theta_{\nu}$ ($\nu = 0,t, 1, \infty$) according
 to the formulae
 $$
 \alpha = \frac{\lb 2\theta_{\infty} - 1\rb^2}{2}, \quad \beta = -2\theta^2_{0}, \quad \gamma = 2\theta_1^2, \quad \delta = \frac{1 - 4\theta_t^2}{2}.
 $$
 An alternative form of the sixth Painlev\'e equation (\ref{PVI0}) can be written in terms of the tau function, or rather its logarithmic derivative. Putting $ \zeta(t) = t(t-1)\frac{d}{dt}\ln\tau $,
 one has
 \beq
 \Bigl(t(t-1)\zeta_{tt}\Bigr)^2 =-2\det
 \begin{pmatrix}2\theta_0^2&t\zeta_t - \zeta&\zeta_t +\theta_0^2  + \theta_t^2 +\theta_1^2  - \theta_{\infty}^2 \cr
 t\zeta_t - \zeta&2\theta_t^2&(t-1)\zeta_t - \zeta\cr
 \zeta_t +\theta_0^2  + \theta_t^2 +\theta_1^2  - \theta_{\infty}^2& (t-1)\zeta_t - \zeta &2\theta_1^2
 \end{pmatrix}.
 \eeq
 \end{subequations}

The space  ${\mathcal M}$ of  monodromy data can be identified 
with the collection of 7-tuples
$$
\mathcal{M}_{\mathrm{PVI}} = \left\{M=\lb p_{01}, p_{0t}, p_{t1}; \theta_0, \theta_t, \theta_1, \theta_{\infty}\rb\in\Cb^7\right\}, 
$$
where $\theta_0$, $\theta_t$, $\theta_1$, $\theta_{\infty}$ are the above Painlev\'e VI parameters, and $p_{01}, p_{0t}, p_{t1}$ belong to the  Jimbo-Fricke cubic hypersurface
 \begin{equation}\label{jimbofricke}
 \begin{gathered}
p_{0t}p_{t1}p_{01} - \left(p_0p_t+p_1p_{\infty}
  \right)p_{0t}-\left(p_tp_1+p_0p_{\infty}\right)p_{t1}
  -\left(p_0p_1+p_tp_{\infty}
    \right)p_{01}+\\
  +p_{0t}^2+p_{t1}^2+p_{01}^2+p_0^2+p_t^2+p_1^2+p_{\infty}^2 +  p_0p_tp_1p_{\infty} =4,
  \end{gathered}
 \end{equation}
encoding the Painlev\'e VI initial conditions. 
In this equation, $p_{\nu}=2\cos2\pi\theta_{\nu}$ with $\nu=0,t,1,\infty$. It is convenient to introduce a similar trigonometric parametrization of $p_{\mu\nu}$,
$$
p_{\mu\nu} = 2\cos 2\pi\sigma_{\mu\nu},  \quad 0\leq\Re \sigma_{\mu\nu} \leq \frac{1}{2},
\qquad \mu\nu =0t,t1,01.
$$
The quantities $\{p_{\mu\nu}\}$, $\{p_{\nu}\}$  represent traces of  monodromy matrices of the fundamental solution $\Phi\lb z\rb$ of (\ref{PVIlinsys0}) along certain loops on $\Cb\backslash\left\{0,t,1\right\}$; see Subsection \ref{subsmon} for more details.

 In this paper, we study the two-parameter family of generic Painev\'e VI transcendents. It is characterized, in addition to 
 the non-resonance condition (\ref{nonresp6}),  by the following
 restrictions on the monodromy data $\{\sigma_{\mu\nu}, \theta_{\nu}\}$:
 \begin{equation}\label{p6cond1}
  \begin{gathered}
 \sigma_{0t},\sigma_{t1} \neq 0, \quad 0\leq\Re\sigma_{0t},
 \Re\sigma_{t1} <\frac{1}{2}, \\
 \theta_0 + \theta_t \pm \sigma_{0t}, \,\,\,  \theta_0 - \theta_t \pm \sigma_{0t}, \,\,\,  \theta_\infty + \theta_1\pm \sigma_{0t},
 \,\,\,  \theta_\infty - \theta_1 \pm \sigma_{0t} \notin \Zb,\\
 \theta_1 + \theta_t \pm \sigma_{t1}, \,\,\,  \theta_1 - \theta_t \pm \sigma_{t1}, \,\,\,  \theta_\infty + \theta_0\pm \sigma_{t1},
 \,\,\,  \theta_\infty - \theta_0 \pm \sigma_{t1} \notin \Zb.
 \end{gathered}
 \end{equation}
 These conditions imply the following asymptotic behavior of the 
 Painlev\'e VI tau function $\tau(t)$ as $ t \to 0, 1$: 
 \begin{subequations} \label{jimfs}
   \beq
   \begin{gathered}
   \label{p6at0}
   \tau(t)=\mathcal{C}_{0}\cdot t^{\sigma^2-\theta_0^2-\theta_t^2}\,\Biggl[ 1-
   \sum_{\epsilon=\pm1}\frac{
   \left(\left(\theta_t-\epsilon\sigma\right)^2-\theta_0^2\right)
   \left(\left(\theta_1-\epsilon\sigma\right)^2-\theta_{\infty}^2
   \right)}{
   4\sigma^2\left(1+2\epsilon\sigma\right)^2}\,\kappa^{\epsilon}
   t^{1+2\epsilon\sigma}+
   \Biggr. \\ 
   \qquad\qquad\qquad\qquad\quad +\frac{
      \left(\sigma^2-\theta_{\infty}^2+\theta_1^2\right)
      \left(\sigma^2-\theta_0^2+\theta_t^2\right)}{2\sigma^2}\,t+
   o\left(t\right)\Biggr],\qquad t \to 0,
   \end{gathered}
   \eeq
 and
      \beq
      \begin{gathered}
      \label{p6at1}
      \tau(t)=\mathcal{C}_{1}\cdot \lb 1-t\rb^{\overline\sigma^2-\theta_1^2-\theta_t^2}\,\Biggl[ 1-
      \sum_{\epsilon=\pm1}\frac{
      \left(\left(\theta_t-\epsilon\overline\sigma\right)^2-
      \theta_1^2\right)
      \left(\left(\theta_0-\epsilon\overline\sigma\right)^2-
      \theta_{\infty}^2
      \right)}{
      4\overline\sigma^2\left(1+2\epsilon\overline\sigma\right)^2}\,
      \overline\kappa^{\,\epsilon}
      \lb 1-t\rb^{1+2\epsilon\overline\sigma}+
      \Biggr. \\ 
      \quad\qquad\qquad\qquad\qquad\qquad\quad +\frac{
         \left(\overline\sigma^2-\theta_{\infty}^2+\theta_0^2\right)
         \left(\overline\sigma^2-\theta_1^2+\theta_t^2\right)}{2
         \overline\sigma^2}
         \lb 1-t\rb +
      o\left(1-t\right)\Biggr], \qquad t \to 1.
      \end{gathered}
      \eeq
\end{subequations}      
 The respective ``Cauchy data'' at $t =0$ and $t=1$, i.e. the pairs $(\sigma, \kappa)$ and $(\overline\sigma, \overline\kappa)$,
can be explicitly related  to the monodromy data $M = (p_{01}, p_{0t}, p_{t1}; \theta_0, \theta_t, \theta_1, \theta_{\infty})$. Indeed,
 $\sigma = \sigma_{0t}$, $\overline{\sigma} = \sigma_{t1}$,
 whereas the expressions for $\kappa$ and $\overline\kappa$ are more involved and are given by the equations (\ref{kappadef}), (\ref{etadef}), (\ref{kappadef1}) and (\ref{etabardef}) of Section \ref{sec_fourpoint}. 
 
 The asymptotic formulae (\ref{jimfs}), together with their explicit parametrization via the monodromy data, are due to M. Jimbo \cite{Jimbo}. 
 These formulae  also come out  within our general Riemann-Hilbert analysis of the 4-point Fuchsian tau function in Section \ref{sec_fourpoint},
 see Propositions \ref{jimbof} and \ref{jimbof1}. Our main new result concerns the rigorous derivation of the ratio  $\Upsilon\lb M\rb:=\frac{\mathcal{C}_{1}}{\mathcal{C}_{0}}$, which has been an open problem since the 80s. It is given by the following theorem.
  \begin{customthm}{A}\label{thp6}Under the assumptions (\ref{nonresp6}) and (\ref{p6cond1}) on the monodromy data, 
 the ratio $\Upsilon\lb M\rb$ of the constant factors in the asymptotic formulae (\ref{jimfs}) is given by 
\begin{equation}\label{p6const}
 \frac{\mathcal{C}_{1}}{\mathcal{C}_{0}}(M)=\prod_{\epsilon,\epsilon'=\pm}
 \frac{G\left( 1+\epsilon\overline\sigma+\epsilon'\theta_t-
     \epsilon\epsilon'\theta_1\right) G\left(1+\epsilon\overline\sigma+\epsilon'\theta_0-
         \epsilon\epsilon' \theta_{\infty}\right)}{
 G\left(1+\epsilon\sigma+\epsilon'\theta_t+\epsilon\epsilon'\theta_0\right)   G\left(1+\epsilon\sigma+\epsilon'\theta_1+\epsilon\epsilon'
     \theta_{\infty}\right)}
\prod_{\epsilon=\pm}
\frac{G(1+2\epsilon\sigma)}{G(1+2\epsilon\overline\sigma)}
\prod_{k=1}^4\frac{\hat{G}(\varsigma +\nu_k)}{ \hat{G}(\varsigma +\lambda_k)},
\end{equation}
 where $G\lb z\rb$ denotes the Barnes G-function, $\hat{G}\lb z\rb = \ds\frac{G\lb 1+z\rb}{G\lb 1-z\rb}$,
 the parameters $\nu_{1\ldots 4}$ and $\lambda_{1\ldots 4}$ are defined by
\beq
  \begin{aligned}\label{nus}
    \begin{array}{ll}
    \nu_1=\sigma+\theta_0+\theta_t, & \qquad\lambda_1=\theta_0+\theta_t+\theta_1+\theta_{\infty},\\
    \nu_2=\sigma+\theta_1+\theta_{\infty},& \qquad\lambda_2=\sigma+\overline\sigma+\theta_0+\theta_1, \\
    \nu_3=\overline\sigma+\theta_0+\theta_{\infty},& \qquad\lambda_3=\sigma+\overline\sigma+\theta_t+\theta_{\infty},\\
    \nu_4=\overline\sigma+\theta_t+\theta_1,& \qquad\lambda_4=0,\\
 \end{array}
    \end{aligned}
 \eeq
 and the quantity $\varsigma$ is determined by
 \beq\label{eqvarsigma}
 e^{2\pi i \varsigma}=\frac{2\cos2\pi\lb \sigma-\overline\sigma\rb 
  -2\cos2\pi\lb\theta_0+\theta_1\rb
  -2\cos2\pi\lb\theta_{\infty}+\theta_t\rb
   +p_{01}}{\sum_{k=1}^4\lb e^{2\pi i \lb\nu_{\Sigma}-\nu_k\rb}
   - e^{2\pi i \lb\nu_{\Sigma}-\lambda_k\rb}\rb},
 \eeq
 with $2\nu_{\Sigma}=\sum_{k=1}^4\nu_k=\sum_{k=1}^4\lambda_k$.
 \end{customthm}
 \noindent The choice of solution of (\ref{eqvarsigma})  for $\varsigma$ is not important, as the expression on the right of (\ref{p6const}) is invariant under integer shifts $\varsigma\mapsto\varsigma+1$. 
 The formula (\ref{p6const}) has been previously conjectured in \cite{ILT13} using a relation between Painlev\'e VI tau function and conformal blocks of the Virasoro algebra.
 
 We now move to the  tau function of the Painlev\'e II equation. Here, one has to deal with non-Fuchsian isomonodromic deformation of the $2\times 2$ linear system with a single irregular singular point 
 of Poincare rank $3$ located at infinity, 
 \begin{equation}\label{PIIinsys0}
\frac{d\Phi}{d z}=A\left(z\right)\Phi,\quad A\left(z\right) = A_{-3}z^2 + A_{-2}z+ A_{-1}.
\end{equation}
We shall also impose a symmetry condition,
$A\lb z\rb = -\sigma_2 A\lb -z\rb\sigma_2$, with 
$\sigma_2=\begin{pmatrix}0& -i\cr i & 0\end{pmatrix}$.
With the help of simple gauge and affine transformations, the system  (\ref{PIIinsys0}) can then be reduced to the following normal form:
\begin{equation}\label{PIIinsys2}
\frac{d\Phi}{dz} = \begin{pmatrix}-4iz^2 -it -2iu^2 & 4izu -v\cr 
-4izu -v &4iz^2 +it +2iu^2\end{pmatrix}\Phi.
\end{equation}
Here $u$, $v$ and $t$ are complex parameters playing the role of coordinates on the space ${\mathcal A}$. In this case,
$$
 {\mathcal A} = \left\{\lb u,v, t\rb\right\},\qquad 
 \mathcal T = \left\{t \in \Cb\right\}.
$$
 The space of monodromy data $\mathcal M$ can  
 be identified with the collection of the triples 
 $$
 \mathcal M_{\mathrm{PII}} = \left\{s=\lb s_1, s_2, s_3\rb\in\Cb^3 : s_1 -s_2 +s_3 + s_1s_2s_3 = 0\right\},
 $$
 where $s_1$, $s_2$, $s_3$ parameterize six Stokes matrices of the fundamental solution $\Phi\lb z\rb$ of (\ref{PIIinsys0}); the details are given in Subsection \ref{subsecPII}. Monodromy preserving deformations of the system (\ref{PIIinsys0}) are described by the equations \cite{FN1}  $v=2u_t$ and
 \begin{equation}\label{PIIin}
 u_{tt} = 2u^3+ tu.
 \end{equation}
 The latter equation is a particular case of the Painlev\'e II equation. The corresponding tau function is determined
 by the relation
 $$
 d\ln\tau = \omega_{\mathrm{JMU}} = (u_t^2 - u^4 -tu^2)dt \equiv Hdt,
 $$
 where $H= \frac{v^2}{4} -tu^2 -u^4$ is  the Hamiltonian of (\ref{PIIin}), considered  as a dynamical system
 on the $u,v$ phase plane with respect to the symplectic form $\Omega = dv\wedge du$.
 
 We are concerned with the two-parameter family of generic solutions of the second Painlev\'e equation~(\ref{PIIin}). It is specified by the following conditions on the monodromy data $s\in\mathcal M_{\mathrm{PII}}$: 
 \begin{subequations}
\label{genmonIN}
 \begin{gather}\label{m1IN}
 s_1s_3\ne 1,\qquad \arg\lb 1-s_1s_3\rb\in \lb-\pi,\pi\rb,\\
\label{m2IN}
s_2\notin \mathbb R,\qquad
 \arg\lb i\sigma s_2\rb\in \lb{-\textstyle\frac{\pi}{2}},\textstyle\frac{\pi}{2}\rb,\qquad \sigma: = \operatorname{sgn}\Re \lb is_2\rb =\pm 1.
 \end{gather}
\end{subequations} 
The condition  (\ref{m1IN}) ensures that the solution $u(t)$ is smooth as $t \to -\infty$ while condition
(\ref{m2IN}) guarantees  its smoothness as $t \to +\infty$. The respective asymptotics and their explicit monodromy parametrization are
due to A. Kapaev \cite{Kap} (see also \cite{IN}, \cite{DZ1}, and \cite{FIKN}). They are presented in detail in Subsection \ref{subsectauPII},
see equations (\ref{atmininfty})--(\ref{inftyparam}).
This asymptotics in turn implies the asymptotics for the tau function,
\beq\label{f44IN}
\tau(t)\simeq
\begin{cases} \ds \mathcal C_-e^{-\frac{4i\mu}{3}\lb-t\rb^{\frac{3}{2}}}\lb -t\rb^{-\frac{3\mu^2}{2}}\bigl[1+o\lb 1\rb\bigr],\qquad & \text{as}\; t\to-\infty,\vspace{0.1cm}\\
\ds \mathcal C_+e^{\frac{t^3}{12}+\frac{2i\sqrt 2}{3}\nu t^{\frac{3}{2}}}t^{-\frac{3\nu^2}{4}-\frac{1}{8}}\bigl[1+o\lb 1\rb\bigr],\qquad & 
\text{as}\; t\to+\infty,
 \end{cases}
\eeq
where
\beq\label{mprp2}
\mu=-\frac{\ln\lb1-s_1 s_3\rb}{2\pi i},\qquad \nu=\frac{\ln\lb i\sigma s_2\rb}{\pi i},\qquad \sigma=\operatorname{sgn}\Re\lb is_2\rb.
\eeq
Our goal is to find the ratio 
$\Upsilon\lb s\rb:=\frac{\mathcal C_+}{\mathcal C_-}$
in terms of monodromy data $s\in \mathcal M_{\mathrm{PII}}$. Here is the answer.
\begin{customthm}{B} \label{theoP2IN} Under the genericity assumptions  \eqref{genmonIN} on the monodromy data, the connection coefficient
 	$\Upsilon\lb s\rb$ for the Painlev\'e II tau function is given by
 	\beq\label{finalp2conIN}
 	\begin{gathered}
 	\frac{\mathcal C_+}{\mathcal C_-}\lb s\rb=2^{\frac1{24}} e^{\zeta'\lb-1\rb+\frac{i\pi}{24}}\,  2^{{3}\mu^2-\frac{7\nu^2}{4}}
 	(2\pi)^{{-\mu-\frac{\nu}{2}}}e^{\frac{\pi i}{4}\lb\eta^2+2\mu^2+2\eta\nu-8\mu\eta\rb } 
 	\frac{G\lb 1-\nu\rb\hat G\lb \eta\rb}{G^2\lb 1-\mu\rb\hat{G}^2\lb\frac{\eta-\nu}{2}\rb}\,,
 	\end{gathered}
 	\eeq
 	where $\mu$, $\nu$, $\sigma$ and $\eta$ are related to Stokes parameters
 	$s_1$, $s_2$, $s_3$ by (\ref{mprp2}) and $e^{i\pi \eta}= -i\sigma s_3^{-1}$, and  $\hat G\lb z\rb  = \ds\frac{G\lb 1+z\rb}{G\lb 1-z\rb}$ is the same combination of Barnes $G$-functions as in Theorem~\ref{thp6}.
 \end{customthm}
 \noindent Here again, the choice of solution for $\eta$ has no importance since the expression on the right of (\ref{finalp2conIN}) is invariant under shifts $\eta\mapsto\eta+2$.

\subsection{Outline of the paper}

Let us now describe the organization of the paper. The next two sections are devoted to the general Fuchsian case of the system (\ref{gensys0}).
The main result of Section~\ref{sec_FS} is Proposition \ref{propomega}. In this proposition we perform the first step of the program: that is, we   construct 
a 1-form $\omega$ which extends the Fuchsian Jimbo-Miwa-Ueno form to the space $ \widetilde{\mathcal T}\times \mathcal{M}$ 
and whose exterior differential, $\Omega := d\omega$, does not depend on the isomonodromic times (in
the Fuchsian case they are just the positions of singular points $a_1, \ldots, a_n$). The form $\hat{\omega}$ is then defined  formally as $\hat\omega := \omega - \omega_0$,
where $\omega_0$ is a 1-form on  ${\mathcal M}$ such that its differential is again $\Omega$. According to the
 scheme outlined above, the form  $\omega_0$ should be determined by analyzing the asymptotics of the forms $\omega$ and~$\Omega$.
The relevant analysis is carried out in Section~\ref{sec_fourpoint} for the case of 4-point Fuchsian systems. It is based
on the Riemann-Hilbert method. In Subsection~\ref{subsec_RHP}, the Riemann-Hilbert problem which represents the inverse monodromy 
map (\ref{mmap}) for the 4-point system is formulated. In Subsections~\ref{subsec_param} and~\ref{subsec_GLA}, its asymptotic
solution is constructed  in terms of  solutions of certain 3-point Fuchsian systems.
The result is used in Subsection~\ref{subsecas} to derive the asymptotics of the forms $\omega$ and $\Omega$ and hence determine
the forms $\omega_0$ (see Lemma~\ref{asmonodromypart}) and  $\hat{\omega}$ (Proposition \ref{propom}).
Solution of the constant problem for the tau function of 4-point  Fuchsian systems thereby reduces to 3-point  inverse monodromy
problems.  Explicit solution of the latter for general  Fuchsian system of rank $N>2$ is not known and therefore Proposition \ref{propom}
 is the best we could do for
the generic 4-point tau function.  For $N=2$, however, the 3-point systems may be solved in terms of contour integrals, i.e.  in terms of   hypergeometric functions. This  yields explicit
solutions of the corresponding  3-point inverse monodromy problems in terms of gamma functions and an explicit solution of the constant problem for the 4-point isomonodromic tau function in terms of Barnes $G$-functions. Computational details are presented in Subsections~\ref{subsmon}--\ref{secconcof}. As it has  already been explained above, in the 4-point $N=2$ Fuchsian case, the general monodromy preserving deformation equation (\ref{isomeq0}) reduces to a single scalar nonlinear $2$nd order ODE --- the sixth Painlev\'e equation (\ref{PVI0}).  Hence Subsections
~\ref{subsmon}--\ref{secconcof}  provide us with the proof of the first of our main results (Theorem \ref{thp6}). An important
difference as compared to  Painlev\'e III ($D_8$) considered in \cite{IP} is that the form $\omega$ in the Painlev\'e~VI case can not be
completely localized, i.e. it can not be expressed exclusively via the solution $u(t)$ of Painlev\'e~VI. This means that one can not just use
already known asymptotics and connection formulae for the Painlev\'e function as in \cite{IP}. One needs to use the
complete information on the  asymptotic behavior of the solution of the corresponding Riemann-Hilbert problem.

The fourth section of the paper is concerned with the non-Fuchsian case. We begin with a detailed description  of
monodromy data for non-Fuchsian systems, that is, general systems (\ref{gensys0}) which allow for the 
presence of irregular singular points. We then introduce
the form $\omega$ in this general case, and check that its exterior differential is a
2-form on $\mathcal M$ independent of isomonodromic times ${\mathcal T}$ (Theorem~\ref{omirrth}). We do not pursue the general case
further. Instead,  in  Subsection~\ref{subsecPII}, we move  to a nontrivial example of  a $2\times2$ non-Fuchsian system  whose isomonodromic deformations are described  by the second Painlev\'e equation. The evaluation of the asymptotics
of the form $\omega$, its further transformation to a closed form $\hat{\omega}$ and the evaluation of the connection constant
for the corresponding Painlev\'e II tau function up to the numerical factor are done in Subsections~\ref{subsectauPII} and \ref{subsecconPII}. This time the constant
in question is the ratio of constant factors corresponding to the asymptotic behaviors of the tau function
$\tau(t)$ along the rays $t \to +\infty$ and $t \to -\infty$.  Unlike in the Painlev\'e III case considered in \cite{IP}, the evaluation of the remaining numerical factor is  not trivial, and it is given in Subsection
\ref{numcoef}. The final formula for the PII constant is presented in Theorem \ref{theoP2IN}  above.
 
\subsection{Remarks} 

There is a very interesting 
additional observation related to the form $\omega$ in the  Painlev\'e II case. Equation (\ref{actiontau}) indicates that the 1-form $\omega$, up to addition of an explicit total differential, is an extension to the space 
$\widetilde{\mathcal T}\times \mathcal{M}$ of the  classical action differential.
The same fact has already been noticed in  \cite{IP} in the case of Painlev\'e III ($D_8$) equation.
 We conjecture that this relation of the extended  tau function (\ref{exttau0})  to the classical action 
is a general fact of the monodromy theory of linear systems. This conjecture is closely related to another 
observation that can be made about  $\omega$. As follows from Remark~\ref{rmksymp} and the calculations in
Subsection 4.2, the 2-form  $\Omega = d\omega$ is nothing but (up to a numerical coefficient and restriction to symplectic leaves) 
the symplectic form on monodromy manifolds of Painlev\'e VI and II, respectively. The same fact has also been  observed in the case of Painlev\'e III ($D_8$) equation
in \cite{IP} and we again conjecture it to be  general. We intend to study this issue in
more detail in a future work.

We would like to close this introduction  with some historical remarks. Since its birth in 1980, the concept of  tau function  has been  playing an increasingly important role in the theory
of integrable systems and its numerous applications. Correlation functions of various exactly solvable
quantum mechanical and statistical models are tau functions associated to  special 
examples of the linear system (\ref{gensys0}). Partition functions of matrix models and 2D quantum gravity, the generating
functions in the intersection theory of moduli spaces of  algebraic curves 
are again special examples of tau functions. Yet more examples arise in the study of Hurwitz spaces and quantum cohomology.  The evaluation of constant terms in the asymptotics of these correlation, distribution and generating functions has always  been
a great analytic challenge.  The first rigorous solution of a constant problem for Painlev\'e equations (a special Painlev\'e~III transcendent
appearing in the Ising model) has been obtained in the work of Tracy \cite{T}.  Other constant problems have been studied in the works \cite{BT,BB,K,E,DIKZ,DIK,DKV,Liso11} and \cite{BBD,BBDI}.  
 
 The tau  functions that appear in the papers quoted above
correspond to very special families of Painlev\'e functions. The first results concerning the general two-parameter families of solutions of Painlev\'e equations have been obtained
only recently in \cite{ILT13,ILT14}. These works  are based
on  {\it conformal block representations}  of isomonodromic tau functions --- see \cite{GIL12,GIL13,ILTe}
and also \cite{BS,Gav,GM,Nagoya} for subsequent developments.  Although very powerful, the conformal block approach still has to be put on  rigorous ground. In this paper, we show that with the help of Riemann-Hilbert techniques the conjectural formula
of \cite{ILT13} for the constant factor in the asymptotics of the Painlev\'e VI tau function can be proven. 
In a sequel, we plan to understand within the Riemann-Hilbert formalism the other key results provided by conformal field theory; 
first of all the novel series representations for isomonodromic tau functions.

 \section{Fuchsian systems\label{sec_FS}}   
     Let us start by fixing the notations. They are slightly different from the ones used in the Introduction. Indeed,in this section we are dealing exclusively 
     with the Fuchsian systems, and the notations can be naturally simplified,e.g., no need for double subscripts for $\Theta_{\nu,0}$. We are going to consider monodromy preserving deformations of rank~$N$ Fuchsian systems with $n+1$ regular singular points $a_1,\ldots,a_n,a_{n+1}= \infty$ on~$\mathbb{P}$:
     \beq\label{lax1fuchs}
     \partial_z\Phi=\sum_{\nu=1}^n \frac{A_{\nu}}{z-a_{\nu}}\, \Phi,\qquad A_{1},\ldots,A_n\in\mathfrak{sl}_N\left(\mathbb C\right).
     \eeq
     Define $A_{\infty}:=-\sum_{\nu=1}^n A_{\nu}$ and for $\nu=1,\ldots,n,\infty$ denote by $\theta_{\nu,k}$  the eigenvalues of $A_{\nu}$.
     
     \begin{ass}\label{assum1} All eigenvalues satisfy a non-resonance condition $\theta_{\nu,k}-\theta_{\nu,l}\notin \mathbb Z$ for $k\ne l$. It implies in particular that all $A_{\nu}$ are diagonalizable.
     \end{ass}

     Fix the diagonalizations of $A_{\nu}$ by introducing 
     the matrices  $G_{\nu}$ , $\Theta_{\nu}$ such that
     \beq\label{adiag}A_{\nu}=G_{\nu}\Theta_{\nu}G_{\nu}^{-1},\qquad \Theta_{\nu}=\operatorname{diag}
     \left\{\theta_{\nu,1},\ldots \theta_{\nu,N}\right\}.
     \eeq
    The choice of $G_{\nu}$ is not unique, as there remains an ambiguity of right multiplication by a diagonal matrix. Local behavior of the fundamental matrix solution near the singular points may  be written~as 
     \beq\label{locbeh}
     \begin{split}
     \Phi(z)=\,G_{\nu}\,&\left[\mathbf{1}+\sum_{m=1}^{\infty}
     g_{\nu,m}\left(z-a_{\nu}\right)^m\right]
     \left(z-a_{\nu}\right)^{\Theta_{\nu}} C_{\nu},\quad |z-a_{\nu}| < r,\\
     \Phi(z)=G_{\infty}&\left[\mathbf{1}+\sum_{m=1}^{\infty}
      g_{\infty,m}z^{-m}\right]z^{-\Theta_{\infty}}C_{\infty}, \quad |z| > R .
      \end{split}
     \eeq     
    Connection matrices $C_{\nu}$ are determined by the Fuchsian system, the initial conditions and the choice of diagonalisations. They also depend on the choice of branch cuts making the solution single-valued, and on the determination of fractional powers $\left(z-a_{\nu}\right)^{\Theta_{\nu}}$.     
    The series (\ref{locbeh}) have non-zero radii of convergence, and their coefficients $g_{\nu,m}$ can be calculated recursively from (\ref{lax1fuchs}). In particular, $g_{\nu,1}$ may be found from
    \ben
     g_{\nu,1} +\left[g_{\nu,1},\Theta_{\nu}\right] = 
    \sum_{\mu\ne\nu}
    \frac{G_{\nu}^{-1}A_{\mu}G_{\nu}}{a_{\nu}-a_{\mu}}.
    \ebn
        
     The idea of isomonodromic deformation is to vary  $a_{\nu}$ and $A_{\nu}$ (with $\nu=1,\ldots,n$) simultaneously keeping constant 
     the local monodromy exponents $\Theta_{\nu}$ and the connection matrices $C_{\nu}$. 
     The matrix $G_{\infty}$ will also be  fixed.
     The singularity at $\infty$ then plays the role of a normalization point of the fundamental matrix solution $\Phi\left(z\right)$.    
     The product $\partial_{a_{\nu}}\Phi\cdot\Phi^{-1}$ is a meromorphic matrix function on $\mathbb{P}$ with poles only possible at $a_1,\ldots,a_n,\infty$. Local analysis shows that
     \beq\label{lax2fuchs}\qquad \partial_{a_{\nu}}\Phi=-\frac{A_{\nu}}{
     z-a_{\nu}}\,\Phi,\qquad \nu=1,\ldots,n.
     \eeq
     The compatibility of (\ref{lax1fuchs}) and (\ref{lax2fuchs})  yields the classical Schlesinger system of nonlinear matrix PDEs:
     \beq\label{schlA}
     \begin{split}
     \partial_{a_{\mu}}A_{\nu}=\frac{[A_{\mu},A_{\nu}]}{
     a_{\mu}-a_{\nu}},\qquad &\mu\neq \nu,\\
     \partial_{a_{\nu}}A_{\nu}=-\sum_{\mu\neq \nu}^n\frac{[A_{\mu},A_{\nu}]}{a_{\mu}-a_{\nu}},\qquad &\nu=1,\ldots,n.
     \end{split}
     \eeq
   A slightly more refined problem is to describe the isomonodromic evolution of diagonalization matrices $G_{\nu}$. It can be addressed using the same linear equations (\ref{lax1fuchs}), (\ref{lax2fuchs}). The result is
     \beq\label{schlG}
     \begin{split}
     \partial_{a_{\mu}}G_{\nu}\cdot G_{\nu}^{-1}&=\quad\frac{A_{\mu}}{
     a_{\mu}-a_{\nu}},\qquad \quad\;\;\mu\neq \nu,\\
     \partial_{a_{\nu}}G_{\nu}\cdot G_{\nu}^{-1}&=-\sum_{\mu\neq \nu}^n\frac{A_{\mu}}{a_{\mu}-a_{\nu}},\qquad \nu=1,\ldots,n.
     \end{split}
     \eeq
  \begin{defin}
  We denote by $\mathcal M= \left(\mathbb C^{\times}\right)^{\left(N-1\right)\left(n+1\right)}\times
  \left(GL_N\left(\mathbb C\right)\right)^{n+1}\times  GL_N\left(\mathbb C\right)$ the space of monodromy data parameterizing local monodromy exponents $\Theta_{\nu}$,
    connection matrices $C_{\nu}$ and normalization matrix $G_{\infty}$. We also introduce  the space of isomonodromic times $\mathcal{T}=\left\{\left(a_1,\ldots,a_n\right)\in\mathbb C^n\,|\,a_{\mu}\neq a_{\nu}\right\}$ and denote by
    $\widetilde{\mathcal T}$ its universal cover.
  \end{defin}        
 For any point in $\mathcal M$ the inverse  monodromy problem for system  (\ref{lax1fuchs}) is locally solvable. This means 
 that for all $(a_1, ...., a_n)$ in an open set  in $\mathcal{T}$ there exists  a unique invertible matrix $\Phi\left(z\right)$ holomorphic on the universal cover of $\mathbb P\backslash\{ a_1,\ldots,a_{n+1}\}$ with singular behavior (\ref{locbeh}) at the branch points (see e.g. \cite{Palmer,AB}). This in turn uniquely determines the local solution
 $\left\{A_1,\ldots,A_n\right\}$,
  $$
 A_1 = A_1(a_1, ..., a_n; M),\,  \ldots\, , A_n = A_n(a_1, ..., a_n; M),
 $$
 of the corresponding Schlesinger system. Solving this system thus amounts to constructing an inverse of the Riemann-Hilbert map
 \ben
 \mathcal{RH}:\;\left\{G_1,\ldots,G_n\right\}
 \mapsto \left\{C_1,\ldots, C_n\right\}
 \ebn  
 for given $a_{\nu}$, $\Theta_{\nu}$, $G_{\infty}$ and $C_{\infty}$.  According to the Malgrange-Miwa theorem mentioned in the introduction, the 
 local solution $\left\{A_1,\ldots,A_n\right\}$,
as well as the solution of the inverse monodromy problem, in fact, admit  meromorphic  continuations  to  the whole $\widetilde{\mathcal{T}}$, and the singularities are located at the zeros of the corresponding Jimbo-Miwa-Ueno tau function  (see again \cite{Malgrange,Miwa,Palmer,AB,Bertola}).
 Also note that the solution of the Schlesinger system remains invariant under the right action $C_{\nu}\mapsto C_{\nu}H$ with $H\in GL_N\left(\mathbb C\right)$ and $\nu=1,\ldots,n,\infty$. Gauge transformations $G_{\nu}\mapsto H G_{\nu}$ (with fixed $C_{\infty}$) preserve the connection matrices $C_1,\ldots, C_n$.
 \begin{prop}\label{propomega}
   Let  $\omega$ be a (possibly meromorphic in $a_{\nu}$) 1-form on $\widetilde{\mathcal{T}}\times\mathcal  M$
  locally defined by
  \beq\label{taudef2}
  \omega=\sum_{\nu<\mu}^n\operatorname{Tr}A_{\mu}A_{\nu}\,d\ln\left(
  a_{\mu}-a_{\nu}\right)+\sum_{\nu=1,\ldots,n,\infty}\operatorname{Tr}\left(\Theta_{\nu}G_{\nu}^{-1}d_{\mathcal M}G_{\nu}\right),
  \eeq
  where $d_{\mathcal{M}}$ denotes the differential with respect to monodromy data. Its exterior differential $\Omega:=d\omega$ is a closed 2-form
  on $\mathcal M$ independent of $a_1,\ldots,a_n$.
 \end{prop}
 \pf
  Straightforward calculation using  Schlesinger equations (\ref{schlA})  shows that $\Omega\left(\partial_{a_{\mu}},\partial_{a_{\nu}}\right)=0$.  Let $M$ be a local coordinate on $\mathcal M$. It can be  deduced from (\ref{schlG}) that
   \beq\label{auxdiffG}
   \begin{split}
   \partial_{a_{\mu}}\left(G_{\nu}^{-1}\partial_{M} G_{\nu}\right)&=\quad\frac{G_{\nu}^{-1}\left(\partial_M A_{\mu}\right) G_{\nu}}{
        a_{\mu}-a_{\nu}},\qquad\qquad\quad\;\; \mu\neq \nu,\\
    \partial_{a_{\nu}}\left(G_{\nu}^{-1}\partial_{M} G_{\nu}\right)&=-\sum_{\mu\neq \nu}\frac{G_{\nu}^{-1}\left(\partial_M A_{\mu}\right)G_{\nu}}{a_{\mu}-a_{\nu}},\qquad\qquad \nu=1,\ldots,n,
    \end{split}
    \eeq 
 which in turn implies that $\Omega\left(\partial_{a_{\mu}},\partial_{M}\right)=0$. Since $\Omega$ is a total differential, it follows that $d_{\mathcal{T}}\Omega\left(\partial_{M_1},\partial_{M_2}\right)$ vanishes for any pair $M_1,M_2$ of monodromy parameters.

 The last assertion can also be checked directly. Indeed, we have
  \begin{align}
  \label{deromega2}
  \Omega\left(\partial_{M_2},\partial_{M_1}\right)&=
  \sum_{\nu} \operatorname{Tr}\left(\Theta_{\nu}\left[
   G_{\nu}^{-1} \partial_{M_1}G_{\nu},
   G_{\nu}^{-1} \partial_{M_2}G_{\nu}\right]+
   \partial_{M_2}\Theta_{\nu}\,G_{\nu}^{-1}\partial_{M_1}G_{\nu}
   -\partial_{M_1}\Theta_{\nu}\,G_{\nu}^{-1}\partial_{M_2}G_{\nu}
   \right) .
  \end{align}    
  The relations (\ref{auxdiffG}) may be repackaged into a more compact expression
  \beq\label{dergcomp}
  d_{\mathcal T}\left(G_{\nu}^{-1}\partial_{M} G_{\nu}\right)=\sum_{\mu\neq \nu}
  G_{\nu}^{-1}\left(\partial_M A_{\mu}\right) G_{\nu}\,d\ln\left(a_{\mu}-a_{\nu}\right),
  \eeq  
  which can be used to differentiate
  $\Omega$. For example:
  \begin{align*}
  & \sum_{\nu} \operatorname{Tr}\left(\Theta_{\nu}\left[
    d_{\mathcal T}\left( G_{\nu}^{-1} \partial_{M_1}G_{\nu}\right),
    G_{\nu}^{-1} \partial_{M_2}G_{\nu}\right]\right)=\\=&
    \sum_{\nu} \sum_{\mu\neq \nu}\operatorname{Tr}\left(\Theta_{\nu}\left[
    G_{\nu}^{-1}\left(\partial_{M_1} A_{\mu}\right) G_{\nu},G_{\nu}^{-1} \partial_{M_2}G_{\nu}\right]\right)
    d\ln\left(a_{\mu}-a_{\nu}\right)=\\
    =&\sum_{\nu} \sum_{\mu\neq \nu}\operatorname{Tr}\left(A_{\nu}\left[
        \partial_{M_1} A_{\mu}, \partial_{M_2}G_{\nu}\cdot G_{\nu}^{-1}\right]\right)
        d\ln\left(a_{\mu}-a_{\nu}\right)=\\
   =&\sum_{\nu} \sum_{\mu\neq \nu}
   \operatorname{Tr}\left(\partial_{M_1} A_{\mu}\left[
     \partial_{M_2}G_{\nu}\cdot G_{\nu}^{-1},A_{\nu}\right]\right)
                d\ln\left(a_{\mu}-a_{\nu}\right)=\\
 =&\sum_{\nu} \sum_{\mu\neq \nu}\operatorname{Tr}\left(\partial_{M_1} A_{\mu}\cdot\partial_{M_2} A_{\nu}-\partial_{M_1} A_{\mu}\cdot G_{\nu}\partial_{M_2}\Theta_{\nu}\,G_{\nu}^{-1}\right)d\ln\left(a_{\mu}-a_{\nu}\right). 
  \end{align*}
 Similarly,
 \ben
 d_{\mathcal T}
 \sum_{\nu}
 \operatorname{Tr}\left(\partial_{M_2}\Theta_{\nu}\,
 G_{\nu}^{-1}\partial_{M_1}G_{\nu}\right)=\sum_{\nu}\sum_{\mu\ne\nu}\operatorname{Tr}
 \left(\partial_{M_2}\Theta_{\nu}\,
  G_{\nu}^{-1}\left(\partial_{M_1} A_{\mu}\right) G_{\nu}\right)
  \,d\ln\left(a_{\mu}-a_{\nu}\right)
 \ebn 
  Since the sum $\sum_{\nu} \sum_{\mu\neq \nu}\operatorname{Tr}\left(\partial_{M_1} A_{\mu}\cdot\partial_{M_2} A_{\nu}\right)d\ln\left(a_{\mu}-a_{\nu}\right)$ of the last two expressions is symmetric with respect to the exchange
  $M_1\leftrightarrow M_2$, we finally  obtain 
  the expec\-ted result $d_{\mathcal T}\Omega
  \left(\partial_{M_2},\partial_{M_1}\right)=0$. 
 \epf

     The first term in (\ref{taudef2}) is the usual definition of the Jimbo-Miwa-Ueno tau function \cite{JMU}, while the second sum incorporates its dependence on monodromy. 
     
  \begin{defin}   
    Let $\omega_0\in\Lambda^1\left(\mathcal M\right)$ be a 1-form such that $d\omega_0=\Omega$.  The extended isomonodromic tau function
    $ \tau_{\omega_0}:\widetilde{\mathcal{T}}\times \mathcal{M}\to \mathbb C$ is defined  by
    \ben
    d\ln\tau_{\omega_0}=\omega-\omega_0 \equiv \hat{\omega}
    \ebn  
  \end{defin}        
         \noindent In the next sections, this construction is explicitly carried out  in the case 
          $n=3$, $N=2$ corresponding to Painlev\'e~VI equation.
      
   \begin{rmk}\label{remarkgauge} Left multiplication of all $G_{\nu}$  by a matrix $H\in GL_N\left(\mathbb C\right)$ possibly depending on monodromy parameters leads to transformation
   $A_{\nu}\mapsto HA_{\nu}H^{-1}$. It obviously preserves the first term in (\ref{taudef2}). Since 
   \ben
   \sum_{\nu}\operatorname{Tr}
   \left(G_{\nu}\Theta_{\nu}G_{\nu}^{-1}H^{-1}d_{\mathcal M}H\right)
   =\operatorname{Tr}
    \left(\sum_{\nu}A_{\nu}
    H^{-1}d_{\mathcal M}H\right)=0
   \ebn
   due to the relation $\sum_{\nu =1}^{n}A_{\nu} + A_{\infty} =0$, the second term also remains invariant. Hence the form $\omega$ is preserved
   by the gauge transformations.
   \end{rmk}  
   
      \begin{rmk} Right $GL_N\left(\mathbb C\right)$-action $C_{\nu}\mapsto C_{\nu}H$ does not affect the solution of the Schlesinger system. Therefore 
      \ben
      d\ln\tau^H_{\omega_0}-d\ln\tau_{\omega_0}=\omega_0-\omega^H_0\in d\left(\Lambda^0\left(\mathcal M\right)\right).
      \ebn
      The corresponding tau functions thus necessarily coincide up to a factor depending only on monodromy data (but not on the isomonodromic times!). In other words, the tau function depends in a nontrivial way only on the conjugacy class of monodromy.
      \end{rmk}

 \section{Four-point tau function\label{sec_fourpoint}}  
 \subsection{Riemann-Hilbert problem\label{subsec_RHP}}  
 It is always possible to explicitly integrate the isomonodromic flows associated to global conformal transformations. This allows to fix $3$ of the singular points at $0$, $1$, and $\infty$. The simplest nontrivial case of isomonodromy equations therefore corresponds to $n=3$ ($4$ regular singularities). The position of the $4$th singular point is the only remaining time variable, to be denoted by $t$. The Fuchsian system (\ref{lax1fuchs}) then acquires the form
 \beq\label{fuchspvi}
 \partial_z\Phi=A\left(z\right) \Phi,\qquad 
 A\left(z\right)
  = \frac{A_{0}}{z}+ \frac{A_{t}}{z-t}+
    \frac{A_{1}}{z-1},
 \eeq
 and the Schlesinger system consists of two matrix ODEs
 \ben
 \frac{dA_0}{dt}=\frac{\left[A_t,A_0\right]}{t},\qquad 
 \frac{dA_1}{dt}=\frac{\left[A_t,A_1\right]}{t-1},
 \ebn
 where $A_{0,t,1}$ satisfy the constraint $ A_0+A_t+A_1=-A_{\infty}$.
 The $1$-form $\omega$ from Proposition~\ref{propomega} becomes
 \beq\label{tauextpvi}
 \omega=\mathcal P\,dt+\sum_{\nu=0,t,1,\infty}
 \operatorname{Tr}\left( \Theta_{\nu} G_{\nu}^{-1}d_{\mathcal{M}}
 G_{\nu}\right),
 \eeq
 where the time part of $\omega$ is defined by
 \ben
 \mathcal{P}=\frac{\operatorname{Tr}A_t A_0}{t}+
  \frac{\operatorname{Tr}A_t A_1}{t-1}=\frac12\operatorname{res}_{z=t}
  \operatorname{Tr}A^2\left(z\right).
 \ebn
 
 One of our tasks is to compute the exterior differential $\Omega=d\omega$. It was already shown to be independent of $t$, therefore it suffices to determine the asymptotics of $\omega$ when two singular points of the Fuchsian system (\ref{fuchspvi}) collide. We will therefore attempt to analyze the  behavior of the fundamental matrix solution $\Phi\left(z\right)$ as $t\to 0$, extract from it the asymptotics of $G_{0,t,1}\left(t\right)$ and  use the latter to calculate $\Omega$.
     
 The most convenient framework for realization of this plan is provided by the Riemann-Hilbert method. Instead of working with a Fuchsian system, $\Phi\left(z\right)$ may be related to the unique solution
 $\Psi\left(z\right)$ of the following Riemann-Hilbert problem (RHP):
 \begin{itemize}
 \item[$\clubsuit$]
 Given an oriented contour $\Gamma_{\Psi}\subset\mathbb  C$ and a prescribed jump matrix $J_{\Psi}:\Gamma_{\Psi}\to GL_N\left(\mathbb{C}\right)$, find a holomorphic matrix $\Psi:\mathbb{C}\backslash\Gamma_{\Psi} \to GL_N\left(\mathbb{C}\right)$ such that its boundary values on $\Gamma_{\Psi}$ satisfy $\Psi_+\left(z\right)=\Psi_-\left(z\right)J_{\Psi}\left(z\right)$  and ${\Psi\left(\infty\right)=G_{\infty}}$. 
 \end{itemize}
 The contour $\Gamma_{\Psi}$ consists of $4$ circles and $3$ segments\footnote{To avoid unnecessary complications it is  assumed that $t\in (0,1)$ and the segments belong to the real line.} represented by solid lines. The jump matrix $J_{\Psi}\left(z\right)$ is defined as  
 \beq\label{jumpspsi}
  \begin{gathered} 
 J_{\Psi}\left(z\right)\bigl|_{\ell_{0t}}=M_0^{-1},\qquad
   J_{\Psi}\left(z\right)\bigl|_{\ell_{t1}}=\left(M_t M_0\right)^{-1},\\
  J_{\Psi}\left(z\right)\bigl|_{\ell_{1\infty}} =\left(M_1M_tM_0\right)^{-1}=M_{\infty},\quad\qquad
  J_{\Psi}\left(z\right)\bigl|_{\gamma_{\infty}}=
  C_{\infty}^{-1}
  \left(-z\right)^{\Theta_{\infty}},\\
  J_{\Psi}\left(z\right)\bigl|_{\gamma_{0}}=C_{0}^{-1} \left(-z\right)^{-\Theta_{0}},\qquad 
 J_{\Psi}\left(z\right)\bigl|_{\gamma_{t}}=C_{t}^{-1} \left(t-z\right)^{-\Theta_{t}},\qquad
 J_{\Psi}\left(z\right)\bigl|_{\gamma_{1}}=C_{1}^{-1} \left(1-z\right)^{-\Theta_{1}},
 \end{gathered}
 \eeq
 where $M_{\nu}$ are counterclockwise monodromies of $\Phi\left(z\right)$ around ${\nu}$ with basepoint chosen on the negative real axis. Fractional powers  will always be understood in terms of their principal branches. Expressions for the connection matrices $C_{t}$, $C_1$ differ on the upper and lower halves of $\gamma_{t}$, $\gamma_1$. We have
 \ben
 \begin{gathered}
   M_0=C_{0}^{-1}e^{2\pi i \Theta_0}C_{0}=C_{t,+}^{-1}C_{t,-},\qquad M_tM_0=C_{t,+}^{-1}e^{2\pi i \Theta_t}C_{t,-}=C_{1,+}^{-1}C_{1,-},\\
 M_{\infty}^{-1}=C_{1,+}^{-1}e^{2\pi i \Theta_1}C_{1,-}=C_{\infty}^{-1}e^{-2\pi i\Theta_{\infty}}C_{\infty}.
 \end{gathered}
 \ebn
 where the indices $\pm$ correspond to $\Im z\gtrless 0$. Below the indices of this type are omitted whenever it may not lead to confusion.
  The solution $\Phi\left(z\right)$ of the Fuchsian system (\ref{fuchspvi}) is given by $\Psi\left(z\right)$  outside the circles $\gamma_{\nu}$ and by $\Psi\left(z\right)J^{-1}_{\Psi}\left(z\right)$ in their interior. We adopt the convention that the interior of the circle of  largest radius (here $\gamma_{\infty}$) is a disk around~$\infty$.

\begin{figure}
 \centering
 \includegraphics[width=6cm]{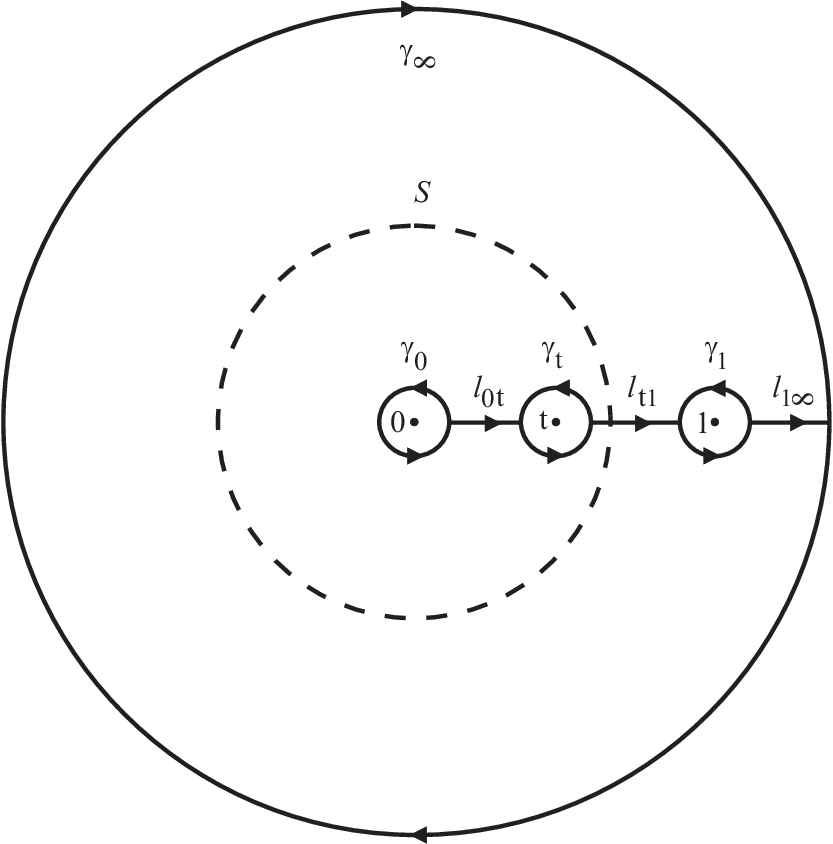}
 \caption{\label{fig1}
 Contour $\Gamma_{\Psi}$ of the Riemann-Hilbert problem for $\Psi\left(z\right)$}
 \end{figure}

 \begin{ass}\label{asssigma}
 The monodromy matrix $M_{t0}:=M_tM_0$, which is incidentally  the monodromy  along the circle $S$
 indicated in  Fig.~\ref{fig1}, is assumed to be diagonalizable. Fix a matrix $C_S\in GL_N\lb\Cb\rb$ such that
 \beq
 M_{t0}=C_S^{-1}e^{ 2\pi i \mathfrak S}C_S,\qquad \mathfrak S=\operatorname{diag}\left\{\sigma_1,\ldots,\sigma_N\right\}\in
 \mathfrak{sl}_N\left(\mathbb C\right).
 \eeq
 The logarithms $\sigma_k$ of the eigenvalues of $M_{t0}$ are assumed to satisfy the conditions  $\left|\Re \left(\sigma_j-\sigma_k\right)\right|< 1$ for $j,k=1,\ldots, N$. It is  furthermore assumed that all $\sigma_k$ are distinct.
 \end{ass}
 
 \begin{rmk}
 The condition $\left|\Re \left(\sigma_j-\sigma_k\right)\right|< 1$ involves almost no loss in generality. Indeed, for any choice of logarithms satisfying $\operatorname{Tr}\,\mathfrak S=0$ let $\sigma_{\text{max}}$ and $\sigma_{\text{min}}$ denote the eigenvalues of $\mathfrak S$ with maximal and minimal real part. If $\Re\left(\sigma_{\text{max}}-\sigma_{\text{min}}\right)>1$, then  replace $\sigma_{\text{max}}\mapsto \sigma_{\text{max}}-1$,  $\sigma_{\text{min}}\mapsto \sigma_{\text{min}}+1$ and iterate the procedure. After a finite number of steps we will reach the situation where $\left|\Re \left(\sigma_j-\sigma_k\right)\right|\leq 1$ for all $j,k=1,\ldots, N$. The values with ${\Re \left(\sigma_j-\sigma_k\right)=\pm1}$ are excluded to avoid some technicalities in what follows.
 \end{rmk}
 
 The RHP could also be formulated directly in terms of $\Phi\left(z\right)$. In this case the circles $\gamma_{\nu}$ are not needed, the contour $\Gamma_{\Phi}$ may be identified with the real line and the jump matrix is piecewise constant:
 \beq\label{jumpphi}
    J_{\Phi}\left(z\right)\bigl|_{ (0,t)}=M_0^{-1},\qquad
    J_{\Phi}\left(z\right)\bigl|_{ (t,1)}=\left(M_{t0}\right)^{-1},\qquad
  J_{\Phi}\left(z\right)\bigl|_{ (1,\infty)}=M_{\infty}.
  \eeq
  However, extra conditions (\ref{locbeh}) should be imposed on the behavior of $\Phi\left(z\right)$ in the vicinity of $0,t,1,\infty$ to ensure the uniqueness of solution. 

  The idea of the Riemann-Hilbert nonlinear steepest descend method of Deift-Zhou  \cite{DZ0} is to transform the original RHP
  into a sequence of simpler RHPs that can be solved exactly or asymptotically.  
  We start by constructing the matrices that mimic the monodromy properties of $\Psi\left(z\right)$ inside and outside an auxiliary circle
  $S=\left\{z\in\mathbb{C}:|z|=\delta \right\}$ of fixed finite radius $\delta\in (t,1)$ represented by dashed line in Fig.~\ref{fig1}. The  respective complex domains will be denoted by $S_i$ and $S_e$.

  \subsection{Parametrices\label{subsec_param}}  
\begin{figure}[h]
   \centering
   \includegraphics[height=7cm]{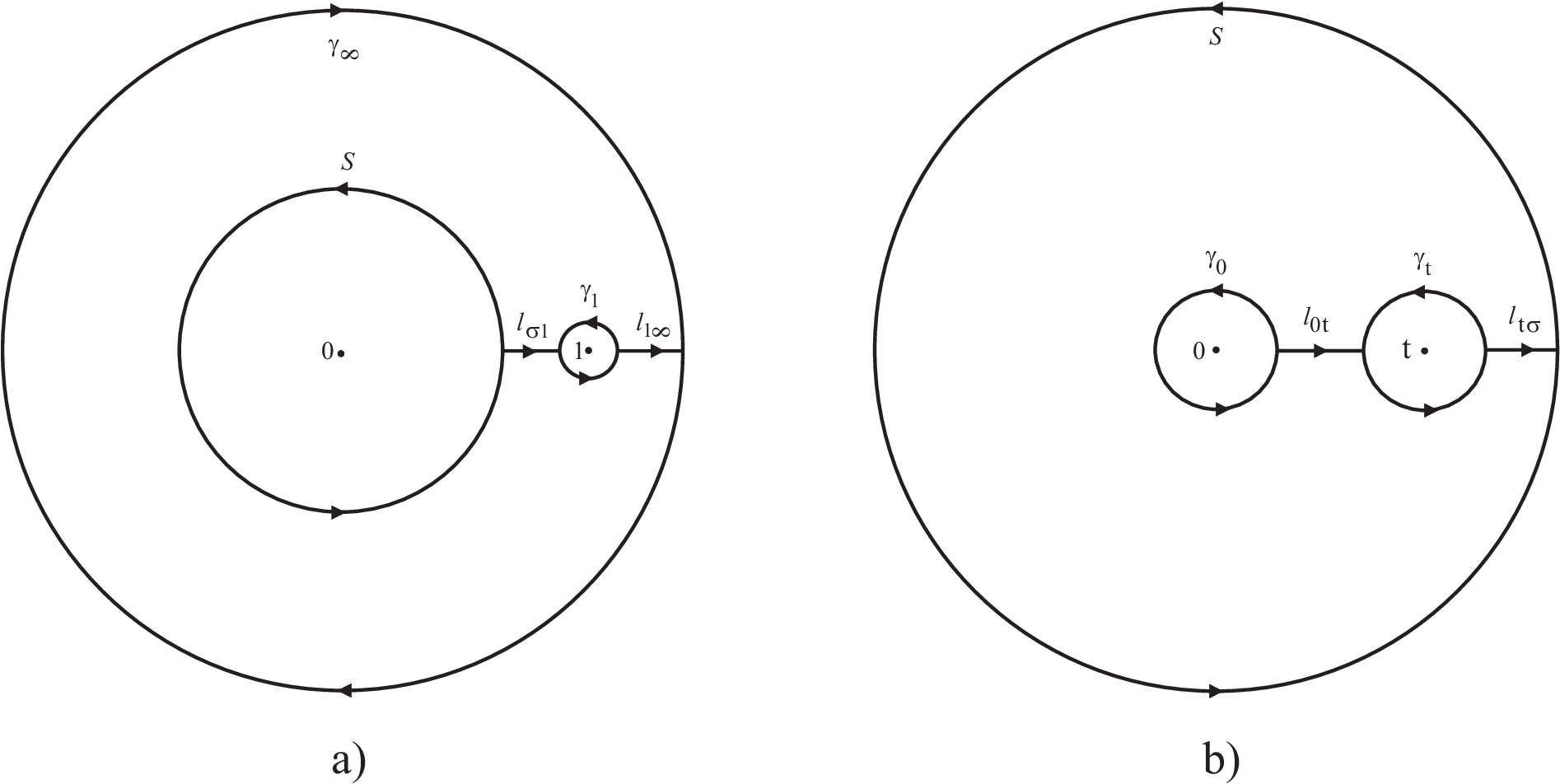}
   \caption{\label{fig2}
   Contours a) $\Gamma_{\Psi}^e$ and b) $\Gamma_{\Psi}^i$ for exterior and interior parametrix}
   \end{figure}

  Inside $S_e$, the matrix $\Psi\left(z\right)$ will be approximated by the solution $\Psi^{e}\left(z\right)$ of the Riemann-Hilbert problem with contour $\Gamma^e_{\Psi}$ shown in Fig.~\ref{fig2}a. The corresponding jump matrix $J^e_{\Psi}\left(z\right)$ is defined by
  \beq\label{mndrext}
  \begin{gathered}
  J^e_{\Psi}\left(z\right)\bigl|_{\gamma_{1}}=C_1^{-1}
  \left(1-z\right)^{-\Theta_1},\qquad
  J^e_{\Psi}\left(z\right)\bigl|_{\gamma_{\infty}}=C_{\infty}^{-1}
  \left(-z\right)^{\Theta_{\infty}},\qquad
  J^e_{\Psi}\left(z\right)\bigl|_{S}=C_S^{-1}
  \left(-z\right)^{-\mathfrak S},\\
  J^e_{\Psi}\left(z\right)\bigl|_{\ell_{\sigma 1}}=\left(M_{t0}\right)^{-1},\qquad
  J^e_{\Psi}\left(z\right)\bigl|_{\ell_{1\infty}}=M_{\infty}.
  \end{gathered}
  \eeq
  Together with the normalization $\Psi^{e}\left(0\right)=\mathbf{1}$, the jumps fix $\Psi^{e}\left(z\right)$ uniquely.   Outside the circles $S$, $\gamma_1$, $\gamma_{\infty}$ (i.e. to the right of the oriented circles  $S$, $\gamma_1$, $\gamma_{\infty}$)  this matrix can be expressed as $\Psi^{e}\left(z\right)=\Phi^e\left(z\right)C_S$ in terms of the solution $\Phi^e\left(z\right)$ of a Fuchsian system with $3$ regular singular points:
  \beq\label{parext}
  \partial_z \Phi^e=A^e\left(z\right)\Phi^e,\qquad 
  A^e\left(z\right)=\frac{A_0^e}{z}+\frac{A_1^e}{z-1},
  \eeq
  normalized as $\Phi^e(z)\simeq 
  \left(-z\right)^{\mathfrak S}$ as $ z \to 0$. In particular,  $A_0^e=\mathfrak{S}$ and the spectra of $A_1^{e}$, $A_{\infty}^{e}:=-A_0^e-A_1^e$ coincide with those of $A_1$ and $A_{\infty}$ of the $4$-point Fuchsian system (\ref{fuchspvi}). Local behavior of $\Phi^e\lb z\rb$ near the singular points $0,1,\infty$ is given by
  \begin{subequations}\label{monodE}
  \beq\label{localexpE}
  \Phi^e\lb z\rb=\begin{cases}
  G_0^e\lb z\rb \lb -z\rb^{\mathfrak S} & \qquad \text{as } z\to0,\\
  G_1^e\lb z\rb \lb 1-z\rb^{\Theta_1}C_1^e & \qquad \text{as } z\to1,\\
  G_{\infty}^e\lb z\rb \lb -z\rb^{-\Theta_{\infty}}C_{\infty}^e & \qquad \text{as } z\to\infty,
  \end{cases}
  \eeq
  where 
  \beq\label{GexpE}
  G_0^e\lb z\rb=\mathbf 1+\sum_{m=1}^{\infty}g_{0,m}^ez^m,\qquad
  G_1^e\lb z\rb=G_1^e\lb\mathbf 1+\sum_{m=1}^{\infty}g_{1,m}^e
  \lb z-1\rb^m\rb,\qquad
   G_{\infty}^e\lb z\rb=G_{\infty}^e\lb\mathbf 1+\sum_{m=1}^{\infty}g_{\infty,m}^e
    z^{-m}\rb,
  \eeq
  connection matrices of $\Phi^e\lb z\rb$ are related to those of $\Phi\lb z\rb$ by 
  \beq\label{connmE}
  C_1^e=C_1C_S^{-1} ,\qquad C_{\infty}^e=C_{\infty}C_S^{-1},
  \eeq
  and $G_1^e$, $G_{\infty}^e$ are diagonalizing transformations for
  $A_1^e$, $A_{\infty}^e$: 
  \beq\label{AmE}
  A_{1}^e=G_{1}^e\Theta_{1}{G_{1}^e}^{-1},\qquad A_{\infty}^e=G_{\infty}^e\Theta_{\infty}{G_{\infty}^e}^{-1}.
  \eeq
   \end{subequations}
It also should be noticed that in the interior of the circles $S$, $\gamma_1$, $\gamma_{\infty}$ 
(i.e. to the left of the oriented circles  $S$, $\gamma_1$, $\gamma_{\infty}$) we have $\Psi^{e}(z) = \Phi^{e}(z) C_{S}J^{e}_{\Psi}(z)$.
     
   Let us next introduce in a similar way an approximation $\tilde{\Psi}^i\left(z\right)$ which reproduces the monodromy properties of $\Psi\left(z\right)$ inside $S_i$. The appropriate contour $\Gamma_{\Psi}^i$ is represented in Fig. \ref{fig2}b and the jump matrix $J_{\Psi}^{i}\left(z\right)$ is
   \beq
   \begin{gathered}
   J^i_{\Psi}\left(z\right)\bigl|_{\gamma_{0}}=C_0^{-1}
     \left(-z\right)^{-\Theta_0},\qquad
     J^i_{\Psi}\left(z\right)\bigl|_{\gamma_{t}}=C_{t}^{-1}
     \left(t-z\right)^{-\Theta_{t}},\qquad
     J^i_{\Psi}\left(z\right)\bigl|_{S}=
     \left(-z\right)^{\mathfrak S}C_S,\\
   J^i_{\Psi}\left(z\right)\bigl|_{\ell_{0t}}=M_0^{-1},
   \qquad J^i_{\Psi}\left(z\right)\bigl|_{\ell_{t\sigma}}=
   M_{t0}^{-1}.
   \end{gathered}
   \eeq
   Outside the circles $\gamma_0,\gamma_t,-S$ (i.e. to the right of the  oriented circles $\gamma_0,\gamma_t$ and to the
   left of the oriented circle $S$), the interior parametrix
   can be written as $\tilde{\Psi}^i\left(z\right)=
   {\Phi}^i\left(\frac{z}{t}\right)C_S$,
   where $\Phi^i\left(z\right)$ is a solution of the Fuchsian system
   \beq
   \label{fuchsint}
     \partial_z \Phi^i= A^i\left(z\right)\Phi^i,
     \qquad  A^i\left(z\right)=\frac{A_0^i}{z}+\frac{A_1^i}{z-1},
   \eeq
   with appropriate monodromy. The matrices $A_0^i$, $A_1^i$ satisfy the constraint $A_{\infty}^i:={-A_0^i-A_1^i}=-\mathfrak S$ and their spectra coincide with those of $A_0$, $A_t$ in (\ref{fuchspvi}). It is convenient to fix the normalization of $\tilde{\Psi}^i\left(z\right)$ by normalizing this solution as $\Phi^i(z)\simeq \left(-z\right)^{\mathfrak S}$ as $z\to \infty$, which amounts to setting $\tilde\Psi^{i}\left(\infty\right)=t^{-\mathfrak S}$. Let us also record for further reference the local expansions
   \begin{subequations}\label{monodI}
   \beq\label{localexpI}
   \Phi^i\lb z\rb=\begin{cases}
    G_0^i\lb z\rb \lb -z\rb^{\mathfrak \Theta_0}C_0^i & \qquad \text{as } z\to0,\\
     G_1^i\lb z\rb \lb 1-z\rb^{\Theta_t}C_1^i & \qquad \text{as } z\to1,\\
     G_{\infty}^i\lb z\rb \lb -z\rb^{\mathfrak S} & \qquad \text{as } z\to\infty,
   \end{cases}
   \eeq
   where
  \beq\label{GexpI}
  G_0^i\lb z\rb=G_0^i\lb\mathbf 1+\sum_{m=1}^{\infty}g_{0,m}^i z^m\rb,\qquad
  G_1^i\lb z\rb=G_1^i\lb\mathbf 1+\sum_{m=1}^{\infty}g_{1,m}^i
  \lb z-1\rb^m\rb ,\qquad
   G_{\infty}^i\lb z\rb=\mathbf 1+\sum_{m=1}^{\infty}g_{\infty,m}^i
    z^{-m}.
  \eeq
  Similarly to the above, 
  connection matrices of $\Phi^i\lb z\rb$ are expressed in terms of connection matrices of the original Fuchsian system (\ref{fuchspvi})  as 
   \beq\label{connmI}
   C_0^i=C_0C_S^{-1},\qquad C_1^i=C_tC_S^{-1},
   \eeq
  whereas $G_0^i$, $G_{1}^i$ diagonalize
    $A_0^i$, $A_{1}^i$:
    \beq\label{AmI}
    A_{0}^i=G_{0}^i\Theta_{0}{G_{0}^i}^{-1},\qquad A_{1}^i=G_{1}^i\Theta_{t}{G_{1}^i}^{-1}.
    \eeq
 \end{subequations}
 It also should be noticed that in the interior of the circles $\gamma_0$, $\gamma_{t}$, we have $\Psi^{i}(z) = \Phi^{i}\left(\frac{z}{t}\right)
  C_{S}J^{i}_{\Psi}(z)$, while  in the interior of the circle $-S$ (i.e., to the right of the oriented circle $S$) we have $\Psi^{i}(z) = \Phi^{i}\left(\frac{z}{t}\right)
  C_{S}J^{i}_{\Psi}(z)^{-1} = \Phi^{i}\left(\frac{z}{t}\right)(-z)^{-{\mathfrak S}}$.

 In the next subsections it will be shown that the asymptotics of the $4$-point isomonodromic tau function at the critical points can be derived given the inverse of the Riemann-Hilbert map for auxiliary $3$-point solutions $\Phi^{i,e}\lb z\rb$. More precisely, it turns out that all one needs is an expression for the matrix coefficients $g^e_{0,1}$, $g^i_{\infty,1}$ in terms of monodromy data:
 see, for instance, Proposition~\ref{propom}.

  \subsection{Global approximation\label{subsec_GLA}}
  Our goal now is to show that the solution $\Psi(z)$ of the 4-point Fuchsian Riemann-Hilbert problem  can be  approximated 
  by the solutions $\Psi^{e}(z)$ and $\tilde{\Psi}^{i}(z)$ of 3-point Riemann-Hilbert problems. To this end, let us  consider the matrix $\Psi^S\left(z\right)$ defined by
  \begin{equation}\label{glpar}
  \Psi^S\left(z\right)=\begin{cases}
  \Psi\left(z\right){\Psi^e\left(z\right)}^{-1},\qquad & z\in S_e,\\
  \Psi\left(z\right){\tilde\Psi^i\left(z\right)}^{-1},\qquad & z\in S_i.
  \end{cases}
  \end{equation}
  It is holomorphic and invertible on $\mathbb P\backslash S$; in particular, it has no jumps on $\Gamma_{\Psi}$. The normalization and the non-constant jump of $\Psi^S\left(z\right)$ on $S$ are given by
  \ben
  J_S\left(z\right)\equiv \Psi^{S}_{-}(z)^{-1}\Psi^{S}_{+}=\Psi^e_-\left(z\right){\tilde\Psi^i_+\left(z\right)}^{-1},\qquad \Psi^S\left(\infty\right)=G_{\infty}{\Psi^e\left(\infty\right)}^{-1}.
  \ebn
  The asymptotic evaluation of the 4-point function $\Psi(z)$ is equivalent to asymptotic solution of the Riemann-Hilbert problem 
  for the function $\Psi^S(z)$ which is generated by the jump matrix $ J_S\left(z\right)$. 
    
  If it were possible to analytically continue $J_S\left(z\right)$ from $S$ to its interior $S_i$, the Riemann-Hilbert problem for $\Psi^S\left(z\right)$ would be trivially solved by 
  \ben
  \Psi^S\left(z\right)=\begin{cases}
    G_{\infty}{\Psi^e\left(\infty\right)}^{-1},\qquad & z\in S_e,\\
    G_{\infty}{\Psi^e\left(\infty\right)}^{-1}J_S\left(z\right),\qquad & z\in S_i.
    \end{cases}
  \ebn  
  The crucial point for our analysis is that this effectively becomes true as $t\to 0$. To see this, we first  notice that 
  on $S$ we have 
   \ben
  \begin{gathered}
  \begin{aligned}
  \tilde\Psi^i_+\left(z\right)= \Phi^{i}\left(\frac{z}{t}\right)C_{S} = G_{\infty}^i\left(\frac zt\right)\left(-\frac{z}{t}\right)^{\mathfrak S}C_S=
   &\left[
  \mathbf 1+\sum_{m=1}^{\infty}g_{\infty,m}^i\left(\frac tz\right)^m\right]\left(-\frac{z}{t}\right)^{\mathfrak S}C_S,\\
  \Psi^e_-\left(z\right)= \Phi^{e}(z)C_{S}=G_0^e\left(z\right)\left(-z\right)^{\mathfrak S}C_S=
   &\left[ \mathbf 1+\sum_{m=1}^{\infty}g_{0,m}^e z^m\right]\left(-z\right)^{\mathfrak S}C_S,
   \end{aligned}
  \end{gathered}
  \ebn
  which in turn implies that
   \beq\label{apprjump}
  J_S\left(z\right)\equiv \Psi^e_-\left(z\right){\tilde\Psi^i_+\left(z\right)}^{-1}= G_0^e\left(z\right)\left(-z\right)^{\mathfrak S}C_S
  \left[G_{\infty}^i\left(\frac zt\right)\left(-\frac{z}{t}\right)^{\mathfrak S}C_S\right]^{-1} = G^e_0\left(z\right)t^{\mathfrak S}G_{\infty}^i\left(\frac zt\right)^{-1}.
  \eeq  
  Here $G_{0}^e\left(z\right)$ and $G_{\infty}^i(z)$ are the matrices that appear in the characterizations (\ref{localexpE}), (\ref{localexpI}) of the asymptotic behavior of $\Phi^e\left(z\right)$ as $z\to 0$ and, respectively, $\Phi^i\left(z\right)$ as $z\to \infty$. It is also worth emphasizing  that both series converge. As $t \to 0$, we have from (\ref{apprjump}) that 
\begin{equation}\label{est00}
J_S(z) = \left\{G^e_0\left(z\right)\left[
  \mathbf 1+\sum_{m=1}^{\infty}t^{m}t^{\mathfrak S}g_{\infty,m}^it^{-\mathfrak S}z^{-m}\right]^{-1} G^e_0\left(z\right)^{-1}\right\}
 G^e_0\left(z\right)t^{\mathfrak S} .
\end{equation}
Observe that
$$
t^m t^{\mathfrak S}g_{\infty,m}^it^{-\mathfrak S} = O\Bigl(t^{m-\mathfrak s}\Bigr),
$$
where
$$
\mathfrak s:=\max_{j,k}\left |\Re\left(\sigma_j-\sigma_k\right)\right|<1,
$$
 and  the  inequality is due to Assumption~\ref{asssigma}. Therefore, from (\ref{est00}) it follows that  indeed,
\begin{equation}\label{appr0}
  J_{S} = \left(\mathbf 1 + O(t^{1-\mathfrak s})\right)J^{0}_S, \qquad J^{0}_S = G^e_0\left(z\right)t^{\mathfrak S},
  \end{equation}
where the   jump matrix $J^{0}_S$ is analytic inside $S_i$. The formula
  \begin{equation}\label{appr1}
 \Psi^{S}_{0}\left(z\right)=\begin{cases}
    G_{\infty}{\Psi^e\left(\infty\right)}^{-1},\qquad & z\in S_e,\\
    G_{\infty}{\Psi^e\left(\infty\right)}^{-1}J^{0}_S\left(z\right),\qquad & z\in S_i,
\end{cases}
\end{equation}
defines the explicit solution of the Riemann-Hilbert problem generated by the jump matrix  $J^{0}_S$.  The estimate (\ref{appr0}) then transforms
into the estimate
$$
\Psi^{S}(z) = \left(\mathbf 1 + O\left(\frac{t^{1-\mathfrak s}}{1 +|z|}\right)\right)\Psi^{S}_0(z), \quad t \to 0, \quad z \in\Cb,
$$
 by the standard arguments  involving the singular integral
 operator  associated with the Riemann-Hilbert problem posed for the ratio $\Psi^{S}(z)\Psi^{S}_0(z)^{-1}$ 
 (see e.g. \cite{DZ0} or  \cite[Chapter 8, Theorem 8.1]{FIKN}). This in turn provides us, in view of (\ref{glpar}),  with the approximation of the 
 4-point function  $\Psi(z)$ by the 3-point functions $\Psi^{e}(z)$ and $\tilde{\Psi}^{i}(z)$. However, the error term is of order 
 $O(t^{1-\mathfrak s})$. We want to do better; in fact, we need to approximate  $\Psi(z)$ up to the error $o(t)$.
 The necessary improvement of (\ref{appr1}) is provided by the following lemma. 
  \begin{lemma}\label{keylemma} Put
    \beq
  \label{apprpsis}
  \begin{aligned}
  &{J_-^S\left(z\right)}=\mathbf 1 +\frac{\mathcal{E}\left(t\right)}{z},\\
  &J_+^S\left(z\right)=\Bigl[\mathbf 1-\mathfrak q\left(z,t\right)\Bigr]\,
  G_0^e\left(z\right) t^{\mathfrak S}.
  \end{aligned}
  \eeq 
  where
  \begin{subequations}\label{eqs318}
   \begin{align} 
   \label{Edef}
   \mathcal{E}\left(t\right)&=\varepsilon\left(t\right)\left[
  \mathbf 1-g^e_{0,1}\varepsilon\left(t\right)\right]^{-1},\\
  \label{smepsdef}
   \varepsilon\left(t\right)&=t^{\mathbf1 +\mathfrak S}g^i_{\infty,1}t^{-\mathfrak S}, \\
   \label{qdef}
   \mathfrak q\left(z,t\right)&=\mathfrak G_1\left(z,t\right)+\mathcal{E}\left(t\right)\mathfrak G_2\left(z,t\right),\\
   \label{G1def}
  \mathfrak G_1\left(z,t\right)&=\frac{G_0^e\left(z\right)
  \varepsilon\left(t\right){G_0^e\left(z\right)}^{-1}-
  \varepsilon\left(t\right)}{z},\\
  \label{G2def}
   \mathfrak G_2\left(z,t\right)&=\frac{\mathfrak G_1\left(z,t\right)
  -\mathfrak G_1\left(0,t\right)}{z},
   \end{align}  
   \end{subequations}
and define
\begin{equation}\label{appr2}
 \Psi^{S}_{0}\left(z\right)=\begin{cases}
    G_{\infty}{\Psi^e\left(\infty\right)}^{-1}J^{S}_-(z),\qquad & z\in S_e,\\
    G_{\infty}{\Psi^e\left(\infty\right)}^{-1}J^{S}_+\left(z\right),\qquad & z\in S_i.
\end{cases}
\end{equation}   
Then,  under the conditions on the spectrum of $\mathfrak S$ specified in Assumption~\ref{asssigma}, the following
uniform estimate takes place
\begin{equation}\label{appr3}
\Psi^{S}(z) = \left(\mathbf 1 + O\left(\frac{t^{2-\mathfrak s}}{1 +|z|}\right)\right)\Psi^{S}_0(z)
=  \Bigl(\mathbf 1 + o(t)\Bigr)\Psi^{S}_0(z), \quad t \to 0, \quad z \in\Cb,
\end{equation}
 \end{lemma}
\pf  First we notice that, under  conventions of Assumption~\ref{asssigma}, 
$$
\varepsilon\left(t\right)=O\left(t^{1-\max_{j,k}\left |\Re\left(\sigma_j-\sigma_k\right)\right|}\right) \equiv 
O\left(t^{1-\mathfrak s}\right)=o\left(1\right)
$$ 
is a small matrix parameter. Secondly, we observe that
the functions $J^S_-(z)$ and $J^S_+(z)$  are analytic in $S_{e}$ and $S_{i}$, respectively. Hence,
equation (\ref{appr2}) indeed determines a holomorphic  and invertible (for sufficiently small $t$) on $\mathbb P\backslash S$ matrix function  
whose jump on $S$ and normalization at infinity are  given by
$$
J^{0}_S\lb z\rb\equiv {\Psi^{S}_{0-}\lb z\rb}^{-1}\Psi^{S}_{0+}  = {J^{S}_-\lb z\rb}^{-1}J_+^{S}\lb z\rb, \qquad \Psi^S_0\left(\infty\right)=G_{\infty}{\Psi^e\left(\infty\right)}^{-1}.
$$ 
 The lemma will be proven if we can show that the new $J^{0}_S(z)$ improves the estimate (\ref{appr0}), i.e.
 \begin{equation}\label{appr4}
  J_{S}\lb z\rb {J^{0}_{S}\lb z\rb}^{-1} = \mathbf 1 + O\lb t^{2-\mathfrak s}\rb \equiv \mathbf 1 + o\lb t\rb, \quad z\in S.
  \end{equation}
 Indeed, suppose we already have (\ref{appr4}). Define 
 $$
 R\lb z\rb = \Psi^{S}\lb z\rb{\Psi_0^{S}\lb z\rb}^{-1}.
 $$
 The  matrix ratio $R\lb z\rb$ satisfies  the Riemann-Hilbert problem posed on $S$ with the jump matrix
   \begin{equation}\label{Rjump}
   J_{R}\lb z\rb = \Psi^{S}_{0-}\left(z\right)J_{S}\lb z\rb{J^{0}_{S}\lb z\rb}^{-1}{\Psi^{S}_{0-}\left(z\right)}^{-1},
   \end{equation}
   and it is normalized at $\infty$ as $R\lb \infty\rb = \mathbf 1$. All the matrix functions which we are dealing with are uniformly bounded
    on $S$ together with their inverses. Therefore, the estimate (\ref{appr4}) would imply that 
     \begin{equation}\label{estGR}
    J_{R}\lb z\rb = \mathbf 1 + O\lb t^{2-\mathfrak s}\rb \equiv \mathbf 1 + o\lb t\rb,
    \end{equation}
   as $t \to 0$ and uniformly for all $z\in S$. Referring again to the general theory of the Riemann-Hilbert problems 
   (\cite{DZ0} or  \cite[Chapter 8, Theorem 8.1]{FIKN}) we could transform the jump matrix estimate (\ref{estGR}) into the 
   solution estimate,
   \begin{equation}\label{appr5}
R\lb z\rb = \mathbf 1 + O\left(\frac{t^{2-\mathfrak s}}{1 +|z|}\right)
=  \mathbf 1 + o\lb t\rb, \quad t \to 0, \quad z \in\Cb,
\end{equation}
 which is equivalent to the statement (\ref{appr3}) of the lemma.
 
 Let us prove (\ref{appr4}).  We begin with recalling once again that all involved matrices are bounded on the circle~$S$ together with their inverses, and that all the series converge uniformly on  $S$.  Moreover, let us make a crucial observation that $\varepsilon^2\left(t\right)={t^2\cdot t^{\mathfrak S}\left(g^i_{\infty,1}\right)^2 t^{-\mathfrak S}}=O\left(t^{2-\mathfrak s}\right)=o\lb t\rb$. It implies that all terms containing $\varepsilon^2\left(t\right)$ can be 
 moved to the error term, $o\lb t\rb$.  At the same time one has to keep e.g.  expressions of the form $\varepsilon\left(t\right)A\,\varepsilon\left(t\right)$ with non-diagonal $A$, since their order can only be estimated as $O\left(t^{2-2\mathfrak s}\right)$. In particular, we have
 \begin{equation}\label{prep1}
 \mathcal{E}^2\lb t\rb = o\lb t\rb, \quad  \left[ \mathbf 1 + \frac{ \mathcal{E}\lb t\rb}{z}\right]^{-1} = \mathbf 1 - \frac{ \mathcal{E}\lb t\rb}{z} +o\lb t\rb,\quad
 G^{e}_0\lb z\rb t^{\mathfrak S}G^{i}_{\infty}\left(\frac{z}{t}\right)t^{-\mathfrak  S}G^{e}_0\lb z\rb^{-1} =
  \mathbf 1 + \frac{ G^{e}_0(z)\varepsilon(t)G^{e}_0(z)^{-1}}{z} + o\lb t\rb,
 \end{equation}
 $$
 \mathfrak q\lb z,t\rb = \frac{G^{e}_0\lb z\rb\varepsilon\lb t\rb G^{e}_0\lb z\rb^{-1} - \varepsilon\lb t\rb}{z}
 + \frac{ \mathcal{E}\lb t\rb}{z}\left\{\frac{G^{e}_0\lb z\rb\varepsilon\lb t\rb{G^{e}_0\lb z\rb}^{-1} - \varepsilon\lb t\rb}{z} -\left[g^{e}_{0,1}, \varepsilon\lb t\rb\right]\right\}=
 $$
 \begin{equation}\label{prep2}
 =  \left[ \mathbf 1 + \frac{ \mathcal{E}\lb t\rb}{z}\right]\frac{G^{e}_0\lb z\rb\varepsilon\lb t\rb G^{e}_0\lb z\rb^{-1} - 
 \varepsilon\lb t\rb}{z} - \frac{ \mathcal{E}\lb t\rb}{z}g^{e}_{0,1} \varepsilon\lb t\rb +o\lb t\rb.
 \end{equation}
 In these formulae, and everywhere below, the error term $o\lb t\rb$ may be understood as $O\left(t^{2-\mathfrak s}\right)$. Our last preparing observation is that 
 (\ref{appr4}) is equivalent to the estimate,
 \begin{equation}\label{appr41}
  J_{S}^{0}\lb z\rb {J_{S}\lb z\rb}^{-1} = \mathbf 1 + O\lb t^{2-\mathfrak s}\rb \equiv \mathbf 1 + o\lb t\rb, \quad z\in S,
  \end{equation}
 and it is this latter estimate which we are going to prove. 
 
 Taking into account (\ref{prep1}) and (\ref{prep2}), we have that
 $$
 J_{S}^{0}\lb z\rb {J_{S}\lb z\rb}^{-1} = \left[ \mathbf 1 + \frac{ \mathcal{E}\lb t\rb}{z}\right]^{-1}\Bigl[ \mathbf 1 - \mathfrak q\lb z,t\rb\Bigr]
  G^{e}_0\lb z\rb t^{\mathfrak S}G^{i}_{\infty}\left(\frac{z}{t}\right)t^{-\mathfrak  S}{G^{e}_0\lb z\rb}^{-1}=
 $$
 \begin{equation}\label{final0}
 =\left[ \mathbf 1 - \frac{ \mathcal{E}\lb t\rb}{z}\right]\Bigl[ \mathbf 1 - \mathfrak q\lb z,t\rb\Bigr]\left(\mathbf 1 + \frac{ G^{e}_0\lb z\rb \varepsilon\lb t\rb {G^{e}_0\lb z\rb}^{-1}}{z} \right)
 +o\lb t\rb.
 \end{equation}
 Using (\ref{prep1}) and (\ref{prep2}) one more time, we conclude that
 $$
  \left[ \mathbf 1 - \frac{ \mathcal{E}\lb t\rb}{z}\right]\Bigl[ \mathbf 1 - \mathfrak q\lb z,t\rb\Bigr] =
  \mathbf 1 - \frac{ \mathcal{E}\lb t\rb}{z} - \frac{G^{e}_0\lb z\rb\varepsilon\lb t\rb{G^{e}_0\lb z\rb}^{-1} - 
 \varepsilon\lb t\rb}{z} + \frac{ \mathcal{E}\lb t\rb}{z}g^{e}_{0,1}\varepsilon\lb t\rb +o\lb t\rb=
 $$
 $$
 = \mathbf 1 -  \frac{ \mathcal{E}\lb t\rb}{z} \Bigl(\mathbf 1 - g^{e}_{0,1}\varepsilon\lb t\rb\Bigr) - \frac{G^{e}_0\lb z\rb\varepsilon\lb t\rb G^{e}_0\lb z\rb^{-1} - 
 \varepsilon\lb t\rb}{z} + o\lb t\rb=
 $$
 \begin{equation}\label{final1}
 = \mathbf 1 -  \frac{ \varepsilon\lb t\rb}{z} - \frac{G^{e}_0\lb z\rb\varepsilon\lb t\rb {G^{e}_0\lb z\rb}^{-1} - 
 \varepsilon\lb t\rb}{z} + o\lb t\rb = \mathbf 1 - \frac{G^{e}_0\lb z\rb\varepsilon\lb t\rb {G^{e}_0\lb z\rb}^{-1}}{z} + o\lb t\rb.
 \end{equation}
 Plugging (\ref{final1}) into (\ref{final0}) and taking into account that
 $$
\lb\frac{G^{e}_0(z)\varepsilon(t)G^{e}_0(z)^{-1}}{z}\rb^2 =  \frac{G^{e}_0\lb z\rb\varepsilon^2\lb t\rb G^{e}_0\lb z\rb^{-1}}{z^2} = o\lb t\rb,
$$
we arrive at the estimate (\ref{appr41}) and hence  complete the proof of the lemma.\epf 

\begin{rmk} Let us explain the appearance of rather 
non-obvious formulae (\ref{apprpsis})--(\ref{eqs318})
 for the improved approximate solution of the $\Psi^{S}$ - RH problem. To this end, 
consider (\ref{apprpsis}) as an ansatz for $J^S_{\pm}\left(z\right)$ where $\mathcal{E}\left(t\right)$ and $\mathfrak q\left(z,t\right)$
  are to be determined. These quantities constitute small corrections including all terms up to $o\left(t\right)$. Comparing ${J_-^S(z)}^{-1}J_+^S(z)$ with the exact jump matrix (\ref{apprjump}), we see that it suffices to solve the equation
  \beq\label{auxeqEq}
  \frac{\mathcal{E}\left(t\right)}{z}+\mathfrak q\left(z,t\right)-
  \frac{\mathcal{E}\left(t\right) \mathfrak q\left(z,t\right)}{z}=
  \frac{G_0^e\left(z\right)
    \varepsilon\left(t\right){G_0^e\left(z\right)}^{-1}}{z}+
    o\left(t\right).
  \eeq 
 We are going 
  to expand $\mathcal{E}\left(t\right)$ and $\mathfrak q\left(z,t\right)$ with respect to the small matrix parameter $\varepsilon(t)$. Let us assign to any product of matrices a degree equal to the number of $\varepsilon\left(t\right)$-factors it contains, and (formally)  decompose $\mathcal{E}\left(t\right)$ and $\mathfrak q\left(z,t\right)$ accordingly as
  \beq
  \label{Eqdec}
  \mathcal{E}\left(t\right)=\sum_{k=1}^{\infty}
  \mathcal{E}_k\left(t\right),\qquad 
  \mathfrak q\left(z,t\right)=\sum_{k=1}^{\infty}\mathfrak q_k\left(z,t\right).
  \eeq 
  Substituting the decompositions (\ref{Eqdec}) into (\ref{auxeqEq}) yields recurrence relations
  \beq
  \label{recEq}
  \begin{aligned}
  \mathcal{E}_k\left(t\right)&=\sum_{m=1}^{k-1}
  \mathcal{E}_{m}\left(t\right)\mathfrak q_{k-m}\left(0,t\right),\\
  \mathfrak q_{k}\left(z,t\right)&=\sum_{m=1}^{k-1}
    \mathcal{E}_{m}\left(t\right)\frac{
    \mathfrak q_{k-m}\left(z,t\right)- \mathfrak q_{k-m}\left(0,t\right)}{z},
  \end{aligned}
  \eeq
  valid for $k>1$ and subject to the initial conditions
  \ben
  \mathcal{E}_1\left(t\right)=\varepsilon\left(t\right),\qquad
  \mathfrak q_{1}\left(z,t\right)=\mathfrak{G}_1\left(z,t\right),
  \ebn
  with $\mathfrak{G}_1\left(z,t\right)$ defined by (\ref{G1def}). This in principle allows to determine all terms in the sums (\ref{auxeqEq}) up to $o\left(t\right)$-corrections. Moreover, similarly to  the derivation of (\ref{prep1}) and (\ref{prep2}) in the proof of Lemma \ref{keylemma}, all terms containing $\varepsilon^2\left(t\right)$ can be dropped out from the sums (\ref{recEq}). Let us write down explicitly a few more terms in the expansion of $\mathcal{E}\left(t\right)$ and $\mathfrak q\left(z,t\right)$:
  \begin{align*}
  \mathcal E_2\left(t\right)&=\mathcal E_1\left(t\right)
  \mathfrak q_1\left(0,t\right)=\varepsilon\left(t\right)\mathfrak G_1\left(0,t\right)=\varepsilon\left(t\right)\left[g_{0,1}^e,\varepsilon\left(t\right)\right]=\varepsilon\left(t\right)g_{0,1}^e
  \varepsilon\left(t\right)+o\left(t\right),\\
  \mathfrak q_2\left(z,t\right)&=\mathcal E_1\left(t\right)\frac{
   \mathfrak q_1\left(z,t\right)- \mathfrak q_1\left(0,t\right)}{z}=
   \varepsilon\left(t\right)\mathfrak{G}_2\left(z,t\right),\\
   \mathcal E_3\left(t\right)&=\mathcal E_1\left(t\right)
    \mathfrak q_2\left(0,t\right)+ \mathcal E_2\left(t\right)
        \mathfrak q_1\left(0,t\right)= \mathcal E_2\left(t\right)
                \mathfrak q_1\left(0,t\right)+o\left(t\right)=
  \varepsilon\left(t\right)g_{0,1}^e
    \varepsilon\left(t\right)g_{0,1}^e
        \varepsilon\left(t\right)+o\left(t\right),\\
 \mathfrak q_3\left(z,t\right)&=\mathcal E_1\left(t\right)\frac{
    \mathfrak q_2\left(z,t\right)- \mathfrak q_2\left(0,t\right)}{z}  +\mathcal E_2\left(t\right)\frac{
        \mathfrak q_1\left(z,t\right)- \mathfrak q_1\left(0,t\right)}{z} =\varepsilon\left(t\right)g_{0,1}^e
          \varepsilon\left(t\right) \mathfrak{G}_2\left(z,t\right) +o\left(t\right),\\
          &\ldots\quad \ldots\quad \ldots
  \end{align*}
  The general pattern now becomes manifest. It remains to employ an inductive argument to show that for $k>1$ the only term in the sums (\ref{recEq})
  which is not of order $o\left(t\right)$ corresponds to $m=k-1$. Indeed, in all other terms $q_{k-m}\left(z,t\right)$ contain $\varepsilon\left(t\right)$ as the leftmost factor and simultaneously $\mathcal E_m\left(t\right)$ has  $\varepsilon\left(t\right)$ on the right. The formulae (\ref{Edef}) and (\ref{qdef}) for $\mathcal{E}\left(t\right)$ and $\mathfrak q\left(z,t\right)$ can be now easily guessed. 
The technique described above allows one to go far beyond the leading order approximation given by
formulae (\ref{apprpsis})--(\ref{eqs318}) and systematically construct a perturbative solution for $\Psi^S(z)$ for small $t$.
\end{rmk}
  
 A direct corollary of Lemma \ref{keylemma} is the following asymptotic representation of the solution $\Psi(z)$ of the 
 4-point Fuchsian Riemann-Hilbert problem in terms of the solutions $\Psi^e(z)$ and $\tilde{\Psi}^i(z)$ of the 3-point problems.
 
    \begin{cor} A uniform approximation for $\Psi\left(z\right)$ as $t\to 0$ is given by
    \beq\label{apprpsi}
  \Psi\left(z\right)=
  \begin{cases}
      G_{\infty}{\Psi^e\left(\infty\right)}^{-1}\left(\mathbf 1 +
      \displaystyle \frac{\mathcal E \left(t\right)}{z}+O\left(\frac{t^{2-\mathfrak s}}{1 +|z|}\right)\right)\Psi^e\left(z\right),
      \qquad & z\in S_e,\vspace{0.1cm}\\
      G_{\infty}{\Psi^e\left(\infty\right)}^{-1}
      \Bigl(\mathbf 1-\mathfrak q\left(z,t\right)+O\left(t^{2- \mathfrak s}\right)\Bigr)
      \,  G_0^e\left(z\right) t^{\mathfrak S}\tilde{\Psi}^i\left(z\right)
     ,\qquad & z\in S_i.
      \end{cases}
    \eeq
  Equivalently, the solution of the Fuchsian system (\ref{fuchspvi}) 
  can be approximated by
     \beq\label{apprphi}
   \Phi\left(z\right)=
   \begin{cases}
       G_{\infty} {G^e_{\infty}}^{-1}\left(\mathbf 1 +
             \displaystyle \frac{\mathcal E\left(t\right)}{z}+
             O\left(\frac{t^{2-\mathfrak s}}{1 +|z|}\right)\right)\Phi^e\left(z\right)C_S,
       \qquad & z\in S_e,\\
       G_{\infty} {G^e_{\infty}}^{-1}
       \Bigl(\mathbf 1-\mathfrak q \left(z,t\right)
               +O\left(t^{2- \mathfrak s}\right)\Bigr)\,
       \Phi^e\left(z\right)\left(-\frac{z}{t}\right)^{-\mathfrak S}{\Phi^i\left(\frac{z}{t}\right)}C_S,\qquad & z\in S_i.
       \end{cases}
     \eeq
  \end{cor}
  \pf The $\Psi$ - approximation (\ref{apprpsi})  follows at once from (\ref{appr3}), (\ref{appr2}) and  (\ref{glpar}) and from the 
  obvious fact that 
  $ \Bigl(\mathbf 1 + o(t)\Bigr) B = B\Bigl(\mathbf 1 + o(t)\Bigr) $
  for any invertible and constant (or bounded in $t$) matrix $B$. In order to pass from (\ref{apprpsi}) to (\ref{apprphi}), one only needs to invoke the relations between the functions $\Psi(z)$, $\Psi^e(z)$, $\tilde{\Psi}^i(z)$ and $\Phi(z)$, $\Phi^e(z)$, $\Phi^i(z)$,
  respectively. For instance,  if $z$ is inside $\gamma_{\infty}$ or $\gamma_1$, i.e. in the neighborhood of $z=\infty$ or 
  $z=1$, we have 
  $$
 \Psi^e(z) = \Phi^e\lb z\rb C_SJ^e_{\Psi}\lb z\rb = G^{e}_{\infty, 1}\lb z\rb,
 $$
 and, in particular, 
 \begin{equation}\label{gge}
 \Psi^e(\infty) = G^e_{\infty}.
 \end{equation}
 Simultaneously, one has
 $ \Psi(z) = \Phi(z)J_{\Psi}(z)$.
 Since $J^e_{\Psi}(z)\bigl|_{\gamma_{\infty}, \gamma_1} = J_{\Psi}(z)\bigl|_{\gamma_{\infty}, \gamma_1}$, 
 it follows that for all $z$ inside  $\gamma_{\infty}$ or~$\gamma_1$ we obtain
 $$
  \Psi^e\lb z\rb J^{-1}_{\Psi}\lb z\rb =  \Phi^e\lb z\rb C_S.
 $$
 Together with (\ref{gge}), this yields the first line in (\ref{apprphi}). 
 If $z$ is outside  of $\gamma_{\infty}$ and $\gamma_1$ but still belongs to $S_e$, we have 
 $ \Psi^e(z) = \Phi^e(z)C_S$ and  $\Psi(z) = \Phi(z)$,
 which gives again the first line in  (\ref{apprphi}). Thus the first line in (\ref{apprphi}) is proved
 for all $z \in S_e$. The second line in (\ref{apprphi}) is proved analogously.
  \epf
     
   We are now going to use this approximation to derive the $t\to 0$ asymptotics of the $1$-form $\omega$ defined by (\ref{tauextpvi}). As a byproduct, we will derive the famous Jimbo asymptotic formula for Painlev\'e VI \cite{Jimbo}. Our main concern is however to describe the dependence of the Painlev\'e VI tau function on monodromy data.

   \subsection{Asymptotics of $\omega$ and $\Omega$\label{subsecas}} 
  The short-distance behavior of the form $\omega$ is described by the following two lemmata. 
     \begin{lemma} \label{astimepart}
     The asymptotics of the time part of $\omega$ as $t\to 0$ is given by
     \beq
     \mathcal{P}=\frac{
     \operatorname{Tr}\left(
     \mathfrak S^2-\Theta_0^2-\Theta_t^2\right)}{2t}+\partial_t \ln\, \operatorname{det}\left(\mathbf 1-g^e_{0,1}t^{1+\mathfrak S}g^i_{\infty,1}t^{-\mathfrak S}\right)+o\left(1\right).
     \eeq
     \end{lemma}
  \pf The estimates (\ref{apprphi}) imply that
        \beq\label{tauest0}
  \begin{gathered}   
  \begin{aligned}
  \mathcal P&=\frac12\operatorname{res}_{z=t}\operatorname{Tr} 
  \left(\partial_z\Phi\cdot\Phi^{-1}\right)^2
  =\frac12\operatorname{res}_{z=t}\operatorname{Tr}
  \left(\frac1{t}{ A^i\left(\frac zt\right)}\right)^2
  +\operatorname{res}_{z=t}\operatorname{Tr}
    \left(\frac1{t}{A^i\left(\frac zt\right)\mathcal G^{-1}}\partial_z\mathcal G\right),
  \end{aligned}
  \end{gathered}    
    \eeq   
  with 
  \ben
  \mathcal G:=
         \Bigl(\mathbf 1-
         \mathfrak q\left(z,t\right)
                 +o\left(t\right)\Bigr)\,
         G^e_0\left(z\right){t}^{\mathfrak S}.
  \ebn       
  
  The first term in (\ref{tauest0}) is readily computed in terms of critical exponents:
  \ben
  \mathcal P_{-1}:=\frac12\operatorname{res}_{z=t}\operatorname{Tr}
    \left(\frac1{t}{A^i\left(\frac zt\right)}\right)^2=\frac{\operatorname{Tr}A_0^iA_1^i}{t}=
    \frac{\operatorname{Tr}\left(\left(A_0^i+A_1^i\right)^2-
    \left(A_0^i\right)^2-\left(A_1^i\right)^2\right)}{2t}=
  \frac{\operatorname{Tr}\left(\mathfrak S^2-
      \Theta_0^2-\Theta_t^2\right)}{2t}.  
  \ebn
  The second term can be transformed as
  \ben
    \begin{gathered}   
    \begin{aligned}
  \mathcal P_{0}:=&\,\operatorname{res}_{z=t}\operatorname{Tr}
      \left(\frac1{t}{A^i\left(\frac zt\right)\mathcal G^{-1}}\partial_z\mathcal G\right)=
      \operatorname{Tr}
      \left(A_1^i\mathcal G^{-1}\partial_z\mathcal G\bigl|_{z=t}\right)=\\=&\,
  \operatorname{Tr}\left(A_1^i t^{-\mathfrak S} 
  \left(g_{0,1}^e-\frac1{\mathbf 1-\mathfrak q\left(0,t\right)}\partial_z\mathfrak q\left( z,t\right)\bigl|_{z=0}\right) t^{\mathfrak S }\right)+
  o\left(1\right). 
  \end{aligned}
  \end{gathered}                  
  \ebn
  Recall that according to Assumption~\ref{asssigma} we have $|\Re \lb\sigma_j-\sigma_k\rb|<1$. Taking into account the expression (\ref{qdef}) for $\mathfrak q\left( z,t\right)$ and $\varepsilon^2\lb t\rb=o\lb t\rb$, it can be deduced that
  \ben
  \begin{aligned}
   & t^{-\mathfrak S} \frac1{\mathbf 1-\mathfrak q\left(0,t\right)}\partial_z\mathfrak q\left( z,t\right)\bigl|_{z=0}
  t^{\mathfrak S}=t^{-\mathfrak S}
  \frac1{\mathbf 1-\mathfrak G_1\lb 0,t\rb}\Bigl(
  \partial_z\mathfrak G_1\lb 0,t\rb+
  \mathcal E\lb t\rb\partial_z\mathfrak G_2\lb 0,t\rb
  \Bigr) t^{\mathfrak S}+o\lb t\rb=\\
  =\,&t^{-\mathfrak S}\frac1{\mathbf 1-g^e_{0,1}\varepsilon\lb t\rb}\partial_z\mathfrak G_1\lb 0,t\rb
  t^{\mathfrak S}+o\lb t\rb=-t^{-\mathfrak S}\frac1{\mathbf 1-g^e_{0,1}\varepsilon\lb t\rb}g^e_{0,1}\varepsilon\lb t\rb g^e_{0,1}
    t^{\mathfrak S}+o\lb t\rb,
  \end{aligned}
  \ebn
  and, consequently,
  \ben
  \mathcal P_{0}=
        \operatorname{Tr}\left(A_1^i t^{-\mathfrak S} 
          \frac1{\mathbf 1-g^e_{0,1}
          \varepsilon\left(t\right)}\,
          g_{0,1}^e t^{\mathfrak S }\right)+
          o\left(1\right).
  \ebn
  
  The quantities $g^{e}_{0,1}$ and $g_{\infty,1}^i$ (the latter matrix appears in the definition of $\varepsilon\left(t\right)$) can be related to the coefficients of the linear systems for $\Phi^e$ and $\Phi^i$, namely:
    \beq\label{relgA}
      \begin{gathered}   
      \begin{aligned}
      g_{0,1}^e+&\bigl[ g_{0,1}^e,\mathfrak S\bigr]+A_1^e=0,\\
      g_{\infty,1}^i-&\bigl[ g_{\infty,1}^i,\mathfrak S\bigr]+A_1^i=0.
      \end{aligned}
        \end{gathered}                  
        \eeq
   This allows to rewrite $\mathcal P_0$ as  
    \ben
    \mathcal P_0=\partial_t \operatorname{Tr}\, \ln \left(\mathbf 1-g^e_{0,1}
    \varepsilon\left(t\right)\right)+o\left(1\right)
    =\partial_t \ln\, \operatorname{det}\left(\mathbf 1-g^e_{0,1}t^{1+\mathfrak S}g^i_{\infty,1}t^{-\mathfrak S}\right)+o\left(1\right),
    \ebn      
  which finishes the proof. \epf

  \begin{lemma} 
  \label{asmonodromypart}
  The asymptotics of the monodromy part of $\omega$ as $t\to 0$ is given by
  \beq\label{monest}
  \begin{gathered}
  \sum_{\nu=0,t,1,\infty}
  \operatorname{Tr}\left(\Theta_{\nu}G_{\nu}^{-1}
  d_{\mathcal M} G_{\nu}\right)=\frac{\ln t}2\, d_{\mathcal M}\operatorname{Tr}\left(\mathfrak{S}^2-\Theta_0^2-\Theta_t^2\right)
  +\omega_0+o\lb 1\rb,
  \end{gathered}
  \eeq
  where $\omega_0\in\Lambda^1\lb\mathcal M\rb$ is a 1-form on $\mathcal M$ defined as
  \beq\label{defomega0}
  \omega_0=\operatorname{Tr}\left(\Theta_0
    {G^i_{0}}^{-1}
    d_{\mathcal M} G^i_{0}+
    \Theta_t {G^i_{1}}^{-1}
      d_{\mathcal M} G^i_{1}+\Theta_1{G_1^e}^{-1}
      d_{\mathcal M}G_1^e+\Theta_{\infty} 
         {G^e_{\infty}}^{-1}d_{\mathcal M} G^e_{\infty}\right).
  \eeq
  \end{lemma}
  \pf From the approximation (\ref{apprphi}) it also follows that
     \beq\label{monest2}
     \begin{gathered}
     \begin{aligned}
     G_0&= G_{\infty}{G^e_{\infty}}^{-1}\Bigl(\mathbf 1 - \mathfrak q \left(0,t\right)
    +o\left(t\right)
    \Bigr)t^{\mathfrak S}G_0^i t^{-\Theta_0},\\
     G_t&= G_{\infty} {G^e_{\infty}}^{-1}
     \Bigl(\mathbf 1 - \mathfrak q\left(0,t\right)+g^e_{0,1}t
       +o\left(t\right)\Bigr)
     t^{\mathfrak S}G_1^i t^{-\Theta_t},\\
       G_{1}&= G_{\infty} {G_{\infty}^e}^{-1}\Bigl(\mathbf 1 +
                    \mathcal E\left(t\right)+o\left(t\right)\Bigr)
                    \,G^e_1.
       \end{aligned}
     \end{gathered}  
     \eeq 
     The statement follows by straightforward computation combining the above estimates with the relations ${A_0^i+A_1^i=\mathfrak S}$ and
     $\mathfrak S + A_1^e=-A^e_{\infty}$.
     \epf
     
   Compatibility of Lemmata~\ref{astimepart} and \ref{asmonodromypart} is manifest. The estimate (\ref{monest}) may seem too rough to give the $O\left(1\right)$ short-distance behavior of $\Omega=d\omega$  by directly computing the differential: naively for that one would need  the asymptotics of the left side up to $o\left(t\right)$. However, we already know from Proposition~\ref{propomega} that $\Omega\left(\partial_t,\partial_M\right)=0$ for any local coordinate $M$ on $\mathcal M$ and that $\Omega$ does not depend on $t$. Therefore this 2-form is completely determined by the $O\lb 1\rb$ asymptotics of the monodromy part of $\omega$.
   \begin{cor}
   The $2$-form  $\Omega$ coincides with $d\omega_0$, where 
   $\omega_0\in\Lambda^1\lb \mathcal M\rb$ is defined by (\ref{defomega0}).
   \end{cor}

  The results of this subsection can now be summarized as follows.
  \begin{prop}\label{propom}
  Given $\omega$  defined by (\ref{tauextpvi}) and $\omega_0$ by (\ref{defomega0}), the difference $\hat{\omega}:=\omega-\omega_0$ is a closed $1$-form on $\widetilde{\mathcal{T}}\times \mathcal{M}$. Its short-distance ($t\to 0$) asymptotics is given by
  \beq\label{auxcomm8}
  \hat\omega=d\ln\left(t^{\frac12
       \operatorname{Tr}\left(
       \mathfrak S^2-\Theta_0^2-\Theta_t^2\right)} \operatorname{det}\left(\mathbf 1-g^e_{0,1}t^{1+\mathfrak S}g^i_{\infty,1}t^{-\mathfrak S}\right)\right)+o\left(1\right).
  \eeq
  \end{prop}
  \begin{defin}\label{taudefin}
  The four-point tau function $\tau_{\omega_0}: \widetilde{\mathcal{T}} \times \mathcal{M}\to \mathbb C$ is defined by
  \beq\label{tau4def}
  d\ln\tau_{\omega_0}=\omega-\omega_0.
  \eeq
  This relation defines $\tau_{\omega_0}$ up to a multiplicative constant independent of monodromy data, including the local monodromy exponents $\Theta_{0,t,1,\infty}$.
  \end{defin}
  
  We have thus expressed $4$-point tau function asymptotics in terms of parameters of two auxiliary $3$-point Fuchsian systems with appropriate monodromy. Given the solutions $\Phi^i\left(z\right)$, $\Phi^e\left(z\right)$ of the inverse monodromy problem for these systems, Proposition \ref{propom} provides an explicit asymptotic expression for $d\ln\tau_{\omega_0}$. Finding $3$-point solutions remains an open problem in general. However, in a number of cases the $3$-point inverse monodromy problem can be solved in terms of generalized hypergeometric functions. Below we discuss the simplest situation of this type which corresponds to generic monodromy in rank $N=2$ and leads to Gauss hypergeometric system.
   
 \subsection{Painlev\'e VI monodromy data \label{subsmon}}
 The space of essential monodromy data of the $4$-point Fuchsian system (\ref{fuchspvi}) is the space of conjugacy classes of triples
 $\left(M_0,M_t,M_1\right)\in {SL_N\left(\mathbb C\right)}^3$. Here the matrices $M_{0,t,1}$ and $M_{\infty}:=\left(M_1M_tM_0\right)^{-1}$ represent monodromy  of $\Phi\left(z\right)$ around the singular points.
 The spectra of $M_{\nu}$  coincide with those of $e^{2\pi i \Theta_{\nu}}$ ($\nu=0,t,1,\infty$) and are considered as fixed. Define
 \ben
 \mathcal M_{4}^{\Theta}=\left\{\left(M_0,M_t,M_1\right)\in
 {SL_N\left(\mathbb C\right)}^3: M_{\nu}\sim e^{2\pi i \Theta_{\nu}}\text{ for }\nu=0,t,1,\infty\right\}\bigl/SL_N\left(\mathbb C\right).
 \ebn
 
 To compute the dimension of $\mathcal M_{4}^{\Theta}$ and introduce on it a convenient set of local coordinates, it is useful to start with a simper case of $3$ points. Having diagonalized one of the two generators of monodromy group, the second is defined up to diagonal conjugation and has fixed spectrum, which gives $\left(N-1\right)^2$ parameters. Fixing the spectrum of monodromy around the third singular point subtracts $N-1$ parameters so that
 \ben
 \operatorname{dim}\mathcal M_{3}^{\Theta}=\left(N-1\right)\left(N-2\right).
 \ebn
 Note in particular that in rank $2$ the conjugacy class of monodromy is completely determined by the local monodromy exponents --- the corresponding Fuchsian system is rigid. In higher rank $N>2$, monodromy data have an even number of nontrivial internal moduli.
   
\begin{figure}[h!]
  \centering
  \includegraphics[width=4cm]{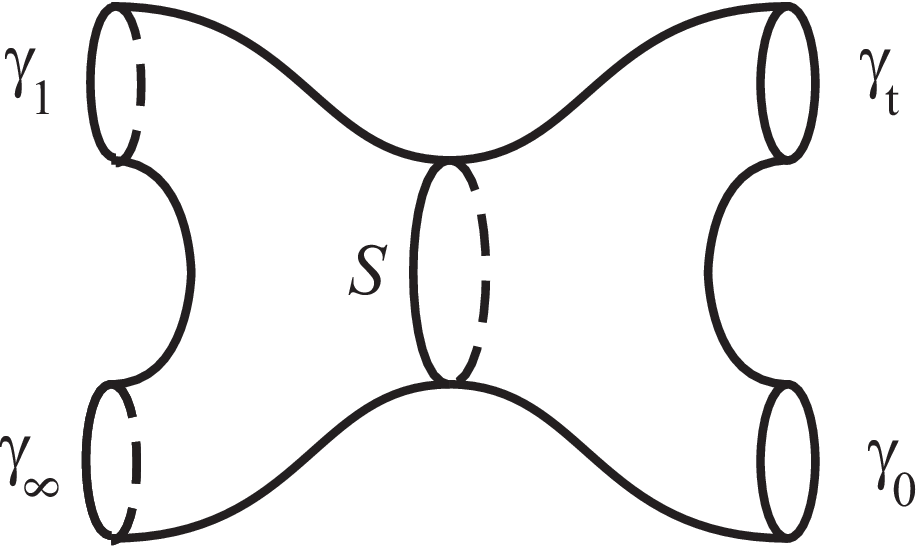}
  \caption{\label{fig3} Pants decomposition of the $4$-punctured Riemann sphere}
  \end{figure}

 In the case of 4 or more poles a convenient parameterization is suggested by decompositions of the punctured sphere into pairs of pants, such as the one represented in Fig.~\ref{fig3}. It is instructive to compare this picture with Figs. \ref{fig1} and \ref{fig2}.  
 
 The spectrum of the product $M_t M_0$ (monodromy along $S$) contains $N-1$ monodromy parameters, encoded in the matrix $\mathfrak S$.  Let us fix in each of the conjugacy classes $\left[\left(M_0,M_t\right)\right]$, $\left[\left(M_1,M_{\infty}\right)\right]$ the representatives 
 \ben
 \tilde M_{\nu}=C_S M_{\nu}C_S^{-1},\qquad \nu=0,t,1,\infty
 \ebn  
 with diagonal $\tilde M_{t} \tilde M_{0}=\left(\tilde M_{\infty}\tilde M_1\right)^{-1}= e^{2\pi i \mathfrak S}$. This involves fixing $\left(N-1\right)\left(N-2\right)$ parameters for each pair of pants and choosing a representative in each of the two $(N-1)$-dimensional orbits of diagonal conjugation. The possibility of simultaneous diagonal conjugation of $\left(\tilde M_0,\tilde M_t,\tilde M_1\right)$ reduces the latter $2(N-1)$ coordinates to $N-1$ parameters of relative twist, to be collectively denoted as $\mathfrak
 T$. The resulting dimension is
 \ben
 \operatorname{dim}\mathcal M_{4}^{\Theta}=\underbrace{\left(N-1\right)}_{\text{spectrum of }\mathfrak S}+
 2\times\underbrace{\left(N-1\right)\left(N-2\right)}_{
 \text{moduli of  pants}}+\underbrace{2\times\left(N-1\right)
 -\left(N-1\right)}_{\text{twist parameters }\mathfrak T}=2\left(N-1\right)^2.
 \ebn

 Let us now make this parameterization completely explicit for rank $N=2$, where $\mathcal M_{4}^{\Theta}$ is two-dimensional.   
 The three-point moduli are absent and two local coordinates are provided by the eigenvalue of $\mathfrak S=\operatorname{diag}\left\{
 \sigma,-\sigma\right\}$ and one twist parameter.  
 Indeed, for $\nu=0,t,1,\infty$ denote the eigenvalues of $\Theta_{\nu}$ by $\pm \theta_{\nu}$. The knowledge
 of $\operatorname{Tr} \tilde M_0=2\cos2\pi \theta_0$ and $\operatorname{Tr} M_0{M_{t0}}^{-1}=\operatorname{Tr} \tilde M_0 e^{-2\pi i \mathfrak S}=2\cos2\pi \theta_t$ determines the diagonal elements of $\tilde M_0$. The unit determinant fixes the product of its off-diagonal elements.  
 
 In addition to Assumption~\ref{asssigma} (implying that $|\Re \sigma|<\frac12$ and
 $\sigma\neq0$), it is convenient to impose further genericity assumptions on the monodromy:
    \begin{ass}\label{assirrmon}
   Parameters $\theta_{0,t,1,\infty}$ and $\sigma$ satisfy 
   \beq
   \theta_0+\theta_t\pm\sigma,\;\;
   \theta_0-\theta_t\pm\sigma,\;\;
   \theta_{\infty}+\theta_1\pm\sigma,\;\;
   \theta_{\infty}-\theta_1\pm\sigma\;\notin\mathbb Z.
   \eeq
     \end{ass}  
 \noindent We can then write
 \begin{subequations}
 \beq
 \label{m0par}
 \tilde M_0=\frac{1}{i\sin2\pi\sigma}\left(\begin{array}{cc}
 e^{2\pi i \sigma}\cos2\pi\theta_0-\cos2\pi\theta_t & {s_i}\left[\cos2\pi\left(\theta_t-\sigma\right)
 -\cos 2\pi\theta_0\right]\\
 {s_i}^{-1}\left[\cos 2\pi\theta_0-\cos2\pi\left(\theta_t+\sigma\right)\right] & \cos2\pi\theta_t-e^{-2\pi i \sigma}\cos2\pi\theta_0
 \end{array}\right).
 \eeq
 Next, from $\tilde M_t=e^{2\pi i \mathfrak S}M_0^{-1}$ it follows that
  \beq\label{mtpar}
  \tilde M_t=\frac{1}{i\sin2\pi\sigma}\left(\begin{array}{cc}
  e^{2\pi i \sigma}\cos2\pi\theta_t-\cos2\pi\theta_0 & {s_i}e^{2\pi i\sigma}\left[\cos 2\pi\theta_0-\cos2\pi\left(\theta_t-\sigma\right)\right]\\
  {s_i}^{-1}e^{-2\pi i\sigma}\left[\cos2\pi\left(\theta_t+\sigma\right)-\cos 2\pi\theta_0\right] & \cos2\pi\theta_0-e^{-2\pi i \sigma}\cos2\pi\theta_t
  \end{array}\right).
  \eeq
  In these two expressions, $s_i$ is a coordinate on the one-dimensional orbit of diagonal conjugation of $\left(\tilde M_0,\tilde M_t\right)$ with fixed product $\tilde M_t\tilde M_0=e^{2\pi i \mathfrak S}$. In a similar fashion, we can write a parameterization of $\tilde M_{1}$, $\tilde M_{\infty}$: it suffices to replace in the above formulas $\theta_0\mapsto\theta_{1}$, $\theta_t\mapsto\theta_{\infty}$, $\sigma\mapsto -\sigma$ so that
  \begin{align}
  \label{m1par}
  \tilde M_1=\frac{1}{i\sin2\pi\sigma}\left(\begin{array}{cc}
  \cos2\pi\theta_{\infty}-e^{-2\pi i \sigma}\cos2\pi\theta_{1} & {s_e}e^{2\pi i\sigma}\left[\cos2\pi\left(\theta_1+
          \sigma\right)-\cos2\pi\theta_{\infty}\right]\\
  {s_e}^{-1}e^{-2\pi i \sigma}\left[\cos 2\pi\theta_{\infty}-\cos2\pi\left(\theta_1-
      \sigma\right)\right] & e^{2\pi i \sigma}\cos2\pi\theta_{1}-\cos2\pi\theta_{\infty}
  \end{array}\right),\\
  \label{mipar}
  \tilde M_{\infty}=\frac{1}{i\sin2\pi\sigma}\left(\begin{array}{cc}
    \cos2\pi\theta_{1}-e^{-2\pi i \sigma}\cos2\pi\theta_{\infty} & {s_e}\left[\cos 2\pi\theta_{\infty}-\cos2\pi\left(\theta_1+
    \sigma\right)\right]\\
    {s_e}^{-1}\left[\cos2\pi\left(\theta_1-
        \sigma\right)-\cos2\pi\theta_{\infty}\right] & e^{2\pi i \sigma}\cos2\pi\theta_{\infty}-\cos2\pi\theta_{1}
    \end{array}\right).\qquad\quad
  \end{align}  
  \end{subequations}
  The possibility of simultaneous conjugation of $\tilde M_{0,t,1,\infty}$ by any diagonal matrix implies that the role of the twist parameter is played by the ratio 
  \beq\label{defeta}
  {s_i}/{s_e}\equiv e^{i\eta}.
  \eeq 
  
  The variables $\left(\sigma,\eta\right)$  provide a pair of convenient local coordinates on~$\mathcal M_4^{\Theta}$ which can be thought of as complexified Fenchel-Nielsen coordinates on the Teichm\"uller space for spheres with 4 punctures. They are closely related to monodromy parameterization in \cite{Jimbo} and can  be written in terms of trace functions. For instance, introduce the Fricke coordinates
  \beq\label{trfunc}
  \begin{aligned}
  p_{\nu}=2\cos2\pi\theta_{\nu}=\operatorname{Tr}M_{\nu},&\qquad \nu=0,t,1,\infty,\\
  p_{\mu\nu}=2\cos2\pi\sigma_{\mu\nu}=\operatorname{Tr}M_{\mu}M_{\nu},&\qquad \mu\nu=0t,t1,01.
  \end{aligned}
  \eeq
   We can identify $\sigma_{0t}=\sigma$. Straightforward calculation also shows that \cite{Jimbo}
   \begin{subequations}\label{trfunc2}
    \begin{align}\label{p1tf}
    \left(4-p_{0t}^2\right)p_{t1}=&\,
    2\left(p_0p_{\infty}+p_tp_1\right)-
    p_{0t}\left(p_0p_{1}+p_tp_{\infty}\right)\\ \nonumber -&\,   \sum_{\epsilon=\pm1}
      \bigl(p_{\infty}-2\cos2\pi\left(\theta_1-
      \epsilon\sigma\right)
      \bigr)
      \bigl(p_0-2\cos2\pi\left(\theta_{t}-\epsilon\sigma\right)
        \bigr)
      \, e^{i\epsilon\eta} ,\\
    \label{p01f}  
       \left(4-p_{0t}^2\right) p_{01}=&\,2\left(p_0p_{1}+p_tp_{\infty}\right)-
         p_{0t}\left(p_0p_{\infty}+p_tp_{1}\right)\\ \nonumber +&\,\sum_{\epsilon=\pm1}
          \bigl(p_{\infty}-2\cos2\pi\left(\theta_1-
          \epsilon\sigma\right)
     \bigr)               \bigl(p_0-2\cos2\pi\left(\theta_{t}-\epsilon\sigma\right)
    \bigr) \,e^{i\epsilon \eta -2\pi i \epsilon \sigma} \, . 
    \end{align}
    \end{subequations}
  The functions (\ref{trfunc}) satisfy a quartic relation (\ref{jimbofricke})
  and generate the algebra of invariant polynomial functions on $\mathcal M_4$.  There exists a canonical quadratic Poisson bracket $\{\cdot,\cdot\}$ of geometric origin on $\mathcal M_4$ \cite{Gol}. Trace functions $p_0,p_t,p_1,p_{\infty}$ are its Casimirs and the nontrivial brackets are given by
  \beq\label{pbrack}
  \begin{aligned}
  \left\{p_{0t},p_{t1}\right\}&=
  2p_{01}+p_{0t}p_{t1}-p_0p_1-p_tp_{\infty},\\
  \left\{p_{t1},p_{01}\right\}&=
    2p_{0t}+p_{t1}p_{01}-p_0p_t-p_1p_{\infty},\\
  \left\{p_{01},p_{0t}\right\}&=
      2p_{t1}+p_{0t}p_{01}-p_tp_1-p_0p_{\infty}.  
  \end{aligned}
  \eeq
  
  \begin{lemma}\label{darboux}
  The pair of local coordinates $\left(\sigma,\eta\right)$ satisfies
  $\left\{\eta,\sigma\right\}=\frac{1}{2\pi}$,
  i.e. $\left(\sigma,\eta\right)$ are canonical Darboux coordinates on the symplectic leaf $\mathcal{M}_4^{\Theta}$  with respect to the Poisson bracket (\ref{pbrack}). 
  \end{lemma}
  \pf Direct calculation. For instance,
  from (\ref{trfunc2}) it follows that
  \begin{subequations}\label{etareltheta}
  \beq
  e^{i\eta}=\frac{2i\sin 2\pi  \sigma\, \left(e^{2\pi i \sigma}p_{t1}+p_{01}\right)+ f\left(\theta,\sigma\right)}{g\left(\theta,\sigma\right)},
  \eeq
  where the quantities $f\left(\theta,\sigma\right)$, $g\left(\theta,\sigma\right)$ are independent of $\eta$:
  \begin{align}
  f\left(\theta,\sigma\right)&=
  \left(p_0p_{\infty}+p_tp_1\right)-
  \left(p_0p_1+p_tp_{\infty}\right)e^{2\pi i \sigma},\\
  g\left(\theta,\sigma\right)&=\bigl(p_{\infty}-
  2\cos2\pi\left(\theta_1-\sigma\right)\bigr)
  \bigl(p_0-2\cos2\pi\left(\theta_{t}-\sigma\right) \bigr).
  \end{align}
  \end{subequations}
  But then (\ref{pbrack}) implies that
  \ben
  \left\{p_{0t},e^{i\eta}\right\}=\frac{2i\sin 2\pi  \sigma}{g\left(\theta,\sigma\right)}
  \left\{p_{0t},e^{2\pi i \sigma}p_{t1}+p_{01}\right\}=
  2i\sin 2\pi  \sigma \,e^{i\eta}.
  \ebn
  which is equivalent to the statement of the lemma.
  \epf
  
  \subsection{Inverse monodromy problem in rank $2$}
  We are now going to find an explicit form of the parametrices 
  $\Phi^e\left(z\right)$, $\Phi^i\left(z\right)$ in the case $N=2$.  
   Recall that $\Phi^e\left(z\right)$ is the solution of the Fuchsian system (\ref{parext}) with the following properties:
  \begin{itemize} 
  \item its monodromy is described by (\ref{monodE}); in particular, the connection matrices at $1,\infty$ are 
  given by ${C^e_1=C_1C_S^{-1}}$, $C^e_{\infty}=C_{\infty}C_S^{-1}$.
  \item the normalization is fixed by $\Phi^e\left(z\right)\simeq \left(-z\right)^{\mathfrak S}$ as $z\to 0$;
  \item in (\ref{parext}),  $A_0^e=\mathfrak S=\operatorname{diag}\left\{\sigma,-\sigma\right\}$; the eigenvalues of $A_1^e$, $A_{\infty}^e=-A_0^e-A_1^e$ are 
  $\pm\theta_1$ and $\pm \theta_{\infty}$, respectively.
  \end{itemize}

  The known spectrum of $A^e_{1,\infty}\in\mathfrak{sl}_2\left(
  \mathbb C\right)$ determines these matrices almost completely. Indeed, computing the trace
  \ben
  \operatorname{Tr}A_0^eA_1^e=\frac12\operatorname{Tr}
  \left(\left(A_{\infty}^e\right)^2-\left(A_0^e\right)^2-\left(A_1^e\right)^2\right)=\theta_{\infty}^2-\theta_1^2-\sigma^2,
  \ebn
  one can find the diagonal elements of $A_1^e$. The known determinant $\operatorname{det}A_1^e=-\theta_1^2$ then gives the product of the off-diagonal elements. As a result of this calculation, one finds
  \beq\label{A1eexp}
  A_1^e=\frac1{2\sigma}\left(\begin{array}{cc}
  \theta_{\infty}^2-\theta_1^2-\sigma^2 & r_e\left(\left(\theta_1+\sigma\right)^2-\theta_{\infty}^2\right)
  \vspace{0.1cm} \\
  r_e^{-1}\left(\theta_{\infty}^2-\left(\theta_1-\sigma\right)^2\right)
  & \sigma^2+\theta_1^2-\theta_{\infty}^2
  \end{array}\right),
  \eeq
  where $r_e\in\mathbb{C}^{\times}$ is the only remaining unknown parameter related to the freedom of simultaneous conjugation of $A^e_{0,1,\infty}$ by a non-degenerate diagonal matrix.
  Let us write the general form of diagonalizing transformations for $A_1^e=G_{1}^e
  \Theta_1{G_{1}^e}^{-1}$ and $A_{\infty}^e=G_{\infty}^e
    \Theta_{\infty}{G_{\infty}^e}^{-1}$  as
  \begin{subequations}  
  \label{diagextG}
  \begin{align}
  \label{diagext1}
     G_{1}^e\, &=\left(\begin{array}{cc}
     r_e & \left(\theta_1+\sigma\right)^2
                 -\theta_{\infty}^2\\ 
      1 & \displaystyle r_e^{-1}\lb\left(\theta_1-\sigma\right)^2
      -\theta_{\infty}^2\rb
     \end{array}\right)
     \lb\begin{array}{cc}
     {c_{1a}^e}^{-1} & 0 \\
     0 &  {c_{1b}^e}^{-1}
     \end{array}\rb,\\
   \label{diagextinf}
   G_{\infty}^e &=\left(\begin{array}{cc}
   \displaystyle r_e\lb\theta_1-\theta_{\infty}+\sigma\rb &  
   \theta_1+\theta_{\infty}+\sigma\\ 
   \theta_1-\theta_{\infty}-\sigma & \displaystyle r_e^{-1}\lb
          \theta_1+\theta_{\infty}-\sigma\rb
   \end{array}\right)     \lb\begin{array}{cc}
        {c_{\infty a}^e}^{-1} & 0 \\
        0 &  {c_{\infty b}^e}^{-1}
        \end{array}\rb.
  \end{align} 
  \end{subequations}
  Our main remaining task is to relate $r_e$ in (\ref{A1eexp})--(\ref{diagextG})  to monodromy parameter $s_e$ in (\ref{m1par}), (\ref{mipar}). This can be done using the explicit solution for $\Phi^e\left(z\right)$.
  \begin{lemma}
  Let $\Phi^e\left(z\right)$ denote the solution of  (\ref{parext}) normalized as $\Phi^e(z)\simeq
    \left(-z\right)^{\mathfrak S}$  as $ z\to 0$,  with $A_0^e=\operatorname{diag}\left\{\sigma,-\sigma\right\}$ and 
  $A^e_1$ parameterized as in (\ref{A1eexp}). Then
  \begin{subequations}
  \label{Phie0}
  \begin{align}
  \Phi^e_{11}\left(z\right)&=\left(-z\right)^{\sigma}
  \left(1-z\right)^{\theta_1}{}_2F_1\Bigl(
  \theta_1+\theta_{\infty}+\sigma,
  \theta_1-\theta_{\infty}+\sigma;2\sigma;z\Bigr),\\
   \Phi^e_{12}\left(z\right)&=\;\, r_e\;\frac{\theta_{\infty}^2-\left(
   \theta_1+\sigma\right)^2
   }{2\sigma\left(
   2\sigma-1\right)}\,\left(-z\right)^{1-\sigma}
    \left(1-z\right)^{\theta_1}{}_2F_1\Bigl(
    1+\theta_1+\theta_{\infty}-\sigma,
    1+\theta_1-\theta_{\infty}-\sigma;2-2\sigma;z\Bigr), \\
 \Phi^e_{21}\left(z\right)&=r_e^{-1}\frac{\theta_{\infty}^2-\left(
 \theta_1-\sigma\right)^2
 }{2\sigma\left(
 2\sigma+1\right)}\left(-z\right)^{1+\sigma}
  \left(1-z\right)^{\theta_1}{}_2F_1\Bigl(
  1+\theta_1+\theta_{\infty}+\sigma,
  1+\theta_1-\theta_{\infty}+\sigma;2+2\sigma;z\Bigr), \\
    \Phi^e_{22}\left(z\right)&=\left(-z\right)^{-\sigma}
    \left(1-z\right)^{\theta_1}{}_2F_1\Bigl(
    \theta_1+\theta_{\infty}-\sigma,
    \theta_1-\theta_{\infty}-\sigma;-2\sigma;z\Bigr), 
  \end{align}
  \end{subequations}
  where $_2F_1\left(\alpha,\beta;\gamma;z\right)$ denotes the Gauss hypergeometric function.
  \end{lemma}
  
  \begin{cor} The relation between $s_e$ and $r_e$ is as follows:
  \beq\label{sere}
  s_e=\frac{
  \Gamma\left(1-2\sigma\right)
  \Gamma\left(1+\theta_{\infty}+\theta_1+\sigma\right)
  \Gamma\left(1-\theta_{\infty}+\theta_1+\sigma\right)}{
  \Gamma\left(1+2\sigma\right)
  \Gamma\left(1+\theta_{\infty}+\theta_1-\sigma\right)
  \Gamma\left(1-\theta_{\infty}+\theta_1-\sigma\right)}\,
  r_e.
  \eeq
  \end{cor}
  \pf As $z\to -\infty$, it becomes convenient to replace (\ref{Phie0}) by an equivalent representation
  \beq
  \Phi^e\left(z\right)=G^{e}_{\infty}\hat{\Phi}^e\left(z\right)
  C^e_{\infty},
  \eeq
  where $G_{\infty}^e$ is defined by (\ref{diagextinf}) and
    \begin{subequations}
    \label{Phie0inf}
    \begin{align}
    \hat\Phi^e_{11}\left(z\right)&=\left(-z\right)^{-\theta_{\infty}}
    \left(1-z^{-1}\right)^{\theta_1}{}_2F_1\Bigl(
    \theta_1+\theta_{\infty}+\sigma,
    \theta_1+\theta_{\infty}-\sigma;2\theta_{\infty};z^{-1}\Bigr),\\
     \hat\Phi^e_{12}\left(z\right)&=\frac{c^e_{\infty a}}{r_e
     c^e_{\infty b}} \frac{
     \left(
          \theta_1+\theta_{\infty}\right)^2-\sigma^2
     }{2\theta_{\infty}\left(
     2\theta_{\infty}-1\right)}\,\left(-z\right)^{\theta_{\infty}-1}
           \left(1-z^{-1}\right)^{\theta_1}{}_2F_1\Bigl(
           1+\theta_1-\theta_{\infty}+\sigma,
           1+\theta_1-\theta_{\infty}-\sigma;
           2-2\theta_{\infty};z^{-1}\Bigr), \\
   \hat\Phi^e_{21}\left(z\right)&=
   \frac{r_e c^e_{\infty b}}{
        c^e_{\infty a}} \frac{
   \left(
      \theta_1-\theta_{\infty}\right)^2-\sigma^2
   }{2\theta_{\infty}\left(
   2\theta_{\infty}+1\right)}\left(-z\right)^{-\theta_{\infty}-1}
    \left(1-z^{-1}\right)^{\theta_1}\!\!{}_2F_1\Bigl(
    1+\theta_1+\theta_{\infty}+\sigma,
    1+\theta_1+\theta_{\infty}-\sigma;2+2\theta_{\infty};z^{-1}\Bigr), \\
      \hat\Phi^e_{22}\left(z\right)&=\left(-z\right)^{\theta_{\infty}}
      \left(1-z^{-1}\right)^{\theta_1}{}_2F_1\Bigl(
      \theta_1-\theta_{\infty}+\sigma,
      \theta_1-\theta_{\infty}-\sigma;-2\theta_{\infty};z^{-1}\Bigr). 
    \end{align}
    \end{subequations}
 The connection matrix $C^e_{\infty}$ can be derived from the standard connection formulas for hypergeometric functions. Explicitly, its expression reads
 \begin{align}
 \label{connmext}
 C_{\infty}^e&=\lb\begin{array}{cc}
         {c_{\infty a}^e} & 0 \\
         0 &  {c_{\infty b}^e}
         \end{array}\rb\left(\begin{array}{cc}
 \displaystyle\frac{ r_e^{-1}
 \Gamma\left(-2\theta_{\infty}\right)\Gamma\left(2\sigma\right)}{
 \Gamma\left(-\theta_1-\theta_{\infty}+\sigma\right)
 \Gamma\left(1+\theta_1-\theta_{\infty}+\sigma\right)} &
 \displaystyle
 \frac{
  \Gamma\left(-2\theta_{\infty}\right)\Gamma\left(-2\sigma\right)}{
  \Gamma\left(-\theta_1-\theta_{\infty}-\sigma\right)
  \Gamma\left(1+\theta_1-\theta_{\infty}-\sigma\right)}
  \vspace{0.2cm}\\
 \displaystyle
 \frac{
  \Gamma\left(2\theta_{\infty}\right)\Gamma\left(2\sigma\right)}{
  \Gamma\left(-\theta_1+\theta_{\infty}+\sigma\right)
  \Gamma\left(1+\theta_1+\theta_{\infty}+\sigma\right)} &
 \displaystyle
 \frac{r_e
  \Gamma\left(2\theta_{\infty}\right)\Gamma\left(-2\sigma\right)}{
  \Gamma\left(-\theta_1+\theta_{\infty}-\sigma\right)
  \Gamma\left(1+\theta_1+\theta_{\infty}-\sigma\right)}  
 \end{array}\right).
 \end{align} 
 Now it suffices to compute the monodromy matrix
 $\tilde M_{\infty}={C^e_{\infty}}^{-1}e^{2\pi i \Theta_{\infty}}C_{\infty}^e$. Comparing the result with  (\ref{mipar}), we deduce the identification
 (\ref{sere}).
  \epf
  \begin{rmk} One can similarly derive an expression for the connection matrix $C^e_1$ associated to the choice (\ref{diagext1}) of diagonalizing matrix $G^e_{1}$. In contrast to $C^e_0$ and $C^e_{\infty}$, this matrix is given by different expressions for 
  $\Im z\gtrless 0$, which are related by $C^e_{1,+}=C^e_{1,-}
  e^{-2\pi i
  \mathfrak S}$. In fact we have
  \beq\label{connm1ext}
  C^e_{1,\pm}=\lb\begin{array}{cc}
           {c_{1 a}^e} & 0 \\
           0 &  {c_{1 b}^e}
           \end{array}\rb
 \lb \begin{array}{cc} 
 \ds \frac{r_e^{-1}\Gamma\lb 2\sigma\rb\Gamma\lb-2\theta_1\rb}{
 \Gamma\lb -\theta_1-\theta_{\infty}+\sigma\rb \Gamma\lb -\theta_1+\theta_{\infty}+\sigma\rb}
  & \ds \frac{\Gamma\lb -2\sigma\rb\Gamma\lb-2\theta_1\rb}{
   \Gamma\lb -\theta_1-\theta_{\infty}-\sigma\rb \Gamma\lb -\theta_1+\theta_{\infty}-\sigma\rb} \vspace{0.2cm} \\ 
  \ds \frac{\Gamma\lb 2\sigma\rb\Gamma\lb2\theta_1\rb}{
   \Gamma\lb 1+\theta_1-\theta_{\infty}+\sigma\rb \Gamma\lb 1+\theta_1+\theta_{\infty}+\sigma\rb}    & 
  \ds
  \frac{r_e
  \Gamma\lb -2\sigma\rb\Gamma\lb 2\theta_1\rb}{
   \Gamma\lb 1+\theta_1-\theta_{\infty}-\sigma\rb \Gamma\lb 1+\theta_1+\theta_{\infty}-\sigma\rb} \end{array}\rb e^{\mp i\pi\mathfrak S}.          
  \eeq
  \end{rmk}

  An explicit expression for the interior parametrix, as well as the relation between monodromy and coefficient matrices of the auxiliary Fuchsian system (\ref{fuchsint}), can be obtained in a completely analogous way. The known spectra  $\operatorname{sp}A_0^i=\left\{\pm\theta_0\right\}$,
   $\operatorname{sp}A_1^i=\left\{\pm\theta_t\right\}$ and the diagonal form of $A_0^i+A_1^i=\mathfrak S=\operatorname{diag}\left\{\sigma,-\sigma\right\}$ allow to parameterize $A_0^i$, $A_1^i$~by setting
   \beq\label{paramintA}
   A_1^i=\frac{1}{2\sigma}
   \left(\begin{array}{cc}
   \sigma^2+\theta_t^2-\theta_0^2 & r_i\left(\theta_0^2-\left(\sigma-\theta_t\right)^2\right)
    \vspace{0.1cm}\\ 
   r_i^{-1}\left(\left(\sigma+\theta_t\right)^2-\theta_0^2\right)
      & \theta_0^2-\theta_t^2-\sigma^2\end{array}\right),
   \eeq
   with $r_i\in\mathbb{C}^{\times}$. The diagonalizing transformations for $A_0^i= G_0^i\Theta_0{G_0^i}^{-1}$ and $A_1^i= G_1^i\Theta_t{G_1^i}^{-1}$ can be chosen as
      \begin{subequations}
      \label{Gidiagtr}
      \begin{align}
      \label{G0idiag}
      G^i_{0}&=\left(\begin{array}{cc} \ds
      r_i\lb\theta_t-\theta_0-\sigma\rb & \theta_t+\theta_0-\sigma \\ 
      \theta_t-\theta_0+\sigma & \ds
      r_i^{-1}\lb\theta_t+\theta_0+\sigma\rb
      \end{array}\right)
      \left(\begin{array}{cc}
      {c^i_{0a}}^{-1} & 0 \\
      0 & {c^i_{0b}}^{-1}
      \end{array}\right),\\
      \label{G1idiag}
      G^i_{1}&=\left(\begin{array}{cc}
            r_i & \left(\theta_t-\sigma\right)^2-\theta_0^2 \\ 1 & \displaystyle
            r_i^{-1}\lb\left(\theta_t+\sigma\right)^2-\theta_0^2\rb
            \end{array}\right)
      \left(\begin{array}{cc}
      {c^i_{1a}}^{-1} & 0 \\
      0 & {c^i_{1b}}^{-1}
      \end{array}\right)            .
      \end{align}
      \end{subequations}
   
   \begin{lemma}
   Let $\Phi^i\left(z\right)$ be the solution of the Fuchsian system (\ref{fuchsint}), normalized as $\Phi^i(z)\simeq \left(-z\right)^{\mathfrak S}$ as
   $z \to -\infty$,  with $A_0^i$, $A_1^i$ defined by (\ref{paramintA}) and $A_1^i=\mathfrak S-A_0^i$. Then
   \begin{subequations}
   \label{Phiintexpl}
   \begin{align}
   \Phi^i_{11}\left(z\right)&=\left(-z\right)^{\sigma}
          \left(1-z^{-1}\right)^{\theta_t}{}_2F_1\Bigl(\theta_t+\theta_0-\sigma,\theta_t-\theta_0-\sigma;-2\sigma;z^{-1}\Bigr),\\
   \Phi^i_{12}\left(z\right)&=\;\, r_i \frac{\theta_0^2-\left(\theta_t-\sigma\right)^2}{
   2\sigma\left(2\sigma+1\right)}
   \left(-z\right)^{-\sigma-1}
       \left(1-z^{-1}\right)^{\theta_t}{}_2F_1\Bigl(1+\theta_t+\theta_0+\sigma,1+\theta_t-\theta_0+\sigma;2+2\sigma;z^{-1}\Bigr),\\
  \Phi^i_{21}\left(z\right)&=r_i^{-1} \frac{\theta_0^2-\left(\theta_t+\sigma\right)^2}{
  2\sigma\left(2\sigma-1\right)}
  \left(-z\right)^{\sigma-1}
    \left(1-z^{-1}\right)^{\theta_t}{}_2F_1\Bigl(
    1+\theta_t+\theta_0-\sigma,
    1+\theta_t-\theta_0-\sigma;2-2\sigma;z^{-1}\Bigr),\\     
   \Phi^i_{22}\left(z\right)&=
   \left(-z\right)^{-\sigma}
       \left(1-z^{-1}\right)^{\theta_t}{}_2F_1\Bigl(\theta_t+\theta_0+\sigma,\theta_t-\theta_0+\sigma;2\sigma;z^{-1}\Bigr).       
   \end{align}
   \end{subequations}
   \end{lemma}
  
   An equivalent form of the solution (\ref{Phiintexpl}) suitable for study of its local behavior as $z\to0$ is given by
   \beq
   \Phi^i\left(z\right)=G^i_{0}\,\hat{\Phi}^i\left(z\right)
   C_0^i,
   \eeq
   where $G_0^i$ is defined by (\ref{G0idiag}) and $\hat{\Phi}^i\left(z\right)$ by
   \begin{subequations}
   \begin{align}
   \hat{\Phi}^i_{11}\left(z\right)&=\left(-z\right)^{\theta_0}
   \left(1-z\right)^{\theta_t}{}_2F_1\Bigl(
   \theta_t+\theta_0+\sigma,\theta_t+\theta_0-\sigma,
   2\theta_0,z\Bigr),\\
   \hat{\Phi}^i_{12}\left(z\right)&=
   r_i^{-1}\frac{c^i_{0a}}{c^i_{0b}}\frac{
      \left(\theta_t+\theta_0\right)^2-\sigma^2}{
      2\theta_0\left(2\theta_0-1\right)}
   \left(-z\right)^{1-\theta_0}
      \left(1-z\right)^{\theta_t}{}_2F_1\Bigl(
      1+\theta_t-\theta_0+\sigma,1+\theta_t-\theta_0-\sigma,
      2-2\theta_0,z\Bigr),\\   
   \hat{\Phi}^i_{21}\left(z\right)&=r_i\frac{c^i_{0b}}{c^i_{0a}}\frac{
    \left(\theta_t-\theta_0\right)^2-\sigma^2}{
   2\theta_0\left(2\theta_0+1\right)}
   \left(-z\right)^{\theta_0+1}
   \left(1-z\right)^{\theta_t}{}_2F_1\Bigl(
   1+\theta_t+\theta_0+\sigma,1+\theta_t+\theta_0-\sigma,
   2+2\theta_0,z\Bigr),   \\
   \hat{\Phi}^i_{22}\left(z\right)&=\left(-z\right)^{-\theta_0}
      \left(1-z\right)^{\theta_t}{}_2F_1\Bigl(
      \theta_t-\theta_0+\sigma,\theta_t-\theta_0-\sigma,
      -2\theta_0,z\Bigr).
   \end{align}
   \end{subequations}
  The connection matrix $C_0^i=C_0C_S^{-1}$ is given by 
  \beq
  C_0^i=\left(\begin{array}{cc}
        {c^i_{0a}} & 0 \\
        0 & {c^i_{0b}}
        \end{array}\right)\left(\begin{array}{cc}
  \displaystyle \frac{r_i^{-1}
  \Gamma\left(-2\theta_0\right)
  \Gamma\left(-2\sigma\right)}{
  \Gamma\left(-\theta_0-\theta_t-\sigma\right)
  \Gamma\left(1-\theta_0+\theta_t-\sigma\right)} & 
   \displaystyle \frac{\Gamma\left(-2\theta_0\right)
    \Gamma\left(2\sigma\right)}{
    \Gamma\left(-\theta_0-\theta_t+\sigma\right)
    \Gamma\left(1-\theta_0+\theta_t+\sigma\right)}
   \vspace{0.2cm}\\
  \displaystyle \frac{\Gamma\left(2\theta_0\right)
   \Gamma\left(-2\sigma\right)}{
   \Gamma\left(\theta_0-\theta_t-\sigma\right)
   \Gamma\left(1+\theta_0+\theta_t-\sigma\right)} &
 \displaystyle \frac{r_i\Gamma\left(2\theta_0\right)
    \Gamma\left(2\sigma\right)}{
    \Gamma\left(\theta_0-\theta_t+\sigma\right)
    \Gamma\left(1+\theta_0+\theta_t+\sigma\right)}   
  \end{array}\right).
  \eeq
  Connection matrices $C_{1,\pm}^i$ are related by
  $C^i_{1,+}=C^i_{1,-}e^{-2\pi i\mathfrak S}$ and can be written as
  \beq\label{connmGi1}
  C_{1,\pm}^i=e^{\pm i\pi\Theta_t}\lb\begin{array}{cc}
  c^i_{1a} & 0 \\ 0 & c^i_{1b}
  \end{array}\rb
  \lb\begin{array}{cc} 
  \ds\frac{r_i^{-1}\Gamma\lb -2\theta_t\rb\Gamma\lb -2\sigma\rb}{
  \Gamma\lb -\theta_0-\theta_t-\sigma\rb
   \Gamma\lb \theta_0-\theta_t-\sigma\rb} & 
   \ds\frac{\Gamma\lb -2\theta_t\rb\Gamma\lb 2\sigma\rb}{
     \Gamma\lb -\theta_0-\theta_t+\sigma\rb
      \Gamma\lb \theta_0-\theta_t+\sigma\rb}\vspace{0.2cm} \\
   \ds\frac{\Gamma\lb 2\theta_t\rb\Gamma\lb -2\sigma\rb}{
    \Gamma\lb 1-\theta_0+\theta_t-\sigma\rb
     \Gamma\lb 1+\theta_0+\theta_t-\sigma\rb}    &
 \ds\frac{r_i\Gamma\lb 2\theta_t\rb\Gamma\lb 2\sigma\rb}{
   \Gamma\lb 1-\theta_0+\theta_t+\sigma\rb
    \Gamma\lb 1+\theta_0+\theta_t+\sigma\rb}    
      \end{array}\rb
  e^{\mp i\pi\mathfrak S}.
  \eeq

  Comparing the monodromy matrix $\tilde M_0={C^i_0}^{-1}e^{2\pi i \Theta_0}C^i_0$ with the parameterization (\ref{m0par}), we obtain the following result:
  \begin{cor}
  The quantities $s_i$ and $r_i$ are related by
    \beq\label{seri}
    s_i=\frac{
    \Gamma\left(1+2\sigma\right)
    \Gamma\left(1+\theta_0+\theta_t-\sigma\right)
    \Gamma\left(1-\theta_0+\theta_t-\sigma\right)}{
    \Gamma\left(1-2\sigma\right)
    \Gamma\left(1+\theta_0+\theta_t+\sigma\right)
    \Gamma\left(1-\theta_0+\theta_t+\sigma\right)}\,r_i.
    \eeq
  \end{cor} 
  
  \begin{cor}
     Results obtained above enable us to compute explicitly the 1-form $\omega_0$ defined by (\ref{defomega0}) for $N=2$. By straightforward calculation, it follows from the parameterizations (\ref{diagextG}), (\ref{Gidiagtr}) that
     \beq
     \begin{gathered}\label{omzero}
     \begin{aligned}
     \omega_0=&\, \theta_{\infty}d_{\mathcal M}\ln 
     \frac{r_e c^e_{\infty b}}{c^e_{\infty a}}+
     \theta_{1}d_{\mathcal M}\ln 
          \frac{r_e c^e_{1 b}}{c^e_{1 a}}+
                   \theta_{t}d_{\mathcal M}\ln 
                        \frac{r_i c^i_{1 b}}{c^i_{1 a}} +\theta_{0}\,d_{\mathcal M}\ln 
         \frac{r_i c^i_{0 b}}{c^i_{0 a}}
              + \sigma   d_{\mathcal M}\ln \frac{r_i}{r_e}   -2d_{\mathcal M}\lb \theta_1+\theta_t\rb.
        \end{aligned}
        \end{gathered}
     \eeq
  \end{cor}
  \begin{rmk}\label{rmksymp} If we consider parameters $\theta_{0,t,1,\infty}$ as fixed, then the differential $\Omega=d\omega_0$ is determined by single term $\sigma d_{\mathcal M}\ln\frac{r_i}{r_e}$ in (\ref{omzero}). In this case it follows from (\ref{sere}), (\ref{seri}) and (\ref{defeta}) that
  $\Omega=id\sigma\wedge d\eta$, i.e. the 2-form $\Omega$ coincides with symplectic form on $\mathcal M_4^{\Theta}$ induced by Goldman bracket (\ref{pbrack}). However, for the computation of the connection constant for Painlev\'e VI tau function in Subsection~\ref{secconcof} we need to keep track of the dependence of $\omega_0$ on the spectra of local monodromies.
  \end{rmk}
   
   \subsection{Jimbo asymptotic formula}
   We are now in a position to express the short-distance asymptotics of the Painlev\'e VI tau function (\ref{tau4def}) in terms of monodromy data of the associated Fuchsian system. Indeed, the asymptotics of the $4$-point tau function in arbitrary rank  is described by Proposition~\ref{propom} and involves  matrices $g^e_{0,1}$, $g^{i}_{\infty,1}$ appearing in the local expansions of $3$-point parametrices $\Phi^e\left(z\right)$, $\Phi^i\left(z\right)$ around $0$ and $\infty$. Relations (\ref{relgA}) express these matrices in terms of matrix coefficients $A_1^e$, $A_1^i$ of the corresponding $3$-point Fuchsian systems. In the case $N=2$, the latter quantities are parameterized by two variables $r_e,r_i\in \mathbb{C}^{\times}$ as indicated in (\ref{A1eexp}) and
   (\ref{paramintA}). The parameters $r_e,r_i$ are related to monodromy of the initial $4$-point Fuchsian system by (\ref{sere}) and (\ref{seri}), see also (\ref{defeta}) and (\ref{trfunc2}).
   Altogether, this leads to the following claim. 
   
   \begin{prop}\label{jimbof} For $N=2$,
   let $\left(\sigma,\eta\right)$ be two local coordinates on the space $\mathcal M_4^{\Theta}$ of monodromy data defined in Subsection~\ref{subsmon}. Let $\tau_{\omega_0}\left(t\right)$ denote the Painlev\'e VI tau function normalized as in (\ref{tau4def}).
   Its $t\to 0$ asymptotics is given by (cf (\ref{p6at0}))
   \beq
   \begin{gathered}
   \label{jimboextf}
   \tau_{\omega_0}(t)=\mathcal{C}_{\omega_0}\cdot t^{\sigma^2-\theta_0^2-\theta_t^2}\,\Biggl[ 1-
   \sum_{\epsilon=\pm1}\frac{
   \left(\left(\theta_t-\epsilon\sigma\right)^2-\theta_0^2\right)
   \left(\left(\theta_1-\epsilon\sigma\right)^2-\theta_{\infty}^2
   \right)}{
   4\sigma^2\left(1+2\epsilon\sigma\right)^2}\,\kappa^{\epsilon}
   t^{1+2\epsilon\sigma}+
   \Biggr. \\ 
   \qquad\qquad\qquad\qquad\quad +\frac{
      \left(\sigma^2-\theta_{\infty}^2+\theta_1^2\right)
      \left(\sigma^2-\theta_0^2+\theta_t^2\right)}{2\sigma^2}\,t+
   o\left(t\right)\Biggr],\quad t \to 0,
   \end{gathered}
   \eeq
   where the coefficient $\kappa$ is defined by
   \beq\label{kappadef}
   \kappa=
   \frac{\Gamma^2\left(1-2\sigma\right)
   \Gamma\left(1+\theta_0+\theta_t+\sigma\right)
   \Gamma\left(1-\theta_0+\theta_t+\sigma\right)
   \Gamma\left(1+\theta_{\infty}+\theta_1+\sigma\right)
   \Gamma\left(1-\theta_{\infty}+\theta_1+\sigma\right)}{
   \Gamma^2\left(1+2\sigma\right)
   \Gamma\left(1+\theta_0+\theta_t-\sigma\right)
   \Gamma\left(1-\theta_0+\theta_t-\sigma\right)
   \Gamma\left(1+\theta_{\infty}+\theta_1-\sigma\right)
   \Gamma\left(1-\theta_{\infty}+\theta_1-\sigma\right)}\,e^{i\eta},
   \eeq
     \begin{equation}\label{etadef}
 e^{i\eta}= \frac{
        2i\sin2\pi \sigma\left( p_{01}+p_{t1}e^{2\pi i
        \sigma}\right)+
        \left( p_0p_\infty+p_tp_1\right)-\left( p_0p_1+p_tp_{\infty}\right)
        e^{2\pi i \sigma}
        }{\left( p_{\infty}-2\cos2\pi\left(\theta_1-\sigma\right)\right)
        \left( p_{0}-2\cos2\pi\left(\theta_t-\sigma\right)\right)},
\end{equation}
   and the constant prefactor $\mathcal C_{\omega_0}$ does not depend on monodromy data.
   \end{prop}
   \pf For $N=2$, one can replace (\ref{auxcomm8}) by the estimate
   \beq\label{auxform1}
   \tau_{\omega_0}\left(t\right)=\mathcal C_{\omega_0}\cdot t^{\sigma^2-\theta_0^2-\theta_t^2}\left[1-\operatorname{Tr}\left(
   g^e_{0,1}t^{1+\mathfrak S}g^i_{\infty,1}t^{-\mathfrak S}\right)+o\left(t\right)\right].
   \eeq
   (In higher rank, one needs to take into account more terms in the determinant expansion). From (\ref{relgA}), (\ref{A1eexp}) and
      (\ref{paramintA}) it follows that
   \begin{align*}
   g^e_{0,1}=&\left(\begin{array}{cc}
   \displaystyle
   \frac{\sigma^2+\theta_1^2-\theta_{\infty}^2}{2\sigma} & 
   \displaystyle r_e\frac{
   \left(\theta_1+\sigma\right)^2-\theta_{\infty}^2}{
   2\sigma\left(2\sigma-1\right)} \vspace{0.1cm}\\ 
   \displaystyle r_e^{-1}\frac{
      \left(\theta_1-\sigma\right)^2-\theta_{\infty}^2}{
      2\sigma\left(2\sigma+1\right)} &
   \displaystyle \frac{\theta_{\infty}^2-\theta_1^2-\sigma^2}{2\sigma}
   \end{array}\right),\\
   g^i_{\infty,1}=&\left(\begin{array}{cc}
      \displaystyle
      \frac{\theta_0^2-\theta_t^2-\sigma^2}{2\sigma} & 
      \displaystyle r_i\frac{
      \left(\theta_t-\sigma\right)^2-\theta_{0}^2}{
      2\sigma\left(2\sigma+1\right)} \vspace{0.1cm}\\ 
      \displaystyle r_i^{-1}\frac{
            \left(\theta_t+\sigma\right)^2-\theta_{0}^2}{
            2\sigma\left(2\sigma-1\right)}  &
      \displaystyle \frac{\theta_t^2-\theta_0^2+\sigma^2}{2\sigma}
      \end{array}\right).
   \end{align*}
   Upon substitution into (\ref{auxform1}), the diagonal parts of $g^e_{0,1}$, $g^i_{\infty,1}$ give the linear contribution in (\ref{jimboextf}). The off-diagonal elements determine the coefficients of $t^{1\pm 2\sigma}$, with identification $\kappa=r_i/r_e$. Finally, the latter coefficient is related to invariant monodromy data by (\ref{sere}), (\ref{seri}) and (\ref{defeta}).
   \epf
   
   \begin{rmk} Proposition~\ref{jimbof} is a slightly upgraded version of Jimbo asymptotic formula for Painlev\'e VI (Theorem~1.1 in \cite{Jimbo}). The improvement concerns the error estimate: e.g. the terms such as $t^{2\pm 4\sigma},t^{3\pm 6\sigma},\ldots\,$, a pri\-ori present in the short-distance asymptotics of $\ln \tau_{\omega_0}$, disappear from the expansion of the tau function itself. This fact was already noticed and played an important role in \cite{GIL12}. Similar nontrivial cancellations have been experimentally observed to happen in higher rank \cite{Gav}. This was one of our motivations for establishing the results of Subsection~\ref{subsecas}, in particular, Proposition~\ref{propom}.
   \end{rmk}

   \subsection{Crossing to $t\to 1$}
   To be able to deal with the connection problem for Painlev\'e VI tau function, we now have to investigate the asymptotics of the form
   (\ref{tauextpvi}) as $t\to 1$. It can be obtained by a suitable exchange of parameters, which is not trivial even for the ``time part'' of the tau function: in the latter case  the initial suggestion of \cite[Theorem 1.2]{Jimbo} should be modified as explained in \cite[Remark 7.1]{Liso11}. The tau function extended to the space of monodromy data needs even more care as here at intermediate steps one has to manipulate with quantities that are not preserved by conjugation.
   
\begin{figure}
    \centering
    \includegraphics[width=6cm]{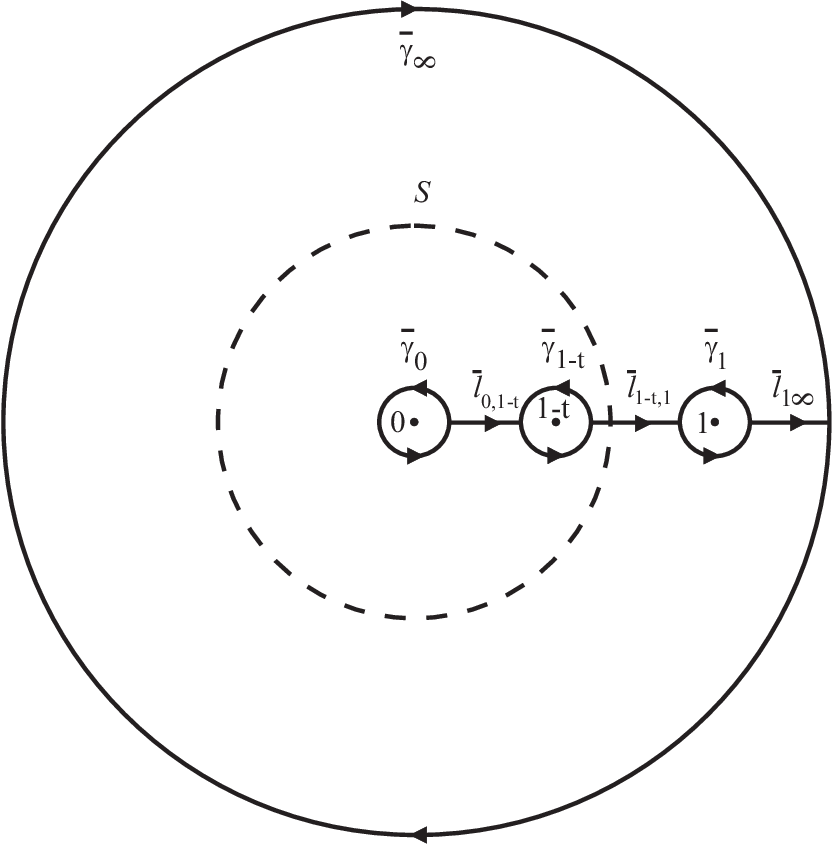}
    \caption{\label{figcross}
    Contour $\Gamma_{\overline\Psi}$ of the Riemann-Hilbert problem for $\overline\Psi\left(z\right)$}
    \end{figure}

   The basic idea is to replace the RHP for $\Psi\lb z\rb$ by a RHP 
   for a new matrix function $\overline\Psi\lb z\rb$. The corresponding contour is represented in Fig.~\ref{figcross} and $\overline\Psi\lb z\rb$ is defined by
   \ben
   \overline{\Psi}\lb z\rb =
   \begin{cases}
    \Psi\lb 1-z\rb M_{\infty},& \qquad z \text{ outside }\bar\gamma_{\nu},\;\Im z>0,\\
    \Psi\lb 1-z\rb& \qquad \text{otherwise}.
   \end{cases}
   \ebn
   This function is designed to have the structure of jump matrix $ J_{\overline\Psi}={\overline\Psi_-}^{\!\!\!\! -1}\overline \Psi_+$  analogous to (\ref{jumpspsi}):
   \beq\label{jumpspsibar}
   \begin{gathered}
    J_{\overline\Psi}\lb z\rb\Bigr|_{\overline{\ell}_{0,1-t}}=
    M_1^{-1},\qquad
    J_{\overline\Psi}\lb z\rb\Bigr|_{\overline{\ell}_{1-t,1}}=
    \lb M_1M_t\rb^{-1},\qquad
    J_{\overline\Psi}\lb z\rb\Bigr|_{\overline{\ell}_{1\infty}}= M_{\infty},\\
    J_{\overline\Psi}\lb z\rb\Bigr|_{\overline{\gamma}_{0}}={\overline{C}_0}^{-1}\lb -z\rb^{-\Theta_1},\qquad 
    J_{\overline\Psi}\lb z\rb\Bigr|_{\overline{\gamma}_{1-t}}={\overline{C}_{1-t}}^{\!\!\!\!\!\! -1}\lb 1- t-z\rb^{-\Theta_t},\\
    J_{\overline\Psi}\lb z\rb\Bigr|_{\overline{\gamma}_{1}}={\overline{C}_1}^{-1}\lb 1-z\rb^{-\Theta_0},\qquad
    J_{\overline\Psi}\lb z\rb\Bigr|_{\overline{\gamma}_{\infty}}=
    {\overline{C}_{\infty}}^{-1}\lb -z\rb^{\Theta_{\infty}}.
    \end{gathered}
   \eeq
   Note that --- in spite of certain similarity --- the first line of (\ref{jumpspsibar}) is \textit{not} obtained from (\ref{jumpspsi}) by exchange ${M_0\leftrightarrow M_1}$ because of reversed order of factors in the corresponding jumps. 
   New connection matrices are expressed in terms of $C_{\nu}$'s in the following way:
   \ben
   \begin{gathered}
   \overline{C}_{0}=e^{i\pi\Theta_1}C_{1,-} M_{\infty}=e^{-i\pi\Theta_1}C_{1,+},\qquad
   \overline{C}_{1-t,+}=e^{i\pi\Theta_t}C_{t,-} M_{\infty},\qquad
   \overline{C}_{1-t,-}=e^{-i\pi\Theta_t}C_{t,+},\\
   \overline{C}_{1,+}=e^{i\pi\Theta_0}C_{0} M_{\infty},\qquad
   \overline{C}_{1,-}=e^{-i\pi\Theta_0}C_{0},\qquad
   \overline{C}_{\infty}=e^{-i\pi\Theta_\infty}C_{\infty} M_{\infty}=e^{i\pi\Theta_\infty}C_{\infty}.
   \end{gathered}
   \ebn   
   The function $\overline{\Psi}\lb1-z\rb$ in the exterior of circles $\gamma_{\nu}$ coincides with the analytic continuation of the solution $\Phi\lb z\rb$ of the initial Fuchsian system from the upper half-plane $\Im z>0$ to the cut Riemann sphere $\mathbb \Pb\backslash [-\infty,1]$.
   
   Approximate solution for $\overline{\Psi}\lb z\rb$ is constructed in the same way as for $\Psi\lb z\rb$. The main building blocks are solutions $\overline{\Phi}^i\lb z\rb$ and $\overline{\Phi}^e\lb z\rb$ of two hypergeometric systems which model the jumps inside and outside the auxiliary circle $S$. To simplify the exposition we consider only sufficiently generic situation described by an analog of 
   Assumptions~\ref{asssigma} and \ref{assirrmon}:
   \begin{ass}\label{assum4}
   The matrix $M_{1t}:=M_1 M_t$ is assumed to be diagonalizable. It will be parameterized as 
   \ben
   M_{1t}=\overline C_{S}^{\,-1} e^{2\pi i \overline{ \mathfrak S}}\;\overline C_{ S},\qquad \overline{\mathfrak S}=
   \operatorname{diag}\left\{\overline\sigma,
   -\overline{\sigma}\right\}.
   \ebn
    It is furthermore assumed that $\left|\Re \overline\sigma\right|<\frac12$, $\overline\sigma\ne0$, and that
   \ben
   \theta_1+\theta_t\pm\overline\sigma,\;\;
   \theta_1-\theta_t\pm\overline\sigma,\;\;
   \theta_{\infty}+\theta_0\pm\overline\sigma,\;\;
   \theta_{\infty}-\theta_0\pm\overline\sigma\;\notin\mathbb Z.
   \ebn
   \end{ass}
   \noindent Referring to the notations of Subsection~\ref{subsmon}, $\overline\sigma$ can be identified with $\sigma_{t1}$. The latter quantity is related to previously used coordinates $\sigma$, $\eta$ by formula (\ref{p1tf}).

  Exterior and interior parametrices are solutions of the Fuchsian systems
   \ben
   \partial_z\overline\Phi^e=\lb\frac{\overline{A}_0^e}{z}   
   + \frac{\overline{A}_1^e}{z-1} \rb\overline\Phi^e,\qquad 
   \partial_z\overline\Phi^i=\lb\frac{\overline{A}_0^i}{z}   
      + \frac{\overline{A}_1^i}{z-1} \rb\overline\Phi^i,
   \ebn
   normalized by the asymptotic conditions $ \overline\Phi^e\lb z\rb\simeq \lb -z\rb^{\overline{\mathfrak S}}$ as $z\to0$, $\overline\Phi^i\lb z\rb\simeq\lb -z\rb^{\overline{\mathfrak S}}$ as $z\to -\infty$. Furthermore, 
   one has $\overline{A}_0^e=\overline{A}_0^i+A_{1}^i=\overline{\mathfrak S}$. The spectra of $\overline{A}_0^i$, $\overline{A}_1^i$, $\overline{A}_1^e$ and $\overline{A}_0^e+\overline{A}_1^e$ are given
   by $\pm\theta_1$, $\pm\theta_t$, $\pm\theta_0$ and $\pm\theta_{\infty}$ respectively. All results of previous subsections have their $t\to1$ counterparts, most of which are obtained by the replacements (the notation should be sufficiently self-explanatory)
   \beq
   \begin{gathered}
   \Phi^{e}\lb z\rb\mapsto \overline\Phi^{\,e}\lb z\rb,\qquad
   \Phi^{i}\lb z\rb\mapsto \overline\Phi^{\,i}\lb z\rb,\\ 
   G_{\nu}^e\lb z\rb\mapsto \overline{G}_{\nu}^e\lb z\rb,\qquad
   G_{\nu}^i\lb z\rb\mapsto \overline{G}_{\nu}^{\,i}\lb z\rb,
   \qquad \nu=0,1,\infty, \\
   G^e_{1,\infty}\mapsto \overline{G}^e_{1,\infty},\qquad
   C^e_{1,\infty}\mapsto \overline{C}^e_{1,\infty},\qquad 
   G^i_{0,1}\mapsto \overline{G}^i_{0,1},\qquad C^i_{0,1}\mapsto \overline{C}^i_{0,1},\\
   \lb\theta_0,\theta_t,\theta_1,\theta_{\infty},\sigma,r_e,r_i\rb
   \mapsto
   \lb\theta_1,\theta_t,\theta_0,\theta_{\infty},\overline\sigma,\overline r_e,\overline r_i\rb,\\
   \lb c^e_{\infty,q},c^e_{1,q},c^i_{0,q},c^i_{1,q}\rb
   \mapsto
   \lb \overline c^e_{\infty,q},
   \overline c^e_{1,q},\overline c^i_{0,q},
   \overline c^i_{1,q}\rb,\qquad q=a,b,
   \end{gathered}
   \eeq
      In particular, one can prove the following statement:
   \begin{lemma} The form $\Omega$ can be alternatively written as $d\overline{\omega}_0$, where $\overline \omega_0\in\Lambda^1\lb \mathcal  M\rb$ is given by
   \begin{subequations}
   \begin{align}
   \overline \omega_0&=\operatorname{Tr}\left(\Theta_1
   {\overline G^i_0}^{-1}d_{\mathcal M}\overline G^i_0+
   \Theta_t
      {\overline G^i_1}^{-1}d_{\mathcal M}\overline G^i_1+
      \Theta_0
            {\overline G^e_1}^{-1}d_{\mathcal M}\overline G^e_1+
   \Theta_{\infty} {\overline G^e_{\infty}}^{-1}d_{\mathcal M}\overline G^e_{\infty}\right)=\\
   &=\theta_{\infty}d_{\mathcal M}\ln 
        \frac{\overline r_e \overline c^e_{\infty b}}{
        \overline c^e_{\infty a}} +\theta_{1}\,d_{\mathcal M}\ln 
            \frac{\overline r_i \overline c^i_{0 b}}{\overline c^i_{0 a}}+
            \theta_{t}d_{\mathcal M}\ln 
                 \frac{\overline r_i \overline c^i_{1 b}}{
                 \overline c^i_{1 a}}+
 \theta_{0}d_{\mathcal M}\ln 
      \frac{\overline r_e \overline c^e_{1 b}}{
         \overline c^e_{1 a}}
                 + \overline \sigma   d_{\mathcal M}\ln \frac{\overline r_i}{\overline r_e}   -2d_{\mathcal M}\lb \theta_0+\theta_t\rb.
   \end{align}
   \end{subequations}
   The first equation remains valid in arbitrary rank and the second is its specialization to $N=2$.
   \end{lemma}
   
   The form $\overline\omega_0$ can be used to define a second tau function
   $\tau_{\overline{\omega}_0}\lb t\rb$ via $d\ln\tau_{\overline{\omega}_0}=\omega-\overline\omega_0$. This tau function is of course proportional to $\tau_{\omega_0}\lb t\rb$ but its asymptotics is normalized at $t=1$ instead of $t=0$. Let us formulate an analog of Proposition~\ref{jimbof} for $\tau_{\overline{\omega}_0}\lb t\rb$:
   \begin{prop}\label{jimbof1}
   Normalized Painlev\'e VI tau function $\tau_{\overline\omega_0}\lb t\rb$ has the following asymptotics as $t\to 1$ (cf (\ref{p6at1})):
      \beq
      \begin{gathered}
      \label{jimboextf1}
      \tau_{\overline\omega_0}(t)=\mathcal{C}_{\overline\omega_0}\cdot \lb 1-t\rb^{\overline\sigma^2-\theta_1^2-\theta_t^2}\,\Biggl[ 1-
      \sum_{\epsilon=\pm1}\frac{
      \left(\left(\theta_t-\epsilon\overline\sigma\right)^2-
      \theta_1^2\right)
      \left(\left(\theta_0-\epsilon\overline\sigma\right)^2-
      \theta_{\infty}^2
      \right)}{
      4\overline\sigma^2\left(1+2\epsilon\overline\sigma\right)^2}\,
      \overline\kappa^{\,\epsilon}
      \lb 1-t\rb^{1+2\epsilon\overline\sigma}+
      \Biggr. \\ 
      \qquad\qquad\qquad\qquad\quad +\frac{
         \left(\overline\sigma^2-\theta_{\infty}^2+\theta_0^2\right)
         \left(\overline\sigma^2-\theta_1^2+\theta_t^2\right)}{2
         \overline\sigma^2}
         \lb 1-t\rb +
      o\left(1-t\right)\Biggr], \quad t \to 1.
      \end{gathered}
      \eeq
      Here the coefficient $\overline\kappa$ is defined by
      \begin{align}\label{kappadef1}
      &\overline\kappa=
      \frac{\Gamma^2\left(1-2\overline\sigma\right)
      \Gamma\left(1+\theta_1+\theta_t+\overline\sigma\right)
      \Gamma\left(1-\theta_1+\theta_t+\overline\sigma\right)
      \Gamma\left(1+\theta_{\infty}+\theta_0+\overline\sigma\right)
      \Gamma\left(1-\theta_{\infty}+\theta_0+\overline\sigma\right)}{
      \Gamma^2\left(1+2\overline\sigma\right)
      \Gamma\left(1+\theta_1+\theta_t-\overline\sigma\right)
      \Gamma\left(1-\theta_1+\theta_t-\overline\sigma\right)
      \Gamma\left(1+\theta_{\infty}+\theta_0-\overline\sigma\right)
      \Gamma\left(1-\theta_{\infty}+\theta_0-\overline\sigma\right)}\,
      e^{i\overline\eta},\\ \label{etabardef}
      &e^{i\overline\eta}= \frac{
        -2i\sin2\pi \overline \sigma\lb p_{01}+p_{0t}e^{-2\pi i \overline
        \sigma}\rb+
        \lb p_0p_t+p_1p_{\infty}\rb-\lb p_0p_1+p_tp_{\infty}\rb
        e^{-2\pi i \overline \sigma}
        }{\lb p_{\infty}-2\cos2\pi\lb\theta_0-\overline\sigma\rb\rb
        \lb p_{1}-2\cos2\pi\lb\theta_t-\overline\sigma\rb\rb},
      \end{align}
      and the constant prefactor $\mathcal C_{\overline\omega_0}$ does not depend on monodromy data.
   \end{prop}
   \pf The only subtlety that has to be taken into account as compared to derivation of Proposition~\ref{jimbof} concerns the quantity $e^{i\overline\eta}$ and is as follows. The coefficient
   $\overline\kappa $ coincides with the ratio $\overline r_i/
   \overline r_e$, and $e^{i\overline\eta}$ with $\overline
   s_i/\overline s_e$. The last ratio is expressed in terms of traces of products of monodromy matrices $\overline{M}_{0,t,1,\infty}$ 
   associated with $\overline\Psi\lb z\rb$ by overlining parameters in (\ref{etareltheta}). These new monodromy matrices can be expressed in terms of $M_{0,t,1,\infty}$ using the first line of (\ref{jumpspsibar}). One has
   \ben
   \overline{M}_0=M_1,\qquad \overline M_{t}=M_1M_tM_1^{-1},\qquad
   \overline{M}_1=M_1M_tM_0M_t^{-1}M_1^{-1},\qquad\overline M_{\infty}=M_{\infty}.
   \ebn
   This implies simple transformation formulas for most of the trace functions:
   \begin{subequations}\label{transformedtrf}
   \beq
   \lb \overline p_0,\overline p_t,\overline p_1,\overline p_{\infty},
   \overline p_{0t},\overline p_{t1}\rb=
   \lb p_1,p_t,p_0,p_{\infty},p_{t1},p_{0t}\rb.
   \eeq
   The only exception is 
   \beq
   \overline p_{01}=\operatorname{Tr}\overline M_0\overline M_1=
   \operatorname{Tr}M_1 M_t M_0 M_t^{-1}=p_0p_1+p_tp_{\infty}-p_{01}-p_{0t}p_{t1}.
   \eeq
   \end{subequations}
   The last equality follows from the fact that any $M\in SL_2\lb 
   \mathbb C\rb$ satisfies skein relation $M+M^{-1}=
   \operatorname{Tr} M\,\cdot \mathbf 1$. Relation~(\ref{etabardef})
   is then obtained from (\ref{etareltheta}) by identification
   (\ref{transformedtrf}).
   \epf

   \subsection{Connection problem for Painlev\'e VI tau function. Proof of Theorem \ref{thp6}\label{secconcof}}
   The tau functions $\tau_{{\omega}_0}\lb t\rb$ and $\tau_{\overline{\omega}_0}\lb t\rb$ can differ only by a constant (i.e. independent of $t$) factor of relative normalization 
   \beq\label{upsilondef1}
   \Upsilon\lb M\rb:=\frac{\mathcal C_{\overline\omega_0}}{\mathcal C_{\omega_0}}\frac{\tau_{{\omega}_0}\lb t\rb}{\tau_{\overline{\omega}_0}\lb t\rb} ,
   \eeq
    which is our main quantity of interest in this section. It can also be written for generic non-normalized Pain\-lev\'e~VI tau function $\tau\lb t\rb$ as
   \beq\label{upsilondef2}
   \Upsilon\lb  M\rb=\frac{\lim_{t\to 1}
   \lb 1-t\rb^{\theta_t^2+\theta_1^2-\overline\sigma^2}
   \tau\lb t\rb}{\lim_{t\to 0}
   t^{\theta_0^2+\theta_t^2-\sigma^2}\tau\lb t\rb} \equiv \frac{\mathcal{C}_1}{\mathcal{C}_0}, 
   \eeq  
   where $\mathcal{C}_{0}$ and $\mathcal{C}_{1}$ are the constants from the equations (\ref{jimfs}).
     The constant   $\Upsilon\lb  M\rb$ is completely determined by Painlev\'e~VI equation and appropriate initial conditions --- that is, it depends only on the conjugacy class of monodromy. The latter property is by no means manifest in the representation 
   \beq\label{upsilondef3}
   d_{\mathcal M}\ln\Upsilon=\overline\omega_0-\omega_0,
   \eeq
   and becomes yet more implicit if we rewrite the right side using previous results: 
   \beq\label{difconcon}
   \begin{gathered}
   \begin{aligned}
   &d_{\mathcal M}\ln\Upsilon= 
    \overline \sigma   d_{\mathcal M}\ln \frac{\overline r_i}{\overline r_e}  - \sigma   d_{\mathcal M}\ln \frac{r_i}{r_e}   
          +2d_{\mathcal M}\lb\theta_1- \theta_0\rb\\
   &+\theta_{\infty}d_{\mathcal M}\ln 
   \frac{\overline r_e \overline c^e_{\infty b}c^e_{\infty a}}{r_e \overline c^e_{\infty a}c^e_{\infty b}}+
        \theta_{1}d_{\mathcal M}\ln 
             \frac{\overline r_i \overline c^i_{0 b}c^e_{1 a}
             }{r_e \overline c^i_{0 a}c^e_{1 b}}+  
             \theta_{t}d_{\mathcal M}\ln 
               \frac{\overline r_i \overline c^i_{1 b} c^i_{1 a}}{r_i \overline c^i_{1 a} c^i_{1 b}} +
   +\theta_{0}\,d_{\mathcal M}\ln 
   \frac{\overline r_e \overline c^e_{1 b}c^i_{0 a}}{r_i    \overline c^e_{1 a} c^i_{0 b}}.                
                 \end{aligned}
    \end{gathered}             
   \eeq
  It is even less obvious that the last expression is a closed $1$-form! Our task is now to rewrite (\ref{difconcon}) in terms of more convenient local coordinates on
   $\mathcal M$, such as $\theta_{0,t,1,\infty}$, $\sigma$ and $\eta$.
  
 The right side of (\ref{difconcon}) is expressed in terms of  parameters 
 \ben
 \theta_{0,t,1,\infty}, \sigma, \overline \sigma,
 r_{e,i}, \overline r_{e,i}, c^e_{\infty q}, 
 c^e_{1 q}, \overline c^e_{\infty q}, \overline c^e_{1 q},
 c^i_{0q}, c^i_{1q}, \overline c^i_{0q}, \overline c^i_{1q},
 \ebn
 which appear in the monodromy and connection matrices. Of course, not all of them are independent.   
 Connection matrices of the original $4$-point problem can be related to $3$-point ones in two different ways:
 \beq
 \begin{gathered}
 C^e_{\infty}=C_{\infty}C_S^{-1},\qquad C^e_{1,+}=C_{1,+}C_S^{-1},\qquad
 C^i_{1,+}=C_{t,+} C_S^{-1},
 \qquad C_0^i=C_0 C_S^{-1},\\
 \overline C_{0}=e^{-i\pi\Theta_1}C_{1,+},\qquad
 \overline C_{1-t,-}=
 e^{-i\pi\Theta_t}C_{t,+},\qquad
 \overline C_{1,-}=e^{-i\pi\Theta_0}C_{0},\qquad
 \overline C_{\infty}=e^{i\pi\Theta_{\infty}}C_{\infty},\\
 \overline C_{\infty}^e=\overline C_{\infty} \overline C_S^{\,-1},
 \qquad
 \overline C^e_{1,-}= \overline C_{1,-} \overline C_S^{\,-1},\qquad
 \overline C^i_{1,-}=\overline C_{1-t,-} \overline C_S^{\,-1},\qquad
 \overline C^i_{0}=\overline C_0 \overline C_S^{\,-1}.
 \end{gathered}
 \eeq
 Here the first line reproduces (\ref{connmE}), (\ref{connmI}), the third lists analogous relations arising in the study of 
 $t\to 1$ tau function asymptotics and the middle one relates the two
 sets of connection matrices. It follows that
 \beq\label{connrels}
 \overline C^e_{\infty}= e^{i\pi\Theta_{\infty}}C^e_{\infty}C_{S\overline S}, \qquad
 \overline C^e_{1,-}= e^{-i\pi\Theta_0}C^i_{0}C_{S\overline S},
 \qquad
 \overline C^i_{1,-}=e^{-i\pi\Theta_t}C^i_{1,+}C_{S\overline S},\qquad
 \overline C^i_0=e^{-i\pi\Theta_1}C^e_{1,+}C_{S\overline S}.
 \eeq
 where $C_{S\overline S}=C_S \overline C_S^{\,-1}$. Since by definition
 $\overline C_S M_1 M_t\overline C_S^{\,-1}=e^{2\pi i\overline{\mathfrak S}}$, the transformation $C_{S\overline S}$ diagonalizes the product~$\tilde M_1\tilde M_t=
  C_{S\overline S}\,
 e^{2\pi i  \overline{\mathfrak S}}\ds
  {C_{S\overline S}}^{\!\!\!-1}$, where $\tilde M_t,\tilde M_1$ are defined in (\ref{mtpar}), (\ref{m1par}). This in turn fixes $C_{S\overline S}$ up to right multiplication by diagonal matrix.  In fact one can write
  \ben
  C_{S\overline S}=\mathsf S_e^{-1} \mathsf C_{S\overline S}\,
  \mathsf{D}_{S\overline S},
  \ebn
  where $\mathsf{D}_{S\overline S}=\operatorname{diag}\left\{d_a,d_b\right\}$, $\mathsf S_e=\operatorname{diag}\left\{1,s_e\right\}$ and
  \beq\label{CSSsf}
  \begin{aligned}
  &\mathsf C_{S\overline S}\equiv 
  \lb\begin{array}{cc}
  \mathsf c_{11} &\mathsf c_{12} \\ \mathsf c_{21} & \mathsf c_{22}
  \end{array}\rb=\lb\begin{array}{cc}
  \alpha & \alpha \\ e^{2\pi i \overline\sigma}-\beta & 
  e^{-2\pi i \overline\sigma}-\beta
  \end{array}\rb,\\
  &4\alpha\sin^22\pi\sigma=\lb p_{\infty}-2\cos2\pi\lb
  \theta_1+\sigma\rb\rb\lb p_0 e^{2\pi i \sigma}
  -p_t\rb-\lb p_{0}-2\cos2\pi\lb
    \theta_t-\sigma\rb\rb\lb p_{\infty} e^{2\pi i \sigma}
    -p_1\rb e^{i\eta},\\
  &4\beta\sin^22\pi\sigma=\lb p_t e^{2\pi i \sigma}
    -p_0\rb \lb p_1 e^{-2\pi i \sigma}
        -p_{\infty}\rb-\lb p_{0}-2\cos2\pi\lb
      \theta_t+\sigma\rb\rb\lb p_{\infty} 
      -2\cos2\pi\lb\theta_1+\sigma\rb\rb e^{-i\eta}.
  \end{aligned}
  \eeq
  The fact that $\mathsf C_{S\overline S}$ depends only on the conjugacy class of monodromy will be important below. On the other hand, explicit form of $d_{a,b}$ does not play any role in subsequent computation.
  
     Recall that the ratios $r_i/r_e$ and $\overline r_i/\overline r_e$ from the first line of (\ref{difconcon}) coincide with $\kappa$ and $\overline \kappa$  defined by (\ref{kappadef}) and (\ref{kappadef1}). They are thus preserved by conjugation.        
    We are now going to check that the ratios from  the second line  share this important property. In order to  derive invariant expressions suitable for subsequent computation of $\Upsilon$, it is crucial to rewrite $3$-point connection matrices in a different form. As we will see in a moment, their nontrivial dependence on monodromy parameters is trigonometric. The key is the following elementary observation.
   
   Let us denote by $\mathfrak M$ the set of $2\times 2$ matrices with non-zero elements. Let $\rho:\mathfrak M\to\mathbb C^\times$ be defined by 
   \ben\rho\lb M\rb :=\ds \frac{M_{11}M_{22}}{M_{12}M_{21}}.
   \ebn
   Two matrices  $M,M'\in\mathfrak M$ will be called $\rho$-equivalent if $\rho\lb M\rb=\rho\lb M'\rb$.
   \begin{lemma}
    Two matrices  $M,M'\in\mathfrak M$ are $\rho$-equivalent iff there exists a pair of  diagonal matrices $D_{L,R}$ such that ${M'=D_L M D_R}$.
   \end{lemma}
   \noindent This basic fact can be used to compute four ratios from the second line of  (\ref{difconcon}). Below we explain the details of the procedure for one of these, namely, $\ds \frac{\overline r_e \overline c^e_{\infty b}c^e_{\infty a}}{r_e \overline c^e_{\infty a}c^e_{\infty b}}$.
   
    The $\rho$-invariant of the connection matrix $C^e_{\infty}$ given by  (\ref{connmext}) is equal to
   \ben
   \rho\lb C^e_{\infty}\rb=\frac{
   \sin\pi\lb \theta_{\infty}+\theta_1-\sigma\rb
   \sin\pi\lb \theta_{\infty}-\theta_1-\sigma\rb}{
   \sin\pi\lb \theta_{\infty}+\theta_1+\sigma\rb
   \sin\pi\lb \theta_{\infty}-\theta_1+\sigma\rb   }.
   \ebn
   It is therefore not surprising that we can parameterize
   $C^e_{\infty}$ as
   \ben
   C^e_{\infty}=\mathsf{D}^e_{\infty,L}
   \mathsf{C}^e_{\infty} \mathsf{D}^e_{\infty,R}\mathsf R_e,
   \ebn
   with
   \beq\label{factorCE}
   \begin{gathered}
   \begin{aligned}
    &  \mathsf{C}^e_{\infty}=\lb
      \begin{array}{cc}
      \sin\pi\lb \theta_{\infty}+\theta_1-\sigma\rb & 
      \sin\pi\lb \theta_{\infty}-\theta_1+\sigma\rb \\
      \sin\pi\lb \theta_{\infty}+\theta_1+\sigma\rb
      & \sin\pi\lb \theta_{\infty}-\theta_1-\sigma\rb
      \end{array}\rb,\qquad
      \mathsf R_e=\operatorname{diag}\left\{1,r_e\right\},\\
   &\mathsf{D}^e_{\infty,L}=\operatorname{diag}
   \left\{\frac{c^e_{\infty a}\Gamma\lb -2\theta_{\infty}\rb
      \Gamma\lb \theta_{\infty}-\theta_1+\sigma\rb}{r_e
      \Gamma\lb 1-\theta_{\infty}+\theta_1+\sigma\rb},
  \frac{c^e_{\infty b}\Gamma\lb 2\theta_{\infty}\rb
        \Gamma\lb -\theta_{\infty}-\theta_1-\sigma\rb}{
        \Gamma\lb 1+\theta_{\infty}+\theta_1-\sigma\rb}    
      \right\},\\
   &\mathsf{D}^e_{\infty,R}=\operatorname{diag}
         \left\{-\frac{\Gamma\lb 2\sigma\rb
            \Gamma\lb 1+\theta_{\infty}+\theta_1-\sigma\rb}{
           \pi \Gamma\lb \theta_{\infty}-\theta_1+\sigma\rb},
        \frac{\Gamma\lb -2\sigma\rb
              \Gamma\lb 1- \theta_{\infty}+\theta_1+\sigma\rb}{\pi
              \Gamma\lb -\theta_{\infty}-\theta_1-\sigma\rb}    
            \right\}.
    \end{aligned}        
   \end{gathered}
   \eeq
   Analogous expressions for $\overline C^e_{\infty}$ is obtained by 
   the exchange $\theta_0\leftrightarrow \theta_1$, $\sigma\to\overline\sigma$, $r_e\to \overline r_e$, $c^{e}_{\infty q}\to \overline c^{e}_{\infty q}$. We can now rewrite the first of 
   relations (\ref{connrels}) in the form
   \ben
   \lb\mathsf{C}^e_{\infty}\mathsf{D}^e_{\infty,R}
       \mathsf R_e \mathsf S_e^{-1}
       \mathsf C_{S\overline S}\,\rb^{-1}
   \lb\begin{array}{cc} x & 0 \\ 0 & y \end{array}\rb 
    {\overline{\mathsf{C}}^{\,e}_{\infty}}= \begin{array}{c}\text{diagonal}
       \\ \text{matrix} \end{array},
   \ebn
   where $\operatorname{diag}\left\{x,y\right\}=
   {\mathsf D^{e}_{\infty,L}}^{-1}
   e^{-i\pi\Theta_{\infty}}
   \overline{\mathsf{D}}^{\,e}_{\infty,L}$. We are  interested in the ratio 
   \begin{align}\label{ratioI1}
   &\frac yx=e^{2\pi i \theta_{\infty}}
   \frac{\Gamma\lb-\theta_1+\theta_{\infty}+\sigma\rb
      \Gamma\lb 1+\theta_1+\theta_{\infty}-\sigma\rb
      \Gamma\lb-\theta_0-\theta_{\infty}-\overline\sigma\rb
      \Gamma\lb 1+\theta_0-\theta_{\infty}+\overline\sigma\rb}{
      \Gamma\lb-\theta_1-\theta_{\infty}-\sigma\rb
         \Gamma\lb 1+\theta_1-\theta_{\infty}+\sigma\rb
         \Gamma\lb-\theta_0+\theta_{\infty}+\overline\sigma\rb
         \Gamma\lb 1+\theta_0+\theta_{\infty}-\overline\sigma\rb}
   \frac{\overline r_e \overline c^e_{\infty b}c^e_{\infty a}}{r_e \overline c^e_{\infty a}c^e_{\infty b}},
   \end{align}
    which is readily computed as
   \beq\label{ratioI2}
   \begin{aligned}
   \frac yx &=-\frac{
    \lb \mathsf C^e_{\infty}\mathsf{D}^e_{\infty,R}
              \mathsf R_e \mathsf S_e^{-1}
              \mathsf C_{S\overline S}\,\rb^{-1}_{11}\lb
              \overline C^{\, e}_{\infty}\rb_{12} }{\lb \mathsf C^e_{\infty}\mathsf{D}^e_{\infty,R}
             \mathsf R_e \mathsf S_e^{-1}
             \mathsf C_{S\overline S}\,\rb^{-1}_{12}\lb
             \overline C^{\, e}_{\infty}\rb_{22}}
  = \frac{
      \lb \mathsf C^e_{\infty}\mathsf{D}^e_{\infty,R}
                \mathsf R_e \mathsf S_e^{-1}
                \mathsf C_{S\overline S}\,\rb_{22}\lb
                \overline C^{\, e}_{\infty}\rb_{12}}{\lb \mathsf C^e_{\infty}\mathsf{D}^e_{\infty,R}
               \mathsf R_e \mathsf S_e^{-1}
               \mathsf C_{S\overline S}\,\rb_{12}\lb
               \overline C^{\, e}_{\infty}\rb_{22}}.                            
   \end{aligned}                   
   \eeq
 All quantities in the last expression are manifestly invariant: $\mathsf C^e_{\infty}$, $\overline{\mathsf{C}}^e_{\infty}$, $\mathsf D^e_{\infty,R}$ are given by (\ref{factorCE}), $\mathsf{C}_{S\overline S}$ by (\ref{CSSsf}), and $ \mathsf R_e \mathsf S_e^{-1}$ by (\ref{sere}). Moreover, up to irrelevant scalar multiple the matrix $C^e_{\infty}\mathsf{D}^e_{\infty,R}
                 \mathsf R_e \mathsf S_e^{-1}$ is given by trigonometric expression
 \beq\label{ratioI3}
 C^e_{\infty}\mathsf{D}^e_{\infty,R} \mathsf R_e \mathsf S_e^{-1}
 \sim \lb
 \begin{array}{cc}
 \cos2\pi\theta_{\infty}-\cos2\pi\lb\theta_1-\sigma\rb &
 \cos2\pi\theta_1-\cos2\pi\lb\theta_{\infty}+\sigma\rb \\
 \cos2\pi\lb\theta_{\infty}+\sigma\rb-\cos2\pi\theta_1 &
 \cos2\pi\lb\theta_1+\sigma\rb-\cos2\pi\theta_{\infty}
 \end{array}\rb.
 \eeq                
 
 After some algebra involving Euler reflection formula
 $\Gamma\lb z\rb\Gamma \lb 1-z\rb=\frac{\pi}{\sin\pi z}$, relations
 (\ref{ratioI1})--(\ref{ratioI3}) yield the following result:
 \begin{subequations}\label{fourratios}
 \begin{lemma}
 One has
 \beq
 \begin{gathered}
 \begin{aligned}
 \frac{\overline r_e \overline c^e_{\infty b}c^e_{\infty a}}{r_e \overline c^e_{\infty a}c^e_{\infty b}}=&\,
 \frac{
 \Gamma\lb 1+\theta_1-\theta_{\infty}-\sigma\rb
 \Gamma\lb 1+\theta_1-\theta_{\infty}+\sigma\rb
 \Gamma\lb 1+\theta_0+\theta_{\infty}-\overline\sigma\rb
 \Gamma\lb -\theta_0+\theta_{\infty}-\overline\sigma\rb}{
 \Gamma\lb 1+\theta_1+\theta_{\infty}+\sigma\rb
 \Gamma\lb 1+\theta_1+\theta_{\infty}-\sigma\rb
 \Gamma\lb 1+\theta_0-\theta_{\infty}-\overline\sigma\rb
 \Gamma\lb -\theta_0-\theta_{\infty}-\overline\sigma\rb}\times \\
 \times&\,e^{-2\pi i \theta_{\infty}}
 \frac{\mathsf c_{12}\sin\pi\lb
 \theta_1-\theta_{\infty}-\sigma\rb-
 \mathsf c_{22}\sin\pi\lb
  \theta_1-\theta_{\infty}+\sigma\rb}{
 \mathsf c_{12}\sin\pi\lb
 \theta_1+\theta_{\infty}-\sigma\rb-
  \mathsf c_{22}\sin\pi\lb
   \theta_1+\theta_{\infty}+\sigma\rb},
 \end{aligned}  
   \end{gathered}
 \eeq
 where matrix elements $\mathsf c_{12}$, $\mathsf c_{22}$ are given by
 (\ref{CSSsf}).
 \end{lemma}
  Explicit invariant expressions for the other three ratios from the  second line of (\ref{difconcon}) can be derived in a completely analogous fashion. The result is
  \begin{align}
 \frac{\overline r_i \overline c^i_{0 b}c^e_{1 a}}{
 r_e \overline c^i_{0 a}c^e_{1 b}}=&\,
 \frac{
 \Gamma\lb-\theta_1+\theta_{\infty}+\sigma\rb
 \Gamma\lb-\theta_1-\theta_{\infty}-\sigma\rb
 \Gamma\lb 1+\theta_1+\theta_t+\overline\sigma\rb
 \Gamma\lb \theta_1-\theta_t+\overline\sigma\rb}{
 \Gamma\lb 1+\theta_1+\theta_{\infty}-\sigma\rb
 \Gamma\lb 1+\theta_1-\theta_{\infty}+\sigma\rb
 \Gamma\lb 1-\theta_1+\theta_t+\overline\sigma\rb
 \Gamma\lb -\theta_1-\theta_t+\overline\sigma\rb}\times\\
 \nonumber\times &\,\frac{e^{2\pi i \theta_1}\bigl(\cos2\pi\theta_1-\cos2\pi
 \lb\theta_{\infty}+\sigma\rb\bigr)\lb\mathsf c_{12}e^{-\pi i\sigma}-
 \mathsf c_{22}e^{\pi i \sigma}\rb}{
 \mathsf c_{12}e^{-\pi i\sigma}\bigl(
 \cos2\pi\lb\theta_1-\sigma\rb-\cos2\pi\theta_{\infty}\bigr)-
  \mathsf c_{22}e^{\pi i \sigma}\bigl(
   \cos2\pi\lb\theta_1+\sigma\rb-\cos2\pi\theta_{\infty}\bigr)},\\  
 \frac{\overline r_i \overline c^i_{1 b} c^i_{1 a}}{
 r_i \overline c^i_{1 a} c^i_{1 b}}=&\,
 \frac{\Gamma\lb\theta_0-\theta_t-\sigma\rb
 \Gamma\lb -\theta_0-\theta_t+\sigma\rb
 \Gamma\lb 1+\theta_1+\theta_t+\overline\sigma\rb
 \Gamma\lb 1-\theta_1+\theta_t+\overline\sigma\rb}{
 \Gamma\lb 1+\theta_0+\theta_t+\sigma\rb
  \Gamma\lb 1-\theta_0+\theta_t-\sigma\rb
  \Gamma\lb \theta_1-\theta_t+\overline\sigma\rb
  \Gamma\lb -\theta_1-\theta_t+\overline\sigma\rb}\times\\
 \nonumber\times &\, 
 \frac{e^{-2\pi i \theta_t}\bigl(\cos2\pi\theta_t-\cos2\pi
  \lb\theta_0-\sigma\rb\bigr)\lb\mathsf c_{12}e^{-\pi i\sigma}-
  \mathsf c_{22}e^{\pi i \sigma+i\eta}\rb}{
  \mathsf c_{12}e^{-\pi i\sigma}\bigl(
  \cos2\pi\lb\theta_t+\sigma\rb-\cos2\pi\theta_0\bigr)-
   \mathsf c_{22}e^{\pi i \sigma+i\eta}\bigl(
    \cos2\pi\lb\theta_t-\sigma\rb-\cos2\pi\theta_0\bigr)},\\
 \frac{\overline r_e \overline c^e_{1 b}c^i_{0 a}}{
 r_i \overline c^e_{1 a} c^i_{0 b}}=&\,
 \frac{
 \Gamma\lb 1-\theta_0+\theta_t+\sigma\rb
 \Gamma\lb 1-\theta_0+\theta_t-\sigma\rb
 \Gamma\lb 1+\theta_0+\theta_{\infty}-\overline\sigma\rb
 \Gamma\lb 1+\theta_0-\theta_{\infty}-\overline\sigma\rb}{
 \Gamma\lb1+\theta_0+\theta_t+\sigma\rb
 \Gamma\lb 1+\theta_0+\theta_t-\sigma\rb
 \Gamma\lb -\theta_0+\theta_{\infty}-\overline\sigma\rb
 \Gamma\lb -\theta_0-\theta_{\infty}-\overline\sigma\rb}\times \\
 \nonumber \times&\,
  \lb -e^{2\pi i \theta_{0}}\rb
  \frac{\mathsf c_{12}\sin\pi\lb
  \theta_0-\theta_t-\sigma\rb-
  \mathsf c_{22}e^{i\eta}\sin\pi\lb
   \theta_0-\theta_t+\sigma\rb}{
  \mathsf c_{12}\sin\pi\lb
  \theta_0+\theta_t+\sigma\rb-
   \mathsf c_{22}e^{i\eta}\sin\pi\lb
    \theta_0+\theta_t-\sigma\rb}.
  \end{align}
  \end{subequations}
  We now have at our disposal all ingredients of the  $1$-form (\ref{difconcon}) expressed in terms of local coordinates  on the space $\mathcal M$ of monodromy data. 
  
  It is still very far from obvious that the right side of (\ref{difconcon}) is indeed a \textit{closed} 1-form. To check that this is the case and to find the antiderivative $\Upsilon\lb M\rb$, we will proceed in several steps. First let us eliminate the gamma function factors appearing in (\ref{fourratios}) and in the ratios $\kappa=r_i/r_e$, $\overline\kappa=\overline r_i/\overline r_e$ given by (\ref{kappadef}), (\ref{kappadef1}) by introducing the notation
  \beq\label{ccredef}
  \Upsilon\lb M\rb=\hat\Upsilon\lb M\rb
  \prod_{\epsilon,\epsilon'=\pm}
   \frac{G\lb 1+\epsilon\overline\sigma+\epsilon'\theta_t-
       \epsilon\epsilon'\theta_1\rb G\lb1+\epsilon\overline\sigma+\epsilon'\theta_0-
           \epsilon\epsilon' \theta_{\infty}\rb}{
   G\lb1+\epsilon\sigma+\epsilon'\theta_t+\epsilon\epsilon'\theta_0\rb   G\lb1+\epsilon\sigma+\epsilon'\theta_1+\epsilon\epsilon'
       \theta_{\infty}\rb}
   \;\prod_{\epsilon=\pm}
     \frac{G(1+2\epsilon\sigma)}{G(1+2\epsilon\overline\sigma)}
  ,
  \eeq
  where $G\lb z\rb$ denotes Barnes $G$-function.
  \begin{lemma} We have
  \begin{align}
  \nonumber & d_{\mathcal M}\ln\hat\Upsilon=
  \overline\sigma d_{\mathcal M}\ln\lb  \frac{
  \sin\pi\lb\theta_0-\theta_{\infty}-\overline\sigma\rb
  \sin\pi\lb\theta_1-\theta_t+\overline\sigma\rb}{
  \sin\pi\lb\theta_0+\theta_{\infty}+\overline\sigma\rb
    \sin\pi\lb\theta_1+\theta_t+\overline\sigma\rb} e^{i\overline\eta}\rb
  -\sigma d_{\mathcal M}\ln\lb \frac{
  \sin\pi\lb\theta_0+\theta_t-\sigma\rb
  \sin\pi\lb\theta_1+\theta_{\infty}-\sigma\rb}{
  \sin\pi\lb\theta_0-\theta_t-\sigma\rb
  \sin\pi\lb\theta_1-\theta_{\infty}+\sigma\rb} e^{i\eta}\rb  
    \\
  \nonumber &\!\!+\theta_{\infty} d_{\mathcal M}\ln\lb\frac{
  \sin\pi\lb\theta_1+\theta_{\infty}-\sigma\rb
  \sin\pi\lb\theta_0-\theta_{\infty}-\overline\sigma\rb}{
  \sin\pi\lb\theta_1-\theta_{\infty}+\sigma\rb
  \sin\pi\lb\theta_0-\theta_{\infty}+\overline\sigma\rb}
   \frac{\mathsf c_{12}\sin\pi\lb
   \theta_1-\theta_{\infty}-\sigma\rb-
   \mathsf c_{22}\sin\pi\lb
    \theta_1-\theta_{\infty}+\sigma\rb}{
   \mathsf c_{12}\sin\pi\lb
   \theta_1+\theta_{\infty}-\sigma\rb-
    \mathsf c_{22}\sin\pi\lb
     \theta_1+\theta_{\infty}+\sigma\rb}\,  e^{-2\pi i \theta_{\infty}}\rb\\
  \nonumber &+\theta_1d_{\mathcal M}\ln\lb
    \frac{\sin\pi\lb\theta_1+\theta_t-\overline\sigma\rb}{
    \sin\pi\lb\theta_1+\theta_t+\overline\sigma\rb}
    \frac{e^{2\pi i \theta_1}\bigl(\cos2\pi\theta_1-\cos2\pi
     \lb\theta_{\infty}-\sigma\rb\bigr)\lb\mathsf c_{12}e^{-\pi i\sigma}-
     \mathsf c_{22}e^{\pi i \sigma}\rb}{
     \mathsf c_{12}e^{-\pi i\sigma}\bigl(
     \cos2\pi\lb\theta_1-\sigma\rb-\cos2\pi\theta_{\infty}\bigr)-
      \mathsf c_{22}e^{\pi i \sigma}\bigl(
      \cos2\pi\lb\theta_1+\sigma\rb-\cos2\pi\theta_{\infty}\bigr)}\rb\\
  \nonumber &+\theta_td_{\mathcal M}\ln\lb
  \frac{\sin\pi\lb\theta_1+\theta_t-\overline\sigma\rb}{
  \sin\pi\lb\theta_1+\theta_t+\overline\sigma\rb}
  \frac{\bigl(\cos2\pi\theta_t-\cos2\pi
    \lb\theta_0-\sigma\rb\bigr)\lb\mathsf c_{12}e^{-\pi i\sigma}-
    \mathsf c_{22}e^{\pi i \sigma+i\eta}\rb e^{-2\pi i \theta_t}}{
    \mathsf c_{12}e^{-\pi i\sigma}\bigl(
    \cos2\pi\lb\theta_t+\sigma\rb-\cos2\pi\theta_0\bigr)-
     \mathsf c_{22}e^{\pi i \sigma+i\eta}\bigl(
      \cos2\pi\lb\theta_t-\sigma\rb-\cos2\pi\theta_0\bigr)}\rb\\
  \label{Upshat}& +\theta_0d_{\mathcal M}\ln\lb
  \frac{\sin\pi\lb\theta_0+\theta_t-\sigma\rb
  \sin\pi\lb\theta_0-\theta_{\infty}+\overline\sigma\rb}{
  \sin\pi\lb\theta_0-\theta_t-\sigma\rb
  \sin\pi\lb\theta_0-\theta_{\infty}-\overline\sigma\rb}
    \frac{\mathsf c_{12}\sin\pi\lb
    \theta_0-\theta_t-\sigma\rb-
    \mathsf c_{22}e^{i\eta}\sin\pi\lb
     \theta_0-\theta_t+\sigma\rb}{
    \mathsf c_{12}\sin\pi\lb
    \theta_0+\theta_t+\sigma\rb-
     \mathsf c_{22}e^{i\eta}\sin\pi\lb
      \theta_0+\theta_t-\sigma\rb}\,e^{2\pi i \theta_{0}}\rb.
  \end{align}
  \end{lemma}
  \pf The crucial point here is that the logarithmic derivative of
  the Barnes $G$-function can be expressed in terms of the digamma function. It is convenient to write the corresponding formula as
  \beq\label{diffg}
  d\ln G\lb1+ z\rb=
  \frac{\ln 2\pi-1}{2}dz-d\lb \frac{z^2}{2}\rb+zd\ln\Gamma\lb1+ z\rb.
  \eeq
  The contributions of the first two terms on the right sum up to zero in the logarithmic derivative of the Barnes function factor from (\ref{ccredef}). The contributions of the third term in combination with gamma factors in (\ref{fourratios}) simplify to trigonometric expressions multiplied by rational combinations of $\theta_{0,t,1,\infty}$, $\sigma$, $\overline\sigma$. The latter rational part rather nontrivially simplifies and cancels out the term
  $2d_{\mathcal M}\lb \theta_1-\theta_0\rb$
  in (\ref{difconcon}).
  \epf
  
  Let us briefly recall the notations used in (\ref{Upshat}):
  \begin{itemize}
  \item Trace functions $p_{t1}$ and $p_{01}$
  are expressed in terms of $\sigma$ and $\eta$  by (\ref{p1tf})
  and (\ref{p01f}); we can take for $\overline\sigma$ any of the two solutions of $p_{1t}=2\cos2\pi\overline\sigma$ satisfying 
  $|\Re\overline\sigma|<\frac12$.
  \item The quantity $e^{i\overline \eta}$ is expressed in terms of $\sigma$, $\eta$, $\overline\sigma$ and $p_{01}$ by (\ref{etabardef}).
  \item The coefficients $\mathsf c_{12}$, $\mathsf c_{22}$ are expressed in terms of $\sigma$, $\eta$ and $\overline\sigma$ by
  (\ref{CSSsf}).
  \end{itemize}
  The dependence of $d_{\mathcal M}\ln\hat\Upsilon$ on monodromy parameters is therefore trigonometric. In particular, one can verify that the right side  of (\ref{Upshat}) is indeed closed by direct (albeit very lengthy) calculation. 
  
  Somewhat cumbersome form of the derivatives and complicated relations between parameters may leave an impression that
  integrating $d_{\mathcal M}\ln\hat\Upsilon$ explicitly is a hopeless task. However, a conjectural answer for $\Upsilon\lb M\rb$ has been already produced in \cite{ILT13}. Hence all we have to do is to compute its logarithmic derivatives and check that they coincide with those given by (\ref{Upshat}).
  
  We start by preparing a convenient notation. Let us introduce an antisymmetrized combination of Barnes functions (it is related to classical dilogarithm but has much nicer analytic properties)
  \beq\label{ghat}
  \hat{G}\lb z\rb=\frac{G\lb 1+z\rb }{G\lb 1-z\rb}.
  \eeq
  Two main properties of $\hat G\lb z\rb$ that will be important for us are its differentiation formula and recursion relation. They can be written as
  \begin{align}
  \label{diffghat}
  d\ln\hat G\lb z\rb=&\, \ln 2\pi \,dz-z\,d\ln\sin\pi z,\\
  \label{recghat}
  \frac{\hat G\lb z+1\rb}{\hat G\lb z\rb}=&\,-\frac{\pi}{\sin\pi z}.
  \end{align}
  Recall the definition of parameters $\nu_{1\ldots 4}$, $\lambda_{1\ldots 4}$, $\nu_{\Sigma}$, $\varsigma$ from Theorem A of the Introduction.
 The equation (\ref{eqvarsigma}) defines $\varsigma$ only up to integer shifts. As we shall see in a moment, this ambiguity turns out to be harmless.
 Note that $\xi=e^{2\pi i \varsigma}$ gives one of the two nontrivial roots of the equation
 \beq\label{quadreq}
 \prod_{k=1}^4\lb 1-\xi e^{2\pi i \nu_k}\rb=\prod_{k=1}^4\lb 1-\xi e^{2\pi i \lambda_k}\rb.
 \eeq
 \begin{lemma} We have
 \beq\label{UpsilonA}
 \hat \Upsilon\lb  M\rb=
 \prod_{k=1}^4\frac{\hat{G}(\varsigma +\nu_k)}{
     \hat{G}(\varsigma +\lambda_k)}.
 \eeq
 \end{lemma}
 \pf Let us first check that the right side of (\ref{UpsilonA}) is well-defined. Indeed, in view of the recurrence relation (\ref{recghat}), the shift $\varsigma\mapsto \varsigma +1$ is equivalent to multiplying the corresponding expression by 
 \beq\label{cyclic}
 \prod_{k=1}^4\frac{\sin\pi\lb\varsigma+\lambda_k\rb}{
 \sin\pi\lb\varsigma+\nu_k\rb}=1,
 \eeq 
 where the equality is nothing but a rewrite of the equation (\ref{quadreq}). 
 
 Computing the differential with the help of (\ref{diffghat}), we get
 \begin{align*}
 d_{\mathcal M}\ln \prod_{k=1}^4
 \frac{\hat{G}(\varsigma +\nu_k)}{
      \hat{G}(\varsigma +\lambda_k)}&=\sum_{k=1}^4
     \Bigl( (\varsigma +\lambda_k)d_{\mathcal M}
     \ln\sin\pi(\varsigma +\lambda_k)
     -(\varsigma +\nu_k)d_{\mathcal M}
          \ln\sin\pi(\varsigma +\nu_k)\Bigr)=\\
 &= \sum_{k=1}^4
      \Bigl( \lambda_k d_{\mathcal M}
      \ln\sin\pi(\varsigma +\lambda_k)
      -\nu_k d_{\mathcal M}
           \ln\sin\pi(\varsigma +\nu_k)\Bigr),        
 \end{align*}
 where the last equality easily follows from (\ref{cyclic}). Taking into account the definition (\ref{nus}) of parameters ${\nu_{k}, \;\lambda_k}$, the last expression can be rewritten as
 \begin{align*}
  d_{\mathcal M}\ln \prod_{k=1}^4 \frac{\hat{G}(\varsigma +\nu_k)}{
       \hat{G}(\varsigma +\lambda_k)}=
 \overline\sigma &\,d_{\mathcal M}\ln\frac{
 \sin\pi\lb \varsigma+\lambda_2\rb
 \sin\pi\lb \varsigma+\lambda_3\rb}{
  \sin\pi\lb \varsigma+\nu_3\rb
  \sin\pi\lb \varsigma+\nu_4\rb}
  \,-\,\sigma\, d_{\mathcal M}\ln\frac{
   \sin\pi\lb \varsigma+\nu_1\rb
   \sin\pi\lb \varsigma+\nu_2\rb}{
    \sin\pi\lb \varsigma+\lambda_2\rb
    \sin\pi\lb \varsigma+\lambda_3\rb}+\\
    +\,\theta_{\infty}&\,d_{\mathcal M}
 \ln\frac{
     \sin\pi\lb \varsigma+\lambda_1\rb
     \sin\pi\lb \varsigma+\lambda_3\rb}{
      \sin\pi\lb \varsigma+\nu_2\rb
      \sin\pi\lb \varsigma+\nu_3\rb}
 +\theta_1 d_{\mathcal M}
  \ln\frac{
      \sin\pi\lb \varsigma+\lambda_1\rb
      \sin\pi\lb \varsigma+\lambda_2\rb}{
       \sin\pi\lb \varsigma+\nu_2\rb
       \sin\pi\lb \varsigma+\nu_4\rb}+\\
     +\,\theta_{t}&\,d_{\mathcal M}
  \ln\frac{
      \sin\pi\lb \varsigma+\lambda_1\rb
      \sin\pi\lb \varsigma+\lambda_3\rb}{
       \sin\pi\lb \varsigma+\nu_1\rb
       \sin\pi\lb \varsigma+\nu_4\rb}
  +\theta_0 d_{\mathcal M}
   \ln\frac{
       \sin\pi\lb \varsigma+\lambda_1\rb
       \sin\pi\lb \varsigma+\lambda_2\rb}{
        \sin\pi\lb \varsigma+\nu_1\rb
        \sin\pi\lb \varsigma+\nu_3\rb}\,.         
 \end{align*}
 To show that this coincides with 1-form in the right side of
 equation (\ref{Upshat}), it suffices to check six trigonometric identities, namely:
 \begin{align*}
 \frac{
  \sin\pi\lb \varsigma+\lambda_2\rb
  \sin\pi\lb \varsigma+\lambda_3\rb}{
   \sin\pi\lb \varsigma+\nu_3\rb
   \sin\pi\lb \varsigma+\nu_4\rb}=&\,\frac{
     \sin\pi\lb\theta_{\infty} -\theta_0+\overline\sigma\rb
     \sin\pi\lb\theta_1-\theta_t+\overline\sigma\rb}{
     \sin\pi\lb\theta_{\infty}+\theta_0+\overline\sigma\rb
       \sin\pi\lb\theta_1+\theta_t+\overline\sigma\rb} e^{i\overline\eta},\\
 \frac{
    \sin\pi\lb \varsigma+\nu_1\rb
    \sin\pi\lb \varsigma+\nu_2\rb}{
     \sin\pi\lb \varsigma+\lambda_2\rb
     \sin\pi\lb \varsigma+\lambda_3\rb}=&\,
     \frac{
       \sin\pi\lb\theta_0+\theta_t-\sigma\rb
       \sin\pi\lb\theta_{\infty}+\theta_1-\sigma\rb}{
       \sin\pi\lb\theta_0-\theta_t-\sigma\rb
       \sin\pi\lb\theta_{\infty}-\theta_1-\sigma\rb} e^{i\eta},\\ 
 \frac{
  \sin\pi\lb \varsigma+\lambda_1\rb
  \sin\pi\lb \varsigma+\lambda_3\rb}{
   \sin\pi\lb \varsigma+\nu_2\rb
   \sin\pi\lb \varsigma+\nu_3\rb}=&\,\frac{
     \sin\pi\lb\theta_{\infty}+\theta_1-\sigma\rb
     \sin\pi\lb\theta_0-\theta_{\infty}-\overline\sigma\rb}{
     \sin\pi\lb\theta_{\infty}-\theta_1-\sigma\rb
     \sin\pi\lb\theta_0-\theta_{\infty}+\overline\sigma\rb}\times\\
           \times&\,\frac{\mathsf c_{12}\sin\pi\lb
      \theta_1-\theta_{\infty}-\sigma\rb-
      \mathsf c_{22}\sin\pi\lb
       \theta_1-\theta_{\infty}+\sigma\rb}{
      \mathsf c_{12}\sin\pi\lb
      \theta_1+\theta_{\infty}-\sigma\rb-
       \mathsf c_{22}\sin\pi\lb
        \theta_1+\theta_{\infty}+\sigma\rb}\,  e^{-2\pi i \theta_{\infty}},\\  
  \frac{
   \sin\pi\lb \varsigma+\lambda_1\rb
   \sin\pi\lb \varsigma+\lambda_2\rb}{
    \sin\pi\lb \varsigma+\nu_2\rb
    \sin\pi\lb \varsigma+\nu_4\rb}=&\,       \frac{\sin\pi\lb\theta_1+\theta_t-\overline\sigma\rb}{
        \sin\pi\lb\theta_1+\theta_t+\overline\sigma\rb}\times\\
              \times\,
        &\frac{\bigl(\cos2\pi
                 \lb\theta_{\infty}-\sigma\rb-\cos2\pi\theta_1\bigr)\lb\mathsf c_{12}e^{-\pi i\sigma}-
         \mathsf c_{22}e^{\pi i \sigma}\rb e^{2\pi i \theta_1}}{
         \mathsf c_{12}e^{-\pi i\sigma}\bigl(
         \cos2\pi\lb\theta_1-\sigma\rb-\cos2\pi\theta_{\infty}\bigr)-
          \mathsf c_{22}e^{\pi i \sigma}\bigl(
  \cos2\pi\lb\theta_1+\sigma\rb-\cos2\pi\theta_{\infty}\bigr)},\\      \frac{
        \sin\pi\lb \varsigma+\lambda_1\rb
        \sin\pi\lb \varsigma+\lambda_3\rb}{
         \sin\pi\lb \varsigma+\nu_1\rb
         \sin\pi\lb\varsigma+\nu_4\rb}=&\,
         \frac{\sin\pi\lb\theta_1+\theta_t-\overline\sigma\rb}{
           \sin\pi\lb\theta_1+\theta_t+\overline\sigma\rb}\times \\
     \times\,& \frac{\bigl(\cos2\pi
          \lb\theta_0-\sigma\rb-\cos2\pi\theta_t\bigr)\lb\mathsf c_{12}e^{-\pi i\sigma}-
             \mathsf c_{22}e^{\pi i \sigma+i\eta}\rb e^{-2\pi i \theta_t}}{
             \mathsf c_{12}e^{-\pi i\sigma}\bigl(
             \cos2\pi\lb\theta_t+\sigma\rb-\cos2\pi\theta_0\bigr)-
              \mathsf c_{22}e^{\pi i \sigma+i\eta}\bigl(
   \cos2\pi\lb\theta_t-\sigma\rb-\cos2\pi\theta_0\bigr)},\\
  \frac{
         \sin\pi\lb \varsigma+\lambda_1\rb
         \sin\pi\lb \varsigma+\lambda_2\rb}{
          \sin\pi\lb \varsigma+\nu_1\rb
          \sin\pi\lb \varsigma+\nu_3\rb}=&\,      \frac{\sin\pi\lb\theta_0+\theta_t-\sigma\rb
            \sin\pi\lb\theta_0-\theta_{\infty}+\overline\sigma\rb}{
            \sin\pi\lb\theta_0-\theta_t-\sigma\rb
    \sin\pi\lb\theta_0-\theta_{\infty}-\overline\sigma\rb}\times \\
     \times &\,\frac{\mathsf c_{12}\sin\pi\lb
              \theta_t-\theta_0+\sigma\rb-
              \mathsf c_{22}e^{i\eta}\sin\pi\lb
               \theta_t-\theta_0-\sigma\rb}{
              \mathsf c_{12}\sin\pi\lb
              \theta_t+\theta_0+\sigma\rb-
               \mathsf c_{22}e^{i\eta}\sin\pi\lb
      \theta_t+ \theta_0-\sigma\rb}\,e^{2\pi i \theta_{0}}.
 \end{align*}
 This can be done by direct algebraic manipulation using
 (\ref{trfunc2}), (\ref{etareltheta}), (\ref{etabardef}), (\ref{CSSsf}) and (\ref{eqvarsigma}). 
 
 The right side of (\ref{UpsilonA}) thus coincides with $\hat{\Upsilon}\lb M \rb$ up to a constant independent on
 monodromy data. To show that this constant is equal to $1$, it suffices to compute it for any Painlev\'e VI solution with monodromy satisfying previous assumptions, see e.g. \cite{KK,LT} for explicit elliptic and algebraic examples. A number of such checks has already been reported in \cite{ILT13,ILST}.
 \epf
 
 The formulae (\ref{ccredef}) and (\ref{UpsilonA}) constitute the statement of Theorem \ref{thp6} which summarizes our solution of the constant problem for the Painlev\'e VI tau function. 
  
 \section{Systems with irregular singularities}\label{secIR}
 
 \subsection{Extended Jimbo-Miwa-Ueno differential}
   In this subsection we sketch the modifications that should be brought to the Fuchsian setup in the general case of systems with $n+1$ irregular singularities at $a_1,\ldots,a_{n},a_{\infty}=\infty$ on $\Pb$.
  In this case, the sys\-tem~(\ref{gensys0}) can be rewritten as 
 \beq\label{irsys}
  \frac{d\Phi}{dz}=A\lb z\rb\Phi, \qquad A\lb z\rb =\sum_{\nu=1}^{n}\sum_{k=1}^{r_\nu+1}
  \frac{A_{\nu,-k+1}}{\lb z-a_\nu\rb^{k}}-\sum_{k=0}^{r_\infty-1}z^kA_{\infty,-k-1} .
  \eeq
  It may be assumed without any loss in generality that $\operatorname{Tr}A\lb z\rb=0$, i.e. that
  $A_{\nu,-k+1},A_{\infty,-k-1}\in \mathfrak{sl}_N\left(\mathbb C\right)$.
 As it has  already been  indicated in the Introduction, we shall also assume  that all highest order matrix coefficients $A_{\nu}\equiv A_{\nu,-r_\nu}$ are diagonalizable
 \ben
 A_{\nu,-r_{\nu}}=G_{\nu}\Theta_{\nu,-r_\nu}G_{\nu}^{-1};\quad \Theta_{\nu,-r_\nu}=\operatorname{diag}\left\{\theta_{\nu,1},\ldots \theta_{\nu,N}\right\},
 \ebn
 and that their eigenvalues are distinct and non-resonant: 
 \ben
 \begin{cases}
 \theta_{\nu,\alpha}\neq \theta_{\nu,\beta} \quad &\mbox{if}\quad r_\nu \geq 1,\quad \alpha\neq \beta,\\
 \theta_{\nu,\alpha}\neq \theta_{\nu,\beta} \mod \mathbb{Z}\quad &\mbox{if}  \quad r_{\nu}=0,\quad \alpha\neq \beta.
 \end{cases}
 \ebn
 If the {\it Poincare index} $r_{\nu} $  of the pole $a_{\nu}$ is greater  or equal to  $1$, then the pole is called an {\it irregular singular point}
 of the system (\ref{irsys}). In the neighborhood of  such a point the asymptotic behavior of solution $\Phi\lb z\rb$ exhibits 
 the {\it Stokes Phenomenon} which is described as follows.
 
 Let $a_{\nu}$ be an irregular singular point of index $r_{\nu}$. For $j=1,\ldots, 2r_{\nu}+1$, let
 \begin{equation}\label{stokssec1}
 \Omega_{j,\nu} = \left\{z :  0<|z-a_{\nu}|<\epsilon,\quad \theta^{(1)}_j< \arg \lb z-a_{\nu}\rb < \theta^{(2)}_{j},
 \quad \theta^{(2)}_j - \theta^{(1)}_j = \frac{\pi}{r_{\nu}} + \delta \right\},
 \end{equation}
 be the {\it Stokes sectors} around $a_{\nu}$ (see, e.g., \cite[Chapter 1]{FIKN} or \cite{Was} for more details). According to the general 
 theory of linear systems, in each sector $ \Omega_{j,\nu}$ there exists a unique {\it canonical solution} 
 $\Phi_j^{(\nu)}\left(z\right)$ of (\ref{irsys}) which satisfies the asymptotic condition 
 \begin{equation}\label{can1}
  \Phi_j^{(\nu)}\left(z\right)\simeq \Phi^{(\nu)}_{\mathrm{form}}\lb z\rb \qquad\text{as } z\to a_{\nu},\quad  z \in \Omega_{j,\nu},\quad j=1,\ldots, 2r_\nu +1,
  \end{equation}
 where  $\Phi^{(\nu)}_{\mathrm{form}}\lb z\rb$ is the formal solution at the point $a_{\nu}$ which has already been mentioned
 in the Introduction. For reader's convenience, we reproduce here the relevant formulae:
 \beq\label{formalstructure}
 \Phi^{(\nu)}_{\mathrm{form}}\lb z\rb=G^{\lb\nu\rb}\left(z\right)e^{
 \Theta_{\nu}\lb z\rb},\qquad 
 G^{\lb\nu\rb}\left(z\right) =
  G_{\nu}\hat{\Phi}^{\lb\nu\rb}\lb z\rb,
 \eeq
 where
 \ben
 \hat{\Phi}^{\lb\nu\rb}\lb z\rb=
 \begin{cases} \mathbf{1}+\sum_{k=1}^{\infty}g_{\nu,k}\left(z-a_{\nu}\right)^k,
 \qquad &\nu = 1, \ldots, n,\\
 \mathbf{1}+\sum_{k=1}^{\infty}g_{\infty,k}z^{-k},\qquad & \nu=\infty,
 \end{cases}
 \ebn
 and  $\Theta_{\nu}(z)$ are diagonal matrix-valued functions,
 \ben
 \Theta_{\nu}(z)=\begin{cases}\ds
 \sum\nolimits_{k=-r_\nu}^{-1}
 \frac{\Theta_{\nu,k}}{k}\lb z-a_\nu\rb^{k}+\Theta_{\nu,0}\ln \lb z-a_\nu\rb,\qquad &\nu=1,\ldots,n\\
 \ds-\sum\nolimits_{k=1}^{r_{\infty}}
 \frac{\Theta_{\infty,-k}}{k}z^{k}-\Theta_{\infty,0}\ln z,\qquad
 & \nu=\infty.
 \end{cases}
 \ebn
 Among the identities that determine $\Theta_{\nu}\lb z\rb$, $\hat{\Phi}^{\lb \nu\rb}\lb z\rb$ and $G^{\lb \nu\rb}\lb z\rb$ in terms of $A\lb z\rb$ and $G_{\nu}$ there is a particularly important family of relations that will be repeatedly used in what follows. Namely, the structure of the formal solution (\ref{formalstructure}) implies that
 \beq\label{Asingular}
 A\lb z\rb-G^{\lb \nu\rb}\lb z\rb\Theta_{\nu}'\lb z\rb
 {G^{\lb \nu\rb}\lb z\rb}^{-1}=\begin{cases}
 O\lb 1\rb,\qquad & \nu=1,\ldots, n,\\
 O\lb z^{-2}\rb,\qquad &\nu=\infty.
 \end{cases}
 \eeq
 Here and below the prime  denotes the derivative with respect to $z$. The matrix $A\lb z\rb$ can thus be reconstructed by taking the sum of principal parts of Laurent series $G^{\lb \nu\rb}\lb z\rb\Theta_{\nu}'\lb z\rb
  {G^{\lb \nu\rb}\lb z\rb}^{-1}$ at $z=a_{\nu}$ (plus a constant part for the point at $\infty$).
 
 Stokes and connection matrices relate the canonical  solutions  $ \Phi_j^{(\nu)}\left(z\right)$ in different Stokes sectors and at different singular points:
 \ben
 \Phi_{j+1}^{(\nu)}=\Phi_{j}^{(\nu)}S_j^{(\nu)},\quad j = 1,\ldots, 2r_{\nu}, \qquad  \Phi_{1}^{(\nu)}=\Phi_{1}^{(\infty)}C_{\nu},\quad
 \nu = 1,\ldots, n.
 \ebn  
 Let us 
 assume as before that the irregular singular points are $\infty$ and the first $m\leq n$ points among the singular points $a_1, \ldots, a_n$. Denote by ${\mathcal S}_{\nu}$ the collection of Stokes matrices at an irregular point $a_{\nu}$, i.e. 
  \begin{equation}\label{sphen}
  {\mathcal S}_{\nu} =\left\{S_1^{(\nu)}, \ldots, S_{2r_{\nu}}^{(\nu)}\right\}.
  \end{equation}
 The space ${\mathcal M}$ of monodromy data of the system (\ref{irsys}) consists of formal monodromy exponents $\Theta_{\nu,0}$,  connection matrices $C_{\nu}$ and Stokes matrices $S_j^{\lb \nu\rb}$. The latter constitute the main difference as compared to the Fuchsian case. More explicitly, 
 \begin{equation}\label{mset}
 {\mathcal M} = \left\{ M \equiv \Bigl(\Theta_{\nu, 0},\,\, \nu = 1, \ldots, n, \infty; \quad  C_{\nu},\,\,  \nu = 1,\ldots, n;\quad  
 {\mathcal S}_{\nu},\,\,  \nu = 1,\ldots, m,  \infty\Bigr) \right\}.
 \end{equation}
 
 Following the notations used in the Introduction, we denote  by $\mathcal{T}$ the set of times
 \begin{equation}\label{istime1}
 a_{1}, \ldots, a_{n}, \quad  (\Theta_{\nu,k})_{ll}, \quad k= -r_\nu,\ldots,-1,\quad \nu = 1, \ldots, m,\, \infty,
 \quad l = 1, \ldots, N.
 \end{equation}
 Observe that in contrast to the Fuchsian case this set contains  much more parameters than just the positions of singular points.
 Let us retain the notation
 $$
 \vec{t} = \lb t_1, \ldots, t_L\rb, \quad L = n + N\lb\sum_{\nu=1}^{m}r_{\nu} + r_{\infty}\rb,
 $$
 for the points $\vec t\in\mathcal T$
 and consider monodromy preserving deformations of the system (\ref{irsys}) with respect to these times.
 We denote by $
 A(z) \equiv A\lb z; \vec{t}; M\rb$
 the isomonodromic family  of the systems (\ref{irsys}) having the same set $M\in\mathcal M$ of monodromy data.
 The isomonodromy implies that the corresponding solution   $\Phi\lb z\rb \equiv \Phi\lb z,\vec{t}\rb$  
 satisfies an overdetermined system 
 \beq
 \begin{cases} \label{isosys}
 \,\,\partial_z\Phi\,\, =  A\lb z,\vec t\rb \Phi\lb z,\vec{t}\rb,\\
 d_{\mathcal{T}}\Phi =U\lb z,\vec t\rb\Phi\lb z,\vec{t}\rb
 \end{cases}
 \eeq
 The coefficients of the matrix-valued differential form $U \equiv \sum_{k=1}^{L}U_k\lb z,\vec t\rb dt_k$ are rational in $z$. Their explicit form may be algorithmically deduced from the expression for $A\lb z\rb$. The compatibility of the system (\ref{isosys}) implies the monodromy preserving deformation equation (\ref{isomeq0}):
 \begin{equation}\label{isomeg00}
 {d_{\mathcal{T}}A}=\partial_z U+[U,A].
 \end{equation}
 
 The construction of an irregular analog of the $1$-form $\omega$ defined in the Fuchsian case by the equation (\ref{taudef2}) is carried out in several steps. Let us first recall once again the standard definition of the Jimbo-Miwa-Ueno differential \cite[equation (5.1)]{JMU},  
 \beq\label{ojmu}
 \omega_{\mathrm{JMU}}=-\sum_{\nu = 1,\ldots, n, \infty} \operatorname{res}_{z=a_\nu} \mathrm{Tr}\left(\hat\Phi^{(\nu)}
 \lb z\rb^{-1}\partial_z\hat\Phi^{(\nu)}\lb z\rb\, d_{\mathcal T}\Theta_{\nu}\lb z\rb\right).
 \eeq
 This $1$-form is  closed on solutions of the  isomonodromy equation (\ref{isomeg00}): 
 \ben d_{\mathcal T}\omega_{\mathrm{JMU}}=0.
 \ebn
 Our goal is to find an extension of it which would be closed on the whole space $ \mathcal{A} \simeq \widetilde{\mathcal T}\times \mathcal{M} $ and which would coincide with (\ref{ojmu}) when restricted to $\mathcal T$.
 To this end, it is convenient to rewrite $\omega_{\mathrm{JMU}}$ in a slightly different way, cf \cite[Remark 5.2]{JMU}.
 \begin{lemma}\label{lemjmu} The $1$-form $\omega_{\mathrm{JMU}}$ can be alternatively written as
 \beq\label{ojmu2}
 \omega_{\mathrm{JMU}}=
 \sum_{\nu=1,\ldots,n,\infty}\operatorname{res}_{z=a_{\nu}}
  \operatorname{Tr}\lb 
  A\lb z\rb \,d_{\mathcal T}G^{\lb \nu\rb}\lb z\rb \,
  {G^{\lb \nu\rb}\lb z\rb}^{-1}\rb.
 \eeq
 \end{lemma}
 \pf Formal series $G^{\lb \nu\rb}\lb z\rb$ appears in the asymptotic behavior of an actual solution $\Phi\lb z\rb$ in some Stokes sector as indicated in (\ref{formalstructure}). The isomonodromy property then implies that $d_{\mathcal T}G^{\lb \nu\rb}\lb z\rb\,
 {G^{\lb \nu\rb}\lb z\rb}^{-1}$ can be replaced by the combination
 \beq\label{ojmu3}
 d_{\mathcal T}G^{\lb \nu\rb}\,
  {G^{\lb \nu\rb}}^{-1}=d_{\mathcal T}\Phi\;\Phi^{-1}-\Phi\,d_{\mathcal T}\Theta_{\nu}\,\Phi^{-1}=U-G^{\lb \nu\rb}d_{\mathcal T}\Theta_{\nu}\,{G^{\lb \nu\rb}}^{-1}.
 \eeq
 in the whole punctured neighborhood of $a_{\nu}$.
 The last equality follows from the second equation of the Lax representation (\ref{isosys}) and the diagonal form of $\Theta_{\nu}\lb z\rb$. Since
 $a_1,\ldots,a_n,\infty$ are the only possible poles of the rational matrix function $A\lb z\rb U\lb z\rb$, we necessarily have $
 \ds\sum\limits_{\nu=1,\ldots,n,\infty}\operatorname{res}_{z=a_{\nu}}
   \operatorname{Tr}\bigl (
   A\lb z\rb U\lb z\rb\bigr)=0$.
   Using this sum rule, the equation (\ref{ojmu3}) and the relation
  \beq\label{AGTh}
   A=\Phi'\Phi^{-1}={G^{\lb\nu\rb}}' {G^{\lb\nu\rb}}^{-1}+
   G^{\lb\nu\rb}\Theta_{\nu}'{G^{\lb\nu\rb}}^{-1},
   \eeq
 the right side of (\ref{ojmu2}) can be rewritten as 
  \ben
  - \sum_{\nu=1,\ldots,n,\infty}\operatorname{res}_{z=a_{\nu}}
    \operatorname{Tr}\lb 
    A G^{\lb \nu\rb}d_{\mathcal T}\Theta_{\nu}\,{G^{\lb \nu\rb}}^{-1}\rb=\omega_{\mathrm{JMU}}-
  \sum_{\nu=1,\ldots,n,\infty}\operatorname{res}_{z=a_{\nu}}
     \operatorname{Tr}\lb \Theta'_{\nu}d_{\mathcal T}\Theta_{\nu} \rb.  
  \ebn 
 All residues in the second sum obviously vanish ($\Theta'_{\nu}$ and $d_{\mathcal T}\Theta_{\nu}$ are Laurent polynomials with only principal part), which proves the statement of the lemma.
 \epf

 The expression (\ref{ojmu2}) is much better adapted for extension to
 $\widetilde{\mathcal T}\times\mathcal M$ than the original formula
 (\ref{ojmu}). Indeed, we have the following result:

 \begin{theo}\label{omirrth} Let $\omega\in\Lambda^1\lb\widetilde{\mathcal T}\times\mathcal M\rb$ be a $1$-form defined by
 \beq\label{jmuext}
 \omega=\sum_{\nu=1,\ldots,n,\infty}\operatorname{res}_{z=a_{\nu}}
   \operatorname{Tr}\lb 
   A\lb z\rb \,dG^{\lb \nu\rb}\lb z\rb \,
   {G^{\lb \nu\rb}\lb z\rb}^{-1}\rb,
 \eeq
 where $d=d_{\mathcal T}+d_{\mathcal M}$. Its exterior differential 
 $\Omega:=d\omega$ is a closed $2$-form on $\mathcal M$ independent
 of isomonodromic times $\vec t\in\mathcal T$ listed in (\ref{istime1}).
 \end{theo}
 \pf Notice that the residues in (\ref{jmuext}) are completely determined by the singular parts of the corresponding Laurent expansions of $A\lb z\rb$.  Using their identification (\ref{Asingular}), we may replace $A\lb z\rb$ by $G^{\lb\nu\rb}\Theta'_{\nu}{G^{\lb\nu\rb}}^{-1}$ in the computation of each residue. This yields another representation of the form $\omega$:
 \beq\label{jmuext2}
  \omega=\sum_{\nu=1,\ldots,n,\infty}\operatorname{res}_{z=a_{\nu}}
    \operatorname{Tr}\lb 
    \Theta'_{\nu}\lb z\rb {G^{\lb \nu\rb}\lb z\rb}^{-1}
    dG^{\lb \nu\rb}\lb z\rb 
    \rb.
 \eeq
 
  Let us now pick an arbitrary isomonodromic time $t_k$ and a parameter $s$ which can be either a local coordinate on the space $\mathcal M$ of monodromy data or another time variable. We are going to show that $\Omega\lb\partial_{t_{k}},\partial_{s}\rb=0$. 
  First, from (\ref{jmuext2}) it follows that 
 \begin{align*}
 &\Omega\lb\partial_{t_{k}},\partial_{s}\rb=\partial_{t_{k}}\omega\lb\partial _s\rb-\partial_{s}\omega\lb\partial _{t_{k}}\rb=\\
 =&\,
 \sum_{\nu=1,\ldots,n,\infty}\operatorname{res}_{z=a_{\nu}}
     \operatorname{Tr}\lb 
     \partial_{t_{k}}\Theta'_{\nu}\, {G^{\lb \nu\rb}}^{-1}
     \partial_s G^{\lb \nu\rb} -
     \partial_{s}\Theta'_{\nu}\, {G^{\lb \nu\rb}}^{-1}
          \partial_{t_{k}} G^{\lb \nu\rb}+\Theta'_{\nu}\left[{G^{\lb \nu\rb}}^{-1}\partial_s G^{\lb \nu\rb},
    {G^{\lb \nu\rb}}^{-1}\partial_{t_k} G^{\lb \nu\rb}      \right]
     \rb.
 \end{align*}
 Thanks to cyclic properties of the trace, we can make in the last  line the following replacements:
 \begin{align*}
 \partial_{s}\Theta'_{\nu}\, {G^{\lb \nu\rb}}^{-1}
           \partial_{t_{k}} G^{\lb \nu\rb}&\,\mapsto
 \partial_{t_{k}} G^{\lb \nu\rb} \partial_{s}\Theta'_{\nu}\, {G^{\lb \nu\rb}}^{-1},\\
 \Theta'_{\nu}\left[{G^{\lb \nu\rb}}^{-1}\partial_s G^{\lb \nu\rb},
     {G^{\lb \nu\rb}}^{-1}\partial_{t_k} G^{\lb \nu\rb}      \right]&\,\mapsto    \partial_{t_{k}} G^{\lb \nu\rb} \left[\Theta'_{\nu},{G^{\lb \nu\rb}}^{-1}\partial_s G^{\lb \nu\rb}\right] {G^{\lb \nu\rb}}^{-1},                
 \end{align*}
 so that the expression under trace becomes
 \ben
 \partial_{t_{k}}\Theta'_{\nu}\, {G^{\lb \nu\rb}}^{-1}
      \partial_s G^{\lb \nu\rb} -\partial_{t_{k}} G^{\lb \nu\rb}\,{G^{\lb \nu\rb}}^{-1}\partial_s\lb
 G^{\lb \nu\rb}\Theta'_{\nu}{G^{\lb \nu\rb}}^{-1}\rb.
 \ebn 
 Using once again the coincidence of singular parts of $A\lb z\rb$ and
 $G^{\lb \nu\rb}\Theta'_{\nu}{G^{\lb \nu\rb}}^{-1}$, let us rewrite the expression for the curvature coefficient $\Omega\lb\partial_{t_{k}},\partial_{s}\rb$ as
 \beq\label{auxjmu1}
 \Omega\lb\partial_{t_{k}},\partial_{s}\rb=\sum_{\nu=1,\ldots,n,\infty}\operatorname{res}_{z=a_{\nu}}
      \operatorname{Tr}\lb \partial_{t_{k}}\Theta'_{\nu}\, {G^{\lb \nu\rb}}^{-1}
            \partial_s G^{\lb \nu\rb}-\partial_{t_{k}} G^{\lb \nu\rb}\,{G^{\lb \nu\rb}}^{-1}\partial_s A \rb.
 \eeq
 
 The next step is to use isomonodromy and the equation $\partial_{t_k}\Phi=U_{t_k}\Phi$ in a way similar to the proof of Lemma~\ref{lemjmu}. Since (cf equation (\ref{ojmu3}))
 \ben
 \partial_{t_{k}} G^{\lb \nu\rb}\,{G^{\lb \nu\rb}}^{-1}=U_{t_k}-
 G^{\lb \nu\rb}\partial_{t_k}\Theta_{\nu}\,{G^{\lb \nu\rb}}^{-1},
 \ebn
 and the sum of residues of rational function $U_{t_k}\lb z\rb\partial_sA\lb z\rb$ at $z=a_1,\ldots,a_n,\infty$ is clearly equal to zero, the expression (\ref{auxjmu1}) can be reduced to
 \beq\label{auxjmu2}
 \Omega\lb\partial_{t_{k}},\partial_{s}\rb=\sum_{\nu=1,\ldots,n,\infty}\operatorname{res}_{z=a_{\nu}}
       \operatorname{Tr}\lb \partial_{t_{k}}\Theta'_{\nu}\, {G^{\lb \nu\rb}}^{-1}
             \partial_s G^{\lb \nu\rb}+G^{\lb \nu\rb}\partial_{t_k}\Theta_{\nu}\,{G^{\lb \nu\rb}}^{-1}\partial_s A \rb.
 \eeq
  Substitute the matrix $A$ in the second term of this identity by its expression (\ref{AGTh}).
  This transforms the expression under trace in (\ref{auxjmu2}) into a sum of six terms:
  \beq\label{auxjmu4}
  \begin{gathered}
  \partial_{t_{k}}\Theta'_{\nu}\, {G^{\lb \nu\rb}}^{-1}
               \partial_s G^{\lb \nu\rb}+G^{\lb \nu\rb}\partial_{t_k}\Theta_{\nu}\,{G^{\lb \nu\rb}}^{-1}\partial_s {G^{\lb\nu\rb}}' {G^{\lb\nu\rb}}^{-1}- G^{\lb \nu\rb}\partial_{t_k}\Theta_{\nu}\,{G^{\lb \nu\rb}}^{-1} {G^{\lb\nu\rb}}' {G^{\lb\nu\rb}}^{-1}
               \partial_s{G^{\lb\nu\rb}} {G^{\lb\nu\rb}}^{-1}+\\
   + G^{\lb \nu\rb}\partial_{t_k}\Theta_{\nu}\,{G^{\lb \nu\rb}}^{-1}\partial_s {G^{\lb\nu\rb}}\Theta_{\nu}'\, {G^{\lb\nu\rb}}^{-1} +  G^{\lb \nu\rb}\partial_{t_k}\Theta_{\nu}\, \partial_s\Theta_{\nu}'\, {G^{\lb\nu\rb}}^{-1} 
  -G^{\lb \nu\rb}\partial_{t_k}\Theta_{\nu}\, \Theta_{\nu}'\, {G^{\lb\nu\rb}}^{-1} \partial_s{G^{\lb\nu\rb}}\,   {G^{\lb\nu\rb}}^{-1}.    
  \end{gathered}
  \eeq
  The fourth and sixth term cancel each other thanks to cyclic property of the trace and the diagonal form of $\Theta_{\nu}$ implying that $\left[\partial_{t_k}\Theta_{\nu},\Theta'_{\nu}\right]=0$.
 
 The last step consists essentially in integration by parts. Namely, we are going to use that 
 \beq\label{auxjmu3}\operatorname{res}_{z=a_{\nu}} f'g=-\operatorname{res}_{z=a_{\nu}} fg'
 \eeq 
 for arbitrary pair of formal Laurent series $f\lb z\rb$, $g\lb z\rb$  around $z=a_{\nu}$. 
 Applying this formula to, say, the first term in (\ref{auxjmu4}) yields two terms which cancel out with the second and third. The only contribution to the curvature thus comes from the fifth term:
 \ben
 \Omega\lb\partial_{t_{k}},\partial_{s}\rb=\sum_{\nu=1,\ldots,n,\infty}\operatorname{res}_{z=a_{\nu}}\operatorname{Tr}\lb
 \partial_{t_k}\Theta_{\nu}\, \partial_s\Theta_{\nu}'\rb.
 \ebn
 It however vanishes by the same argument as in the proof of Lemma~\ref{lemjmu}.  
 We have thus shown that $\Omega$ is a $2$-form on 
 $\mathcal M$ only. Since in addition $\Omega$ is a total differential, it cannot depend on isomonodromic times, as otherwise $d\Omega$ would be non-zero.
 \epf
 \begin{rmk}
 The Fuchsian $1$-form $\omega$ defined by (\ref{taudef2}) is a specialization of the general formula (\ref{jmuext}). This becomes completely manifest in the representation (\ref{jmuext2}), since in
 the Fuchsian case  $\Theta'_{\nu}\lb z\rb$  reduces to a simple pole
 contribution $\Theta_{\nu,0}/\lb z-a_{\nu}\rb$.
 \end{rmk}
 \begin{rmk}\label{bertola_malgrange}
 A related extension of the Jimbo-Miwa-Ueno form has been proposed by 
 M. Bertola in the work \cite{Bertola} generalizing previous results of B. Malgrange \cite{Malgrange}. It has been defined for solution
 $\Psi\lb z\rb$ of a general Riemann-Hilbert problem with contour $\Gamma$ and jump matrix $J\lb z\rb$ as a contour integral
 \beq\label{MBf}
 \omega_{\mathrm{MB}}\lb\partial \rb=\frac1{4\pi i}\int_{\Gamma}\operatorname{Tr}\lb
 \Psi_-^{-1}\Psi_-' \partial J\,J^{-1}+
 \Psi_+^{-1}\Psi_+'J^{-1} \partial J\rb dz.
 \eeq
 The fact that in the isomonodromic setting this Malgrange-Bertola form could localize (i.e. the integral can be evaluated in terms of $\Psi$ and its derivatives with respect to times and monodromy parameters) and become our form $\omega$ was first realized 
 in the paper \cite{IP} by two of the authors in the context of Painlev\'e III ($D_8$). Shortly after M. Bertola pointed out how the localization should be carried out in general, see \cite[Remark 3]{IP}. The result coincides with $\omega$ introduced above up to addition of a monodromy- and contour-dependent term. An intriguing feature of the contour integral set-up is that it may allow for the computation of the differential $d\omega_{\mathrm{MB}}$ in terms of monodromy data \cite{Bertola1} bypassing asymptotic analysis. It has thus a strong potential of simplifying the study of connection problems for isomonodromic tau functions, which we hope to explore in a future work. Also, we plan to relate the constructions of this
 section to the general Hamiltonian formalism of isomonodromic deformations developed by I. Krichever in \cite{Kri}. It can be expected that the technique of \cite{Kri}
 could provide another ``asymptotics free'' derivation of the monodromy representation of the form $\Omega$,  alternative to the  contour integral method of  \cite{Bertola1}.
 \end{rmk}

\subsection{Painlev\'{e} II case \label{subsecPII}}
The first nontrivial case of a non-Fuchsian system is a $2\times2$ linear system with one irregular
singular point of Poincar\'e rank 3. We shall place the singular point at infinity, so that 
the matrix $A(z)$ in (\ref{irsys})  becomes a $2$nd order polynomial in $z$,
\begin{equation}\label{p2lax0}
\partial_z\Phi= A\lb z\rb\Phi, \qquad A\lb z\rb = A_{-3}z^2 + A_{-2}z +  A_{-1}.
\end{equation}
With the help of trivial affine and gauge transformations the system can be reduced to the following
normal form:
\begin{equation}\label{laxp21}
\partial_z\Phi = \begin{pmatrix}-4iz^2 -it -2iuw & 4izu -2x\cr 
-4izw -2y&4iz^2 +it +2iuw\end{pmatrix}\Phi.
\end{equation}
Here $t, u, w, x, y$ are complex parameters playing the role of coordinates on the space ${\mathcal A}$ 
of systems (\ref{p2lax0}). 

The formal solution of this system at $z =\infty$
is given by 
\begin{equation}\label{formalP2}
\Phi^{\lb \infty\rb}_{\text{form}}(z)\equiv \Phi_{\text{form}}(z) = 
\Bigl(\mathbf{1}+\sum\limits_{m=1}^{\infty}
      g_{m}z^{-m}\Bigr)\,e^{-\left(\frac{4i}{3}z^3 + itz +\kappa\ln z\right)\sigma_3}, 
\end{equation}
where $\sigma_3=\operatorname{diag}\left\{1,-1\right\}$ and $\kappa = wx - uy$.
The non-formal behavior of solutions of (\ref{laxp21})
is described by seven canonical solutions uniquely
specified by the following asymptotic conditions, cf (\ref{can1}),
\begin{equation}\label{p2cond1}
\Phi_{j}\lb z\rb \simeq \Phi_{\text{form}}(z) \qquad\text{as }\; z \to \infty, \quad  z \in \Omega_{j},\quad j = 1, \ldots, 7,
\end{equation}
where the Stokes sectors are given by
$$
\Omega_j = \left\{z:  \frac{\pi( j-2)}{3} < \arg z < \frac{\pi j}{3} \right\}.
$$
The canonical solutions satisfy the formal monodromy condition
\begin{equation}\label{cyclic1}
\Phi_{7}\lb z\rb = \Phi_{1}\lb z\rb e^{-2\pi i \kappa \sigma_3}.
\end{equation}
There are six  Stokes matrices $S_{1}, \ldots, S_{6}$  defined by the equations
$$
S_j = \Phi^{-1}_{j}\lb z\rb\Phi_{j+1}\lb z\rb, \qquad j = 1,\ldots, 6.
$$ 
These matrices have the familiar triangular structure (see \cite[chapter 2, section 1.6]{FIKN})
$$
S_{2l +1} =\left(\begin{array}{cc} 1\;\; & 0 \\
s_{2l+1} & 1\end{array}\right), \quad S_{2l} =\lb\begin{array}{cc} 1&s_{2l}\\
0 & 1\end{array}\rb,
$$
and satisfy cyclic relation
\begin{equation}\label{cyclic2}
S_1S_2S_3S_4S_5S_6 = e^{-2\pi i \kappa \sigma_3},
\end{equation}
which follows from (\ref{cyclic1}).
A single matrix equation (\ref{cyclic2}) implies three scalar equations
\beq
\begin{gathered}
1+ s_1s_2 = (1+s_4s_5)e^{2\pi i\kappa}, \quad 1+ s_2s_3 = (1+s_5s_6)e^{-2\pi i\kappa},\\
\label{cyclic3}
s_1 +s_3 + s_1s_2s_3 =-s_5e^{2\pi i\kappa}.
\end{gathered}
\eeq
The space ${\mathcal M}$ of the monodromy data of system (\ref{laxp21}) is parametrized by 
seven parameters $\kappa$, $s_1,\ldots, s_6$ subject to  three constraints (\ref{cyclic3}). 
Hence 
\begin{equation}\label{dim}
\dim{\mathcal M} = 4 = \dim{\mathcal A} -1.
\end{equation}

The exponential factor in the right side of the formula (\ref{formalP2})  shows that the
 parameter $t$ is the only remaining  time variable
(cf with $4$-point Fuchsian  case considered in the previous sections). The fact that isomonodromic families 
in the case under consideration are one-parameter also follows from (\ref{dim}). The second linear differential equation in the  Lax pair (\ref{isosys}) is given by
\begin{equation}\label{laxP22}
\partial_t\Phi = U\lb z\rb\Phi \equiv \lb 
\begin{array}{cc} -iz& iu \\ 
-iw & iz\end{array}\rb\Phi.
\end{equation}
and the corresponding isomonodromy equation (\ref{isomeg00}) yields the following system of nonlinear ODEs satisfied by the scalar functions $u = u\lb t\rb$, 
$w = w\lb t\rb$, $x = x\lb t\rb$ and $y = y\lb t\rb$,
\beq
\begin{gathered}
 x = u_t, \quad y = w_t,\qquad wu_t - uw_t = \mbox{const} \equiv \kappa,\\
\label{P34}
x_{t} = tu +2u^2w, \quad y_{t} = tw +2w^2u.
\end{gathered}
\eeq
In what follows we will only consider the  reduction
\begin{equation}\label{P2red}
u \equiv w, \quad \kappa =0,
\end{equation}
under which the system (\ref{P34}) reduces to a  special case of the second Painlev\'e equation
{\footnote{In the general case, the system (\ref{P34}) reduces to the Painlev\'e XXXIV  equation for
the product $uw$ and  to a pair of Painlev\'e II equations for  functions
$u_t/u$ and $w_t/w$, with parameters
$\frac{1}{2} - \kappa$  and $\frac{1}{2} + \kappa$, respectively; see, e.g. \cite[Chapter 4, Section 2.5]{FIKN}}},
\beq\label{P2}
u_{tt}=2u^3+tu,
\eeq
 and the linear equations (\ref{laxp21}) and (\ref{laxP22})  form the Flaschka-Newell Lax pair for this equation \cite{FN1}.

At the level of Stokes parameters, the reduction (\ref{P2red}) is equivalent to imposing the following constraints (see again \cite{FIKN}),
\ben
s_4 = -s_1, \quad s_5 = -s_2, \quad s_6 = -s_3.
\ebn
The space of monodromy data then becomes two-dimensional, and is given by an affine cubic in $\Cb^3$ (see also Subsection~\ref{subsec_results} of the Introduction):
\begin{equation}\label{monP2}
{\mathcal M}_{\mathrm{PII}} = \left\{ s \equiv \lb s_1, s_2, s_3\rb \in {\Bbb C}^3: s_1 -s_2 +s_3 + s_1s_2s_3 = 0\right\}.
\end{equation}
Solutions $u(t) \equiv u\lb t;s\rb$of the second Painlev\'e equation (\ref{P2}) are parametrized by the points $s\in{\mathcal M}_{\mathrm{PII}}$
via the inverse monodromy map, ${\mathcal M}_{\mathrm{PII}} \to {\mathcal A}$. The latter is realized as follows.

\begin{figure}[h]
\centering
\def\svgwidth{0.3\columnwidth}
{\input{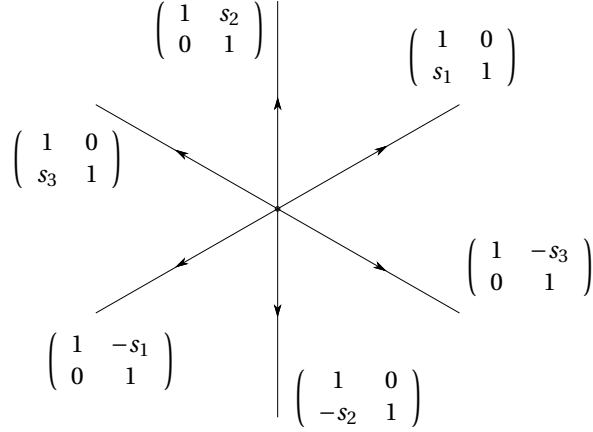}}
\caption{Contour and jump matrices of the PII-RH problem}
\label{figPII}
\end{figure}

For $j=1,\ldots, 6$, let $\Gamma_j$ denote the rays
$$
\Gamma_j = \left\{z\in\Cb : \arg z = \frac{\pi(2j -1)}6\right\}.
$$
oriented towards infinity, and let $\Omega_j^{(0)}$ be the sectors between the rays
$\Gamma_{j-1}$ and $\Gamma_j$. Note that $\overline{\Omega^{(0)}_j} \subset \Omega_j$.
Define a piecewise analytic function $\Psi(z)$ by the relations
$$
\Psi\lb z\rb = \Phi_{j}\lb z\rb \qquad \text{for }\; z \in \Omega_j^{(0)}.
$$
The function $\Psi(z)$ satisfies the following Riemann-Hilbert problem posed on 
the contour $\Gamma = \bigcup_{j=1}^{6}\Gamma_j$:
\begin{itemize}
\item $\Psi\lb z\rb$ is analytic for $z \in {\Bbb C} \setminus \Gamma$;
\item $\Psi_+\lb z\rb = \Psi_-\lb z\rb J\lb z\rb$ for $z\in \Gamma$, where $J\lb z\in \Gamma_j\rb= S_j$;
\item $ \Psi\lb z\rb = \left(\mathbf{1}+O\left(\frac{1}{z}\right)\right) e^{-\left(\frac{4i}{3}z^3 + itz +\kappa\ln z\right)\sigma_3}$ as $z\to \infty$.
\end{itemize}
The contour of the Riemann-Hilbert problem and the associated piecewise constant jump matrices are depicted in Figure \ref{figPII}. Recall that Stokes parameters $s_{1,2,3}$ are subject to the condition 
$s_1-s_2+s_3+s_1s_2s_3=0$.
We shall refer to this Riemann-Hilbert problem as the PII-RH problem.

The PII-RH problem is meromorphically solvable. This means (see \cite{BIK} and  \cite{FIKN}, appendix A)
that for every given $s \in {\mathcal M}_{\mathrm{PII}}$ there exists a discrete set ${\mathcal K}_{s} = \left\{t_k\right\}_{k=1}^{\infty}$ of points in the complex $t$-plane such that for all $t\notin {\mathcal K}_{s}$ the solution $\Psi(z) \equiv \Psi(z;t)$ exists and is meromorphic in $t$ with
 ${\mathcal K}_{s}$ being the set of its poles. The corresponding solution $u\lb t;s\rb$ of the second Painlev\'e equation is given
 by the formula,
 \begin{equation}\label{rhrep}
 u = 2\lb g_1\rb_{12},
 \end{equation}
where $g_1$ is the first matrix coefficient in the asymptotic expansion
\begin{equation}\label{rhrep2}
 \Psi(z) \simeq \left[\mathbf{1}+\sum_{k=1}^{\infty}
      g_{k}z^{-m}\right]e^{-\left(\frac{4i}{3}z^3 + itz +\kappa\ln z\right)\sigma_3}, \quad  z\to \infty.
\end{equation}

Let us now proceed to the calculation of the forms $\omega_{\mathrm{JMU}}$ and  $\omega$ corresponding to the linear system
(\ref{laxp21}) under reduction (\ref{P2red}).
Since there is only one singular point, we can globally define  $G\lb z\rb =\Psi\lb z\rb e^{\lb \frac{4iz^3}{3}+iz t\rb\sigma_3}$ and the sum in
(\ref{jmuext}) contains only one term,
\begin{equation}\label{omegap2}
\omega = \operatorname{res}_{z=\infty}\operatorname{Tr}\left(
 A\lb z\rb dG\lb z\rb {G\lb z\rb}^{-1}\right).
\end{equation}
We shall also restrict our consideration to the part of ${\mathcal M}_{\mathrm{PII}}$ where Stokes parameters $s_1$ and $s_2$
can be taken as  coordinates. This means that 
$$
dG = \frac{\partial G}{\partial t}dt + \frac{\partial G}{\partial s_1}ds_1 + \frac{\partial G}{\partial s_2}ds_2.
$$
Observe that
\ben
\begin{aligned}
\operatorname{res}_{z=\infty}
\operatorname{Tr}
\left(A_{-3}z^2 
dG\,G^{-1}
\right)=&\,-\operatorname{Tr}\left(A_{-3}\lb dg_3-dg_2\cdot g_1-dg_1\cdot g_2+dg_1\cdot g_1^2\rb\rb, \\
\operatorname{res}_{z=\infty}
\operatorname{Tr}\left(A_{-2}z dG\,G^{-1}\right)=&\,-\operatorname{Tr}\left(A_{-2}\lb d g_2-d g_1\cdot g_1\rb\right),
\\ \operatorname{res}_{z=\infty}
\operatorname{Tr}\left(A_{-1}dG\,G^{-1}\right)=&\,-\operatorname{Tr}\left(A_{-1}d g_1\rb.
\end{aligned}
\ebn
Substitution of series (\ref{rhrep2}) into the equation (\ref{laxp21}) yields the
following formulae  for the coefficients  $g_1$, $g_2$, $g_3$:
\ben
\begin{gathered}
g_1=\frac{u}{2}\sigma_1-\frac{iH}{2}\sigma_3 \equiv \alpha_1\sigma_1+\alpha_3\sigma_3,\\
g_2=\frac{u^2 - H^2}{8}\mathbf 1 -\frac{2uH+v}{8}\sigma_2\equiv \beta_1 \mathbf 1+\beta_2\sigma_2,\\
g_3= -\frac{uH^2 + u^3 +vH + 2ut}{16}\sigma_1 + \frac{i\lb H^3 +2tH -3u^2H - uv\rb}{48}\sigma_3
 \equiv \gamma_1 \sigma_1+\gamma_3\sigma_3,
 \end{gathered}
\ebn
where 
\begin{equation}\label{momham}
H =  \frac{v^2}{4} - tu^2-u^4, \quad v = 2u_t,
\end{equation}
and
$$
  \sigma_1=\left(
\begin{array}{cc}
0&1\\
1&0\\
\end{array}\right), \quad  \sigma_2=\left(
\begin{array}{cc}
0&-i\\
i&0\\
\end{array}\right), \quad  \sigma_3=\left(
\begin{array}{cc}
1&0\\
0&-1\\
\end{array}\right).
$$

These equations lead to the following formula for the form
$\omega$,
\ben
\begin{gathered}
\omega=8\beta_2d\alpha_1-8i\beta_1d\alpha_3-8i\alpha_3d\beta_1-8\alpha_1d\beta_2+8id\gamma_3+\\
8i\lb \alpha_1^2+\alpha_3^2\rb d\alpha_3+8iu\alpha_3d\alpha_1-8iu\alpha_1d\alpha_3+8ud\beta_2+\lb 2it+4iu^2\rb d\alpha_3+4u_td\alpha_1.
\end{gathered}
\ebn
It can be rewritten in a completely localized form, i.e. directly in terms of $u\equiv u\lb t;s\rb$ and its derivatives in the form of a linear combination of the differentials $dt$, $ds_1$, $ds_2$. The result is
\beq
\begin{gathered}
{\omega}=\left(u_t^2-u^4-tu^2\right)dt+\frac23
\left(2u_t u_{s_1}-4u^3 t u_{s_1}-uu_{ts_1}+2tu_tu_{ts_1}-2ut^2u_{s_1}\right)ds_1\\
\label{omegaloc}
+\frac23\left(2u_t u_{s_2}-4u^3 t u_{s_2}-uu_{ts_2}+2tu_tu_{ts_2}-2ut^2u_{s_2}\right)ds_2.
\end{gathered}
\eeq
Simultaneously, we see that
\begin{equation}\label{p2jmu}
\omega_{\mathrm{JMU}} = \left(u_t^2-u^4-tu^2\right)dt \equiv Hdt,
\end{equation}
and hence the PII tau function is given by
\begin{equation}\label{p2tau}
\frac{\partial\ln\tau(t; s)}{\partial t} = u_t^2-u^4-tu^2 \equiv H.
\end{equation}
 \begin{rmk}\label{hamilton}
Equation \eqref{P2} admits a Hamiltonian reformulation. In fact, the quantity $H$ from (\ref{momham}) is exactly the corresponding  non-autonomous Hamiltonian, with the canonical coordinate and momentum given by $u$ and $v=2u_t$. It is straightforward to verify that the Hamiltonian satisfies the following relation:
\ben
4H-{2}tH_t-2vu_t+uv_t=0.
\ebn
Using this equation, we can rewrite ${\omega}$ as a natural extension of the classical action (i.e. up to addition of a total differential),
\begin{equation}\label{actiontau}
{\omega}=vdu-Hdt+d\left(\frac{2Ht-uv}{3}\right).
\end{equation}
Let us define
$$
F := \ln\tau\Bigl|_{t=t_1}^{t= t_2} - \frac{2}{3}\Bigl(Ht - uu_t\Bigr)\Bigl|_{t=t_1}^{t= t_2},
$$
where the objects on the right and left side are evaluated on solutions of \eqref{P2}. The formula (\ref{actiontau}) then  implies that 
\begin{equation}\label{difF}
\partial_{s_j} F = 2u_t\partial_{s_j}u\,\Bigl|_{t=t_1}^{t= t_2}, \qquad j = 1,2.
\end{equation}
This relation can, in principle, provide us with an alternative approach to evaluation of the asymptotics of the tau function.
We shall discuss this issue in more detail in a sequel to this paper. It also should be mentioned that, in the special case of the Ablowitz-Segur one-parameter family of solutions of \eqref{P2}, the relation (\ref{difF}) has been already observed 
in \cite[Proposition 6]{BoI}.  
\end{rmk}

\subsection{Tau function asymptotics \label{subsectauPII}}
We will analyze the open set in the space of solutions of the second Painlev\'e 
equation (\ref{P2}) characterized by the genericity assumptions  (\ref{genmonIN}) 
on  monodromy data.
The asymptotics of $u\lb t\rb$ as $t\to \pm\infty$ is
given by the formulae from \cite{Kap} (see also \cite{IN}, \cite{DZ1}, and \cite{FIKN}),
\beq\label{atmininfty}
u\lb t\rb = a^+_{0,0}e^{\frac{2i}{3}\lb-t\rb^\frac{3}{2}}\lb-t\rb^{\frac{3\mu}{2}-\frac{1}{4}} 
+ a^-_{0,0}e^{-\frac{2i}{3}\lb-t\rb^{\frac{3}{2}}}\lb-t\rb^{-\frac{3\mu}{2}-\frac{1}{4}}  
+ O\left(t^{3|\Re\mu| - 1}\right), \qquad t \to -\infty,
\eeq
\beq\label{mininftyparam}
\begin{aligned}
 \mu = -\frac{\ln\lb 1-s_1s_3\rb}{2\pi i},&\qquad a^+_{0,0}a^-_{0,0}=\dfrac{i\mu}{2},\\
 a^+_{0,0} = \dfrac{\sqrt{\pi}\,
 2^{3\mu}e^{-\frac{i\pi\mu}{2}-\frac{i\pi}{4}
 } }{s_1\Gamma(\mu)},&\qquad a^-_{0,0} = \dfrac{\sqrt{\pi}\,2^{-3\mu} e^{-\frac{i\pi\mu}{2}+\frac{i\pi}{4}} }{s_3\Gamma(-\mu)},
 \end{aligned}
 \eeq
and
 \beq
\sigma u\lb t\rb = i\sqrt{\frac{t}{2}}+b^+_{1,1}e^{\frac{2i\sqrt{2}}{3}t^\frac{3}{2}}t^{-\frac{3\nu}{2}-\frac{1}{4}}+ b^-_{1,1}e^{-\frac{2i\sqrt{2}}{3}t^{\frac{3}{2}}}t^{\frac{3\nu}{2}-\frac{1}{4}}
\label{atinfty}
+ O\left(t^{3|\Re\nu| - 1}\right), \qquad t \to +\infty,
\eeq
\beq
\begin{aligned}
  \nu = \frac{\ln\lb i\sigma s_2\rb}{\pi i},&\qquad b^+_{1,1}b^-_{1,1}=\dfrac{i\nu}{4\sqrt{2}},\\
\label{inftyparam}
 b^{+}_{1,1} = \dfrac{\sqrt{\pi}\,2^{-\frac{7\nu}{2}-\frac{3}{4}}e^{\frac{i\pi\nu}{2}-\frac{i\pi}{4}} }{{\lb 1+s_2s_3\rb}\Gamma\lb -\nu\rb},&\qquad b^{-}_{1,1} =- \dfrac{\sqrt{\pi}\,2^{\frac{7\nu}{2}-\frac{3}{4}}e^{\frac{i\pi\nu}{2}+\frac{i\pi}{4}}}{{\lb 1+s_1s_2\rb}\Gamma\lb\nu\rb}.
 \end{aligned}
 \eeq

From the previous section we already know that the 2-form $d\omega$ must be time-independent.  This fact can be
also  established  by a direct differentiation of the equation (\ref{omegaloc}). Indeed, after straightforward
though a bit tedious computation which involves using Painlev\'e~II equation (\ref{P2}), we obtain
\begin{equation}\label{simpp21}
 d{\omega}= \lb{v_{s_1}u_{s_2}-v_{s_2} u_{s_1}}\rb ds_1\wedge ds_2.
 \end{equation}
  From the equation (\ref{P2}) it also follows that
 \ben
 \frac{d}{dt}\lb {v_{s_1}u_{s_2}-v_{s_2} u_{s_1}}\rb = 0,
 \ebn
 which implies the time  independence of $d\omega$. Also, we  can observe that
 \begin{equation}\label{simpp22}
 d\omega = \lim_{t\to -\infty}d{\omega} = 4i da^-_{0,0}\wedge da_{0,0}^+=\lim_{t\to +\infty}d{\omega}=4i\sqrt{2} db^+_{1,1}\wedge db^-_{1,1}.
 \end{equation}
These relations indicate that the form $\omega_0$ for the second Painlev\'e equation  can be  identified
with the form $-4ia_{0,0}^+da_{0,0}^-$, so  that the $1$-form 
\beq
\hat{\omega}:=\omega -\omega_0 = 
\omega+4ia_{0,0}^+da_{0,0}^-
\eeq 
 is closed (with $\omega$ given by (\ref{omegaloc})). We can therefore extend the definition of the tau function as 
 \beq\label{taudef4}
\tau\lb t;s\rb :=\exp{\int\hat{\omega}}.
 \eeq

In order to proceed with  evaluation of the asymptotics of the tau function (\ref{taudef4}),
we will need more terms in the asymptotics of $u\lb t\rb$.  Denote 
 \ben
 p=e^{\frac{2i}{3}\lb-t\rb^\frac{3}{2}}(-t)^{\frac{3\mu}{2}},\qquad \zeta=\lb-t\rb^{-\frac{1}{4}}. 
 \ebn
 We have the following formal asymptotic expansion at $t = -\infty$ :
\ben
u\lb t\rb = \sum_{l\geq k\geq 0,\,\epsilon=\pm}a^\epsilon_{k,l}p^{\epsilon\lb 2k+1\rb}\zeta^{6l+1}.
\ebn
A few first terms are
\ben
u(t)=\lb a_{0,0}^+p+a_{0,0}^-p^{-1}\rb\zeta
+\lb a_{0,1}^+p+a_{0,1}^-p^{-1}
+a_{1,1}^+p^3+a_{1,1}^-p^{-3}\rb\zeta^7
+\ldots,
\ebn
where 
\ben
a_{0,1}^{\pm}=\frac{ia_{0,0}^\pm\lb\mp102\mu^2+36\mu\mp5\rb}{48},
\qquad a_{1,1}^{\pm}=-\frac{\bigl(a_{0,0}^{\pm}\bigr)^3}{4}.
\ebn
Similarly, denoting
 \ben
 q=e^{\frac{2i\sqrt{2}}{3}t^\frac{3}{2}}t^{-\frac{3\nu}{2}},\qquad \xi=t^{-\frac{1}{4}}, 
 \ebn 
 we have a  formal asymptotic expansion at $t=+\infty$ :
\ben
u\lb t\rb = \sum_{l\geq k\geq 0,\,\epsilon=\pm}b^\epsilon_{2k+1,6l+1}q^{\epsilon\lb 2k+1\rb}\xi^{6l+1}+\sum_{l\geq k\geq 0,\,\epsilon=\pm}b^\epsilon_{2k,6l-2}q^{2\epsilon k}\xi^{6l-2}.
\ebn
Let us record its several first terms:
\ben
\begin{gathered}
\sigma u(t)=\frac{i\xi^{-2}}{\sqrt{2}} + \lb b_{1,1}^+q+b_{1,1}^-q^{-1}\rb\xi+\lb b_{0,4}+b_{2,4}^+q^{2}+b_{2,4}^-q^{-2}\rb\xi^4+\lb b_{1,7}^+q+b_{1,7}^-q^{-1}+b_{3,7}^+q^3+b_{3,7}^-q^{-3}\rb\xi^7+\\+\lb b_{0,10}+b_{2,10}^+q^{2}
+b_{2,10}^-q^{-2}+b_{4,10}^+q^{4}+b_{4,10}^-q^{-4}\rb\xi^{10}+\ldots,
\end{gathered}
\ebn
where 
\ben
\begin{gathered}
b_{0,4}=-\frac{3\nu}{4},\qquad
b_{2,4}^{\pm}=-\frac{i\sqrt{2}}{2}\left({b_{1,1}^\pm}\right)^2,
\\
b_{1,7}^{\pm}={b_{1,1}^\pm}\frac{i\sqrt{2}}{6}\left(\mp \frac{51}{8}\nu^2-\frac{3}{2}\nu\mp \frac{17}{16}\right),\qquad
b_{3,7}^{\pm}=-\frac{\left({b_{1,1}^\pm}\right)^3}{2},\\
b_{0,10}=\frac{i\sqrt{2}}{2}\left(\frac{51}{32}\nu^2+
\frac{1}{8}\right),\qquad
b_{2,10}^{\pm}= \left(\mp\frac{17}{8}\nu^2-\frac{11}{8}\nu\mp\frac{41}{48}\right)\lb{b_{1,1}^\pm}\rb^2,\qquad
b_{4,10}^{\pm}=\frac{i\sqrt{2}}{4}\left({b_{1,1}^\pm}\right)^4.
\end{gathered}
\ebn

Justification of this asymptotics can be done using the Riemann-Hilbert approach, cf \cite{DZ2}. Plugging it into \eqref{omegaloc}, we get for sufficiently small $|\Re\mu|$ and $|\Re\nu|$ the following behaviors:
\begin{subequations}
\begin{align}
\label{omminf}
{\omega}=&\,d\left(-\frac{4i\mu}{3}\lb -t\rb^\frac{3}{2}-\frac{3\mu^2}{2}\ln\lb -t\rb-\mu^2 \right)+2i\lb a_{0,0}^-da_{0,0}^+-a_{0,0}^+da_{0,0}^-\rb+o\lb 1\rb,\quad 
&t \to -\infty,
\\
\label{ompinf}
{\omega}=&\,d\left(\frac{t^3}{12}+\frac{2
i\sqrt{2}}{3}\nu t^\frac{3}{2}-\frac{6\nu^2+1}{8}\ln t-\frac{\nu^2}{2} \right)+2i\sqrt{2}\lb b_{1,1}^+db_{1,1}^--b_{1,1}^-db_{1,1}^+\rb+o\lb 1\rb,\quad &t \to +\infty.
\end{align}
\end{subequations}
This in turn yields the asymptotics of $\tau\lb t\rb$ defined by \eqref{taudef4},
\begin{subequations}
\begin{align}\label{tauminf}
\ln\tau\lb t\rb=&\,-\frac{4i\mu}{3}\lb -t\rb^{\frac{3}{2}}-\frac{3\mu^2}{2}\ln\lb -t\rb-\mu^2-\mu+ c_1+o\lb 1\rb,\qquad &t\to-\infty,
\\ \nonumber
\ln\tau\lb t\rb=&\,\frac{t^3}{12}+\frac{2i\sqrt{2}\,\nu}{3} t^{\frac{3}{2}}-\frac{6\nu^2+1}{8}\ln t+4i\int\lb a_{0,0}^+da_{0,0}^-+\sqrt{2}b_{1,1}^+db_{1,1}^-\rb+&\\ \label{taupinf} &\,+\frac{\nu-\nu^2}{2}+c_2+o\lb1\rb,\qquad& t\to+\infty,
\end{align}
\end{subequations}
up to numerical constants $c_{1}$, $c_2$ independent of $s_1$ and $s_2$.
\begin{rmk}\label{ham0}
As follows from (\ref{simpp21}) and also from (\ref{actiontau}), the form $\Omega = d\omega$ coincides  with the symplectic
form for the second Painlev\'e equation (\ref{P2}). From (\ref{simpp22}) one can obtain an expression of this form in terms
of  either the asymptotic data at $t= -\infty$ or at $t=+\infty$. For instance, in the former case we have
\begin{equation}\label{simpp23}
\Omega = 4i da^-_{0,0}\wedge da_{0,0}^+.
\end{equation}
The last equation can in turn be transformed into an expression  of the symplectic  form $\Omega$ 
in terms of  monodromy data $s\in \mathcal M_{\mathrm{PII}}$. Indeed, from (\ref{mininftyparam}) we see that 
the amplitudes $a_{0,0}^{\pm}$ can be written in the form
\begin{equation}\label{a00f}
a_{0,0}^{+} = \frac{f_{+}\lb \mu\rb}{s_{1}},\qquad
a_{0,0}^{-} = \frac{f_{-}\lb \mu\rb}{s_{3}}.
\end{equation}
where $f_{\pm}\lb \mu\rb$ are functions of a single argument which can of course be written 
explicitly using (\ref{mininftyparam}). From these explicit formulae we will only need one relation
\begin{equation}\label{ffmu}
\frac{f_+\lb\mu\rb f_-\lb \mu\rb}{s_1s_3} = \frac{i\mu}{2}.
\end{equation}
Let us also notice that (see again (\ref{mininftyparam})),
\begin{equation}\label{ssmu}
s_1s_3 = 1 - e^{-2\pi i \mu}.
\end{equation}
Now, differentiating (\ref{a00f}), we get 
\begin{align*}
da^{+}_{0,0}=&\,\left(-\frac{f_+}{s^2_1} + \frac{ f'_+}{s_1} \dfrac{\partial \mu}{\partial s_1}\right)ds_1
+\frac{f'_+}{s_1}\dfrac{\partial \mu}{\partial s_3} ds_3,\\
da^{-}_{0,0}=&\,\left(-\frac{f_-}{s^2_3} + \frac{f'_-}{s_3}  \dfrac{\partial \mu}{\partial s_3}\right)ds_3
+\frac{f'_-}{s_3}\dfrac{\partial \mu}{\partial s_1} ds_1.
\end{align*}
Substituting these two equations into (\ref{simpp23}) and using (\ref{ffmu}), we arrive at the equation
\begin{equation}\label{Omega3}
\Omega = 4i\left(-\frac{i\mu}{2s_1s_3}  +\frac{f_-f'_+}{s_1s_3}\cdot\frac{1}{s_3}
\dfrac{\partial \mu}{\partial s_1}
+ \frac{f'_-f_+}{s_1s_3}\cdot \frac{1}{s_1}\dfrac{\partial \mu}{\partial s_3}\right)ds_1\wedge ds_3.
\end{equation}
Noticing that
\ben
\dfrac{1}{s_3}\dfrac{\partial \mu}{\partial s_1}=\dfrac{1}{s_1}\dfrac{\partial \mu}{\partial s_3}=\dfrac{1}{2\pi i\lb 1-s_1s_3\rb},
\ebn
the equation (\ref{Omega3}) is transformed into
\begin{equation}\label{Omega4}
\Omega = 4i\left(-\frac{i\mu}{2s_1s_3}  +\frac{\lb f_-f_+\rb'}{2\pi is_1s_3\lb 1-s_1s_3\rb}\right)ds_1\wedge ds_3.
\end{equation} 
Finally, differentiating (\ref{ffmu}) with respect to $\mu$, taking into account that $\lb s_1s_3\rb' = 2\pi i \lb 1-s_1s_3\rb$
in view of (\ref{ssmu}), we conclude that
$$
-\frac{i\mu}{2s_1s_3}  +\frac{\lb f_-f_+\rb'}{2\pi is_1s_3\lb 1-s_1s_3\rb} = \frac {1}{4\pi \lb 1-s_1s_3\rb},
$$
which means that we end up with the following expression of the symplectic form $\Omega$ in terms 
of monodromy parameters:
\begin{equation}\label{Omegafinal}
d\omega = \Omega = \dfrac{i}{\pi}\dfrac{ds_1\wedge ds_3}{1-s_1s_3}.
\end{equation}
This expression coincides with the  one obtained for the curvature $d\omega_{\mathrm{MB}}$  in \cite{Bertola1} where Painlev\'e II 
was also used to exemplify the general  Malgrange-Bertola  form (\ref{MBf}). It should be
mentioned that the first derivation of this formula was done by H. Flaschka and A. Newell in \cite{FN2}. Also note that 
in our asymptotics-based derivation of (\ref{Omegafinal}) we mimic the  methodology used in \cite{BFT} for the evaluation of
the KdV symplectic form in terms of the relevant scattering data --- the PDE analog of the monodromy data.
\end{rmk}

\subsection{Connection coefficient. Towards the proof of Theorem \ref{theoP2IN}}\label{subsecconPII}
The asymptotic equations (\ref{tauminf}) and (\ref{taupinf}) can be rewritten in the form \eqref{f44IN}:
\beq\label{f44}
\tau(t)\simeq
\begin{cases} \ds \mathcal C_-e^{-\frac{4i\mu}{3}\lb-t\rb^{\frac{3}{2}}}\lb -t\rb^{-\frac{3\mu^2}{2}}\bigl[1+o\lb 1\rb\bigr],\qquad & \text{as}\; t\to-\infty,\vspace{0.1cm}\\
\ds \mathcal C_+e^{\frac{t^3}{12}+\frac{2i\sqrt 2}{3}\nu t^{\frac{3}{2}}}t^{-\frac{3\nu^2}{4}-\frac{1}{8}}\bigl[1+o\lb 1\rb\bigr],\qquad & 
\text{as}\; t\to+\infty.
 \end{cases}
\eeq
The connection problem we want to solve for the Painlev\'e II tau function concerns the evaluation of the ratio $\Upsilon\lb s\rb:=\frac{\mathcal C_+}{\mathcal C_-}$
in terms of monodromy data $s\in \mathcal M_{\mathrm{PII}}$.

From \eqref{tauminf}, \eqref{taupinf} we have
\begin{equation}\label{CC1}
\ln \Upsilon\lb s\rb=-\frac{\nu^2}{2}+\frac{\nu}{2}+\mu^2+\mu+4i\int \lb a_{0,0}^+da_{0,0}^-+\sqrt{2}\,b_{1,1}^+db_{1,1}^-\rb.
\end{equation}
Hence, our task  is to evaluate the integral in the right hand side of (\ref{CC1}).
To this end, it is convenient to introduce new monodromy parameters $\rho$ and $\tilde\eta$ by the equations
\beq\label{newparP2}
(1+s_1s_2)^{-1}=e^{i\pi\rho},\qquad
s_3^{-1}=e^{i\pi\tilde \eta}.
\eeq
This transforms the integral in question into
\begin{equation}\label{inta00b11}
4i\intop a_{0,0}^+da_{0,0}^-+\sqrt{2}b_{1,1}^+db_{1,1}^-=-2\intop\mu \lb\ln a_{0,0}^-\rb'_\mu d\mu - \intop\nu \lb\ln b_{1,1}^-\rb'_\nu d\nu - i\pi\intop \lb \nu d\rho+2\mu d\tilde\eta\rb.
\end{equation}
The first two integrals on the right can be rewritten as
\begin{subequations}
\begin{align}\label{inta00}
\intop\mu \lb\ln a_{0,0}^-\rb'_\mu d\mu=&\,-\frac{6\ln 2+i\pi}{4}\,\mu^2-\intop\mu\, d\ln \Gamma \lb-\mu\rb,\\
\label{intb11}
\intop\nu \lb \ln b_{1,1}^-\rb'_\nu d\nu=&\,\frac{7\ln 2+i\pi}{4}\nu^2-\intop\nu\, d\ln \Gamma\lb\nu\rb.
\end{align}
\end{subequations}
In order to simplify the third integral,  we first notice that due to cyclic relation
$s_1-s_2+s_3 + s_1s_2s_3 = 0$ satisfied by the Stokes parameters
we may write
\beq\label{murho}
\mu=-\frac{1}{2\pi i }\ln \left(\dfrac{1-e^{-2\pi i\eta}}{1-e^{i\pi\lb\nu-\eta\rb}}\right),\qquad
\rho=-\frac{1}{\pi i}\ln\left(\dfrac{1-e^{2\pi i\nu}}{1-e^{i\pi\lb \nu-\eta\rb}}\right),
\eeq
where $\eta=\tilde\eta-\frac{\sigma}{2}$.
These formulae allow one to express the third  integral  in (\ref{inta00b11}) in terms of dilogarithms. 
We have (cf. similar calculations in \cite{ILT14})
\beq\label{intetarho}
\begin{gathered}
\intop 2\mu d\tilde\eta +\nu d\rho  -\nu\rho
=\intop 2\mu d\eta-\rho d\nu
=\frac{1}{2\pi^2}\left[ \operatorname{Li}_2\lb e^{-2i\pi\eta}\rb+\operatorname{Li}_2\lb e^{2\pi i\nu}\rb-2\operatorname{Li}_2\lb e^{i\pi\lb\nu-\eta\rb}\rb\right],
\end{gathered}
\eeq
where $\operatorname{Li}_2\lb z\rb$ denotes the dilogarithm function
\ben
\operatorname{Li}_2\lb z\rb=-\int_0^z\dfrac{\ln(1-x)}{x}dx.
\ebn

The formulae (\ref{inta00}), (\ref{intb11}), and (\ref{intetarho}) enable us to complete the evaluation of $\Upsilon\lb s\rb$ in terms
of Barnes $G$-functions. Indeed, it suffices to use  the  classical formula (\ref{diffg})
for the integrals  (\ref{inta00}), (\ref{intb11}), and another classical formula
\ben\label{4}
\operatorname{Li}_2\lb e^{2\pi i z}\rb=-2\pi i \ln \frac{G\lb 1+z\rb}{G\lb 1-z\rb}-2\pi i z\, \ln \dfrac{\sin\pi z}{\pi}-\pi^2 z\lb 1-z\rb+\dfrac{\pi^2}{6},
\ebn
to rewrite the integral (\ref{intetarho}). Skipping some straightforward though tedious calculations, we arrive at the following representation:
\beq\label{finalp2}
\begin{gathered}
\Upsilon\lb s\rb=\Upsilon_0\,\cdot\,  2^{{3}\mu^2-\frac{7\nu^2}{4}}
(2\pi)^{{-\mu-\frac{\nu}{2}}}e^{\frac{\pi i}{4}\lb\eta^2+2\mu^2+2\eta\nu-8\mu\eta\rb } 
\frac{G\lb 1-\nu\rb\hat G\lb \eta\rb}{G^2\lb 1-\mu\rb\hat{G}^2\lb\frac{\eta-\nu}{2}\rb}
\dfrac{ }{ }\,,
\end{gathered}
\eeq
The remaining task is to determine the numerical (i.e. independent of monodromy data) constant $\Upsilon_0$.

\begin{rmk}
In order to bring the final answer to the compact form (\ref{finalp2}), one has  to use the relations 
\beq\label{trigo}
e^{\frac{i\pi}{2}\lb4\mu-\eta-\nu\rb}=
\dfrac{
\sin\frac{\pi\lb\eta-\nu\rb}{2}}{\sin\pi\eta},\qquad
e^{\frac{i\pi}{2}\lb 2\rho+\eta+\nu\rb}=
\dfrac{\sin\frac{\pi\lb\nu-\eta\rb}{2}}{\sin \pi\nu}.
\eeq
which can be verified with the help of  \eqref{murho}. This allows to get rid of all sine functions in the final formula.
\end{rmk}

\begin{rmk} Strictly speaking, we have derived (\ref{finalp2}) under assumption that $|\Re\mu|$ and $|\Re\nu|$ are sufficiently small.
However, in the final result we can lift this restriction by noticing that both sides of (\ref{finalp2}) are analytic functions of 
monodromy/Riemann-Hilbert data. (For the ratio $\Upsilon\lb s\rb$ this is a corollary of the general Birkhoff-Grothendieck-Malgrange theory).
\end{rmk} 

 \subsection{Numerical constant. End of the proof of Theorem \ref{theoP2IN}}\label{numcoef}
  It would suffice to calculate the numerical constant $\Upsilon_0$ for a particular solution corresponding to admissible monodromy. In contrast to Painlevé VI equation, which has families of explicit algebraic and elliptic solutions, the Painlevé II equation \eqref{P2} has only trivial rational solution $u=0$. Being associated to non-generic Stokes data, it is not suitable for our purposes. 
 
 Another possibility is to consider the transcendental Hastings-McLeod solution $u_{\mathrm{HM}}\lb t\rb$. It has the following  asymptotics
 (non-generic as well) on the real axis:
 \beq 
 u_{\mathrm{HM}}\lb t\rb\simeq\begin{cases}\sqrt{\dfrac{-t}{2}}+O\left(\lb-t\rb^{-\frac{1}{4}}e^{-\frac{2}{3}\sqrt{2}\lb -t\rb^{\frac{3}{2}}}\right), \quad & t\to-\infty,\\
 \dfrac{t^{-\frac{1}{4}}e^{-\frac{2}{3}t^{\frac{3}{2}}}}{2\sqrt{\pi}}+O\left(t^{-\frac{7}{4}}e^{-\frac{2}{3}t^{\frac{3}{2}}}\right), \quad &t\to+\infty.
 \end{cases}
 \eeq 
 Denote by $H_{\mathrm{HM}}\lb t\rb$ the corresponding Hamiltonian. Plugging the asymptotics of $u_{\mathrm{HM}}\lb t\rb$ into the  definition \eqref{momham}, one finds that
 \ben
 H_{\mathrm{HM}}\lb t\rb=\begin{cases}
 \dfrac{t^2}{4}-\dfrac{1}{8t}+O\left(\lb -t\rb^{-\frac{1}{4}}e^{-\frac{2}{3}\sqrt{2}\lb-t\rb^{\frac{3}{2}}}\right),\quad & t\to -\infty,\vspace{0.1cm}\\
 O\left(t^{-1}e^{-\frac{4}{3}t^{\frac{3}{2}}}\right),\quad & t\to +\infty.
 \end{cases}
 \ebn
 The rapid decay of $H_{\mathrm{HM}}$ as $t\to +\infty$ allows to normalize the tau function associated to the Hastings-McLeod solution  by setting
 \beq\label{thm}
 {\tau}_{\mathrm{HM}}\lb t\rb:=\mathrm{exp}\left\{-\int_t^{+\infty}H_{\mathrm{HM}}(s)ds\right\}.
 \eeq
 Its asymptotics is then given by
 \beq\label{tauhmA}
 {\tau}_{\mathrm{HM}}\lb t\rb\simeq 
 \begin{cases}{\Upsilon}_{\mathrm{HM}} \lb-t\rb^{-\frac{1}{8}}e^{\frac{t^3}{12}},\quad &
  t\to-\infty,\\
 1, \quad & t\to +\infty.  
 \end{cases}
 \eeq
 The coefficient ${\Upsilon}_{\mathrm{HM}}$ represents the finite part of the integral in (\ref{thm}) as $t\to-\infty$. It turns out to be a close relative of the quantity $\Upsilon_0$ that we are after. The former constant has been evaluated in \cite{DIK} and the result reads 
 \beq\label{YHMDIK}
 {\Upsilon}_{\mathrm{HM}}=2^\frac{1}{24}e^{\zeta'\lb-1\rb},
 \eeq
 where $\zeta\lb s\rb$ denotes the Riemann zeta function. Alternatively, ${\Upsilon}_{\mathrm{HM}}$ can be expressed in terms of the Glaisher-Kinkelin constant $ A=e^{\frac1{12}-\zeta'\lb-1\rb}$ or in terms of the special value $G\lb\frac12\rb = 2^{\frac{1}{24}}\pi^{-\frac14}e^{\frac32\zeta'\lb-1\rb}$ of the Barnes function introduced above.

 The Hastings-McLeod solution is associated, via the Riemann-Hilbert correspondence, to the following point $s\in\mathcal M_{\mathrm{PII}}$ in the space of Stokes data:
  $$
  s_1=-i,\quad s_2=0, \quad s_3=i.
  $$
 Although these parameters do not satisfy genericity conditions  \eqref{genmonIN}, the apparent difficulty can be overcome using the $\mathbb Z_3$-symmetry of the PII-RH problem. More precisely, the solutions of \eqref{P2} verify the periodicity relation
 $$
 u\lb t;s_1,s_2,s_3\rb=e^{\frac{2\pi i}{3}}u\lb te^{\frac{2\pi i }{3}};s_3,-s_1,-s_2\rb,
 $$
 in which we explicitly indicate the dependence of solutions on monodromy. Introducing a ``rotated'' Hastings-McLeod solution
 $$\tilde{u}_{\mathrm{HM}}\lb t\rb:=e^{\frac{2\pi i}{3}}u_{\mathrm{HM}}\lb t e^{\frac{2\pi i}{3}};-i,0,i\rb,$$
 one may check that  $\tilde{u}_{\mathrm{HM}}\lb t\rb$ satisfies Painlevé II equation \eqref{P2} and corresponds to the Stokes data
 $$
 s_1=0,\quad s_2=-i, \quad s_3=-i. $$
 These new parameters do satisfy the conditions  \eqref{genmonIN}. In the above notations, we have
 $\sigma=1$ and ${\mu= \eta= \nu=0}$, which implies that
 $a_{0,0}^-=b_{1,1}^-=0$.
 One may also rewrite $a_{0,0}^+$, $b_{1,1}^+$ in (\ref{mininftyparam}), (\ref{inftyparam}) as
 $$
 a_{0,0}^+=\dfrac{2^{3\mu-1}e^{-\frac{3\pi i}{4}}e^{\frac{i\pi \mu}{2}}\Gamma\lb 1-\mu\rb s_3}{\sqrt{\pi}}, \qquad b_{1,1}^+=\dfrac{2^{-\frac{7\nu}{2}-\frac{7}{4}}e^{-\frac{3\pi i}{4}}e^{-\frac{i\pi\nu}{2}}\Gamma\lb 1+\nu\rb\lb 1+s_1s_2\rb}{\sqrt{\pi}},
 $$ 
 so that for $\tilde{u}_{\mathrm{HM}}\lb t\rb=u\lb t;0,-i,-i\rb$ we get
  $$
   a_{0,0}^+=\dfrac{e^{\frac{3\pi i}{4}}}{2\sqrt{\pi}},\qquad  b_{1,1}^+=\dfrac{2^{-\frac{7}{4}}e^{-\frac{3\pi i}{4}}}{\sqrt{\pi}}.
 $$
 The asymptotics of $u_{\mathrm{HM}}\lb t\rb$ may be continued inside the sectors $-\frac{\pi}{3}\leq\arg t\leq 0$, $\frac{2\pi}{3}\leq\arg t\leq \pi$, see \cite{FIKN}.
 We record for later use more terms in the relevant asymptotics of $H_{\mathrm{HM}}\lb t\rb$ as $|t|\to \infty$:
 \begin{equation}
 \label{hama}
 H_{\mathrm{HM}}(t)\simeq\begin{cases}
 \dfrac{t^2}{4}-\dfrac{1}{8t}-\dfrac{2^{-\frac{7}{4}}e^{-\frac{2}{3}\sqrt{2}(e^{-i\pi}t)^{\frac{3}{2}}}}{\sqrt{\pi}\lb e^{-i\pi}t\rb^{\frac{1}{4}}}+\dfrac{e^{-\frac{4}{3}\sqrt{2}\lb e^{-i\pi}t\rb^{\frac{3}{2}}}}{16{\pi}t}+O\left(|t|^{-\frac{7}{4}}\right),\qquad &\arg t\in\left[\frac{2\pi}{3}, \pi\right],
 \\
 \dfrac{e^{-\frac{4}{3}t^{\frac{3}{2}}}}{8\pi t}+O\left(|t|^{-\frac{5}{2}}\right),\quad |t|\to \infty,\qquad &\arg t\in\left[-\frac{\pi}{3}, 0\right].
 \end{cases}
 \end{equation}

 Let $\tilde{H}_{\mathrm{HM}}\lb t\rb$ denote the Hamiltonian corresponding to the rotated solution $\tilde u_{\mathrm{HM}}(t)$. The tau function associated to this solution may be defined as
 \beq\label{tib}
 \tilde{\tau}_{\mathrm{HM}}(t)=\exp\left\{\int_{-\infty}^t\tilde{H}_{\mathrm{HM}}\lb s\rb ds\right\}.
 \eeq
 Its asymptotics contains the so far unknown coefficient $\Upsilon_0$: 
 \ben
 \tilde{\tau}_{\mathrm{HM}}\lb t\rb\simeq \begin{cases}
 1,\qquad & t\to-\infty,\\
 \Upsilon_0 t^{-\frac{1}{8}} e^{\frac{t^3}{12}},\qquad 
 & t\to+\infty.
 \end{cases}
 \ebn
 The main idea of our computation of $\Upsilon_0$ is to relate the integrals \eqref{thm} and \eqref{tib}. To this end let us substitute ${e^{\frac{2\pi i}{3}}s=y}$ into \eqref{tib} and take into account that
 $\tilde{H}_{\mathrm{HM}}\lb t\rb=e^{\frac{2\pi i}{3}}{H}_{\mathrm{HM}}\lb t e^{\frac{2\pi i}{3}}\rb$. This yields
 \ben
 \tilde{\tau}_{\mathrm{HM}}\lb t\rb=\exp\left\{\int_{-\infty e^{\frac{2\pi i}{3}}}^{te^{\frac{2\pi i}{3}}}H_{\mathrm{HM}}(y)dy\right\},
 \ebn
 where the integral is taken along the line $e^{\frac{2\pi i}{3}}\mathbb R$. 
 
 From (\ref{tauhmA}) it follows that
 $$
 \ln\Upsilon_{\mathrm{HM}}=\lim_{t\to+\infty}\left(\ln\tau_{\mathrm{HM}}\lb -t\rb+\dfrac{t^3}{12}+\dfrac{\ln t}{8}\right)=\lim_{t\to+\infty}\left(-\int_{-t}^{+\infty}H_{\mathrm{HM}}\lb s\rb ds +\dfrac{t^3}{12}+\dfrac{\ln t}{8}\right).
 $$
 Since the above integral converges, we may write for $a>0$
 \begin{align*}
 \ln\Upsilon_{\mathrm{HM}}=&\,\lim_{t\to+\infty}\left(-\int_{-a}^{t}H_{\mathrm{HM}}(s)ds-\int_{-t}^{-a}H_{\mathrm{HM}}(s)ds +\dfrac{t^3}{12}+\dfrac{\ln t}{8}\right)=\\
 =&\,\lim_{t\to+\infty}\left(-\int_{-a}^{t}H_{\mathrm{HM}}\lb s\rb ds-\int_{-t}^{-a}\left[H_{\mathrm{HM}}\lb s\rb-\frac{s^2}{4}+\frac{1}{8s}\right]ds +\dfrac{a^3}{12}+\dfrac{\ln a}{8}\right).
 \end{align*}
 Here the branch cut for the logarithm is chosen to be the negative imaginary axis so that $-\frac{\pi}{2}\leq\arg z<\frac{3\pi}{2}$. For $\ln \Upsilon_0$, one similarly  obtains
 \begin{align*}
 \ln\Upsilon_0=&\,\lim_{t\to+\infty}\left(\ln \tilde\tau_{\mathrm{HM}}\lb t\rb-\dfrac{t^3}{12}+\dfrac{\ln t}{8}\right)=\lim_{t\to+\infty}\left(\int_{-\infty e^{\frac{2\pi i}{3}}}^{te^{\frac{2\pi i}{3}}}H_{\mathrm{HM}}\lb s\rb ds -\dfrac{t^3}{12}+\dfrac{\ln t}{8}\right)=\\
 =&\,\lim_{t\to+\infty}\left(\int_{-t e^{\frac{2\pi i}{3}}}^{a e^{\frac{2\pi i}{3}}} H_{\mathrm{HM}}\lb s\rb ds+\int_{ae^{\frac{2\pi i}{3}}}^{te^{\frac{2\pi i}{3}}}H_{\mathrm{HM}}\lb s\rb ds -\dfrac{t^3}{12}+\dfrac{\ln t}{8}\right)=\\
 =&\,\lim_{t\to+\infty}\left(\int_{-t e^{\frac{2\pi i}{3}}}^{a e^{\frac{2\pi i}{3}}} H_{\mathrm{HM}}\lb s\rb ds+\int_{ae^{\frac{2\pi i}{3}}}^{te^{\frac{2\pi i}{3}}}\left[H_{\mathrm{HM}}\lb s\rb-\frac{s^2}{4}+\frac{1}{8s}\right]ds -\dfrac{a^3}{12}+\dfrac{\ln a}{8}\right).
 \end{align*}
  We would like to deform the contours in the two integrals so as to connect $\ln\Upsilon_0$ with $\ln\Upsilon_{\mathrm{HM}}$. The relevant deformations are represented in Fig. \ref{contour} below.
 \begin{figure}[h!]
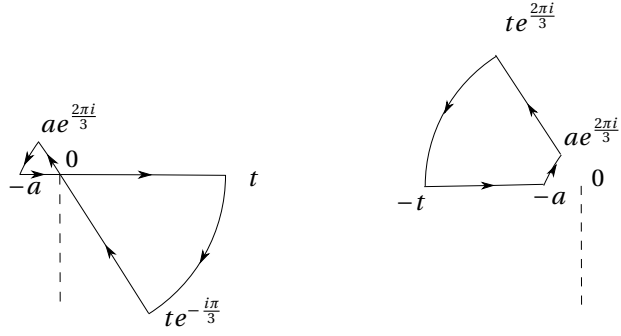

 \begin{center}
 \begin{minipage}{0.4\textwidth}
 	\begin{center} 	
 	\def\svgwidth{0.9\columnwidth}
 	{\input{contour.tex}}
 	\end{center}
\end{minipage}
 \begin{minipage}{0.4\textwidth} 	
 	\begin{center}
 	\def\svgwidth{1\columnwidth}
 	{\input{contour2.tex}}
 	\label{contour2}
 	\end{center}
 \end{minipage}
 	\caption{Contour deformation for the first (left) and second (right) integral
 		\label{contour}}  	
 		\end{center}
 \end{figure} 
 
 The crucial observation is that the integrals along the contours shown in Fig. \ref{contour}  are equal to zero. The reason for their vanishing is the absense of poles in the Hastings-Mcleod solution inside the sectors $\arg t\in\left[-\frac{\pi}{3}, 0\right]\cup \left[\frac{2\pi}{3},\pi\right]$, see \cite{HXZ}. It follows that
 \begin{align*}
 \ln\Upsilon_0=\lim_{t\to+\infty}\left\{\int_{t}^{-a} H_{\mathrm{HM}}\lb s\rb ds+\int_{-a}^{ae^{\frac{2\pi i}{3}}} H_{\mathrm{HM}}\lb s\rb ds+\int_{te^{-\frac{\pi i}{3}}}^{t} H_{\mathrm{HM}}\lb s\rb ds+\int_{-a}^{-t}\left[H_{\mathrm{HM}}\lb s\rb-\frac{s^2}{4}+\frac{1}{8s}\right]ds\right.\\
 \left.+\int_{ae^{\frac{2\pi i}{3}}}^{-a}\left[H_{\mathrm{HM}}(s)-\frac{s^2}{4}+\frac{1}{8s}\right]ds+\int_{-t}^{te^{\frac{2\pi i}{3}}}\left[H_{\mathrm{HM}}(s)-\frac{s^2}{4}+\frac{1}{8s}\right]ds -\dfrac{a^3}{12}+\dfrac{\ln a}{8}\right\}.
 \end{align*}
  Using the asymptotics \eqref{hama} and appropriate version of the Jordan's lemma, one may show that the limits of the integrals over two arcs of the big circle are equal to zero. Therefore we get
  $$
  \ln\Upsilon_0=\lim_{t\to+\infty}\left\{\int_{t}^{-a} H_{\mathrm{HM}}\lb s\rb ds+\int_{-a}^{-t}\left[H_{\mathrm{HM}}\lb s\rb-\frac{s^2}{4}+\frac{1}{8s}\right]ds+\dfrac{a^3}{12}+\dfrac{\ln a}{8}+\dfrac{i\pi}{24}\right\}=\ln\Upsilon_{\mathrm{HM}}+\dfrac{i\pi}{24}.
  $$
  In combination with (\ref{YHMDIK}), this gives us the unknown constant in the connection
  coefficient (\ref{finalp2}):
 \beq\label{Ups00}
 \Upsilon_0=2^{\frac1{24}} e^{\zeta'\lb-1\rb+\frac{i\pi}{24}}.
 \eeq
 This evaluation reproduces the experimentally observed numerical value $\Upsilon_0\approx 0.865+0.114i$ and, together 
 with (\ref{finalp2}), completes the proof of Theorem \ref{theoP2IN}.

\subsection{Quasi-periodicity of the connection constant \label{subsecconPII2}}
As in the case of previously studied Painlev\'e equations, the asymptotic expressions \eqref{f44} of the Painlev\'e II  tau function $\tau(t)$ can be upgraded to full Fourier-type  series.
Similarly to the Painlev\'e~III ($D_8$) equation considered in \cite{ILT14}, by examining higher terms of the asymptotic expansions one may obtain two conjectural representations 
for the tau function $\tau\lb t\rb$. 

The first representation is given by
\begin{subequations}\label{left}
    \beq\label{fourL}
     \tau\left(t\right)=\chi_-\sum_{n\in\mathbb Z}e^{in\rho_-}
      \mathcal{B}\left(\nu_-+n,t\right),      
    \eeq
 where the two parameters $\lb \nu_-,\rho_-\rb$ are related to Stokes data by
 \beq\label{rhoL}
 \nu_-=-\mu=\frac{\ln\lb 1-s_1s_3\rb}{2\pi i},\qquad e^{i\rho_-}=-\frac{e^{2\pi i \lb \mu-\eta\rb}}{2\pi}=\frac{s_3^2}{2\pi\lb 1-s_1s_3\rb},
 \eeq 
 and the function $\mathcal{B}\left(\alpha,t\right)$ admits the following asymptotic expansion as $t\to -\infty$:
 \beq
 \mathcal{B}\left(\alpha,t\right)=6^{-\alpha^2}e^{\frac{i\pi\alpha^2}{2}} G^2\left(1+\alpha\right)r^{-\alpha^2}
       e^{i\alpha r}
       \left[1+\sum_{k=1}^{\infty}\frac{B_k\left(\alpha\right)}{r^k}\right],\qquad r=\frac43\lb -t\rb^{\frac32}.
 \eeq
 \end{subequations}
 Its first few coefficients are given by
      \begin{align*}
      B_1\left(\alpha\right)=
      -\frac{i\alpha\left(34\alpha^2+1\right)}{18},\qquad
      B_2\left(\alpha\right)=
      -\frac{\alpha^2\lb1156\alpha^4+2318\alpha^2+271\rb}{648},\qquad
      \ldots
      \end{align*} 
      
 The second conjectural representation is
 \begin{subequations}\label{right}
 \beq\label{fourR}
 \tau\lb t\rb=\chi_{+}
       t^{-\frac{1}{8}}e^{\frac{t^3}{12}}
       \sum_{n\in\mathbb Z}e^{in\rho_+}
       \mathcal{D}\left(\nu_++n,t\right),
 \eeq 
 with $\lb\nu_+,\rho_+\rb$ expressed in terms of monodromy as
 \beq\label{rhoR}
 \nu_+=\nu=\frac{\ln\lb -s_2^2\rb}{2\pi i},\qquad e^{i\rho_+}=\frac{e^{-i\pi\rho}}{\sqrt{2\pi}}=
 \frac{1+s_1s_2}{\sqrt{2\pi}}.
 \eeq
 The asymptotic expansion of the Fourier coefficients $\mathcal D\lb \alpha,t\rb$ as $t\to+\infty$ has the form
 \beq
 \mathcal{D}\left(\alpha,t\right)=12^{-\frac{\alpha^2}{2}}e^{-\frac{i\pi\alpha^2}{4}} G\left(1+\alpha\right)r^{-\frac{\alpha^2}{2}}
       e^{i\alpha r}\left[1+\sum_{k=1}^{\infty}
       \frac{D_k\left(\alpha\right)}{r^k}\right],\qquad r=\frac{2\sqrt2 }{3} t^{\frac32},
 \eeq
 \end{subequations}
 and its first coefficients are given by
 \ben
        D_1\left(\alpha\right)=-
        \frac{i\alpha\left(34\alpha^2+31\right)}{72},\qquad
        D_2\left(\alpha\right)=-\frac{289}{2592}\alpha^6-
        \frac{413}{648}\alpha^4-
        \frac{11509}{10368}\alpha^2-
        \frac{1}{24},\qquad \ldots
 \ebn

  The coefficient $\chi\lb s\rb:=\chi_+/\chi_-$, $s\in\mathcal M_{\mathrm{PII}}$ of relative normalization of the Fourier series (\ref{fourL}) and (\ref{fourR}) is related to the connection coefficient $\Upsilon\lb s\rb$ considered above by
   \ben
   \Upsilon\lb s\rb=\chi\lb s\rb\cdot 2^{3\nu_-^2-7\nu_+^2/4}
   e^{-\frac{i\pi\lb\nu_+^2+2\nu_-^2\rb}{4}}
   \frac{G\lb1+\nu_+\rb}{
   G^2\lb1+\nu_-\rb}.
   \ebn
   In combination with Theorem~\ref{theoP2IN}, this relation implies that
   \beq\label{chiexp}
   \chi\lb s\rb=\Upsilon_0\cdot \lb2\pi\rb^{\nu_--\frac{\nu_+}{2}}
    e^{i\pi\lb\frac{\lb\eta+\nu_+\rb^2}4+\nu_-^2+2\nu_-\eta\rb}
    \frac{
    \hat{G}\lb \eta\rb}{\hat{G}\lb \nu_+\rb
    \hat{G}^2\lb \frac{\eta-\nu_+}{2}\rb},
   \eeq
   with the same numerical constant $\Upsilon_0$ given by (\ref{Ups00}).
  On the other hand, analogously to Painlevé VI \cite{ILT13} and Painlev\'e III \cite{ILT14} equations, the Fourier  series (\ref{fourL}) and (\ref{fourR}) would imply the 
following quasi-periodic properties of the
constant $\chi$ as a function of monodromy data $\lb\nu_-, \nu_+\rb$:
 \beq\label{periodic}
 \left\{
 \begin{aligned}
 \chi\lb\nu_-+1,\nu_+;\eta\rb&=\,e^{-i\rho_-}\chi\lb\nu_-,\nu_+;
 \eta\rb,\\ \chi\lb\nu_-,\nu_++1;\eta+1\rb&=\,\;\;e^{i\rho_+}\;\chi\lb\nu_-,\nu_+;
 \eta\rb.
 \end{aligned}\right.
 \eeq

Equations (\ref{left})--(\ref{right}) are conjectures. However, the explicit formula (\ref{chiexp})
is rigorous. It can be checked directly that it indeed satisfies the quasi-periodic relations (\ref{periodic}). 
This can be considered as a confirmation of conjectures (\ref{left})--(\ref{right}). In a future work,
we hope to produce their complete proof by generalizing the recent approach of \cite{GL} and constructing proper Fredholm determinant representations for the Painlev\'e II tau function on the canonical rays.

 \vspace{0.5cm}
 
 \noindent
 {\small {\bf Acknowledgements}. We would like to thank M. Bertola, P. Gavrylenko and N. Iorgov for illuminating discussions and correspondence. Special thanks to J. Roussillon for numerical checks of the statement of Theorem~\ref{theoP2IN}. A. Its was supported by the National Science Foundation grants DMS-1361856, DMS-1700261 and
 Russian Science Foundation grant No.17-11-01126, A. Prokhorov was supported by the National Science
 Foundation grants DMS-1637087, DMS-1361856, DMS-1700261 and Russian Science Foundation
 grant No.17-11-01126, O. Lisovyy was supported by the CNRS/Projet International de Coopération
 Scientifique (PICS) project ``Isomonodromic deformations and conformal field theory''.}


\newpage

  \begin{thebibliography}{1000}
  
  \bibitem[AS]{AS}
  M. J. Ablowitz, H. Segur, {\it Asymptotic solutions of the Korteweg-de Vries equation}, Stud. Appl. Math.~\textbf{57}, No. 1, (1977),  13--44.
  
  \bibitem[AB]{AB}
  D. V. Anosov, A. A. Bolibruch, \textit{The Riemann-Hilbert problem: A publication from the Steklov Institute
  of Mathematics}, Aspects of Mathematics \textbf{E22}, Braunschweig; Wiesbaden: Vieweg, (1994).
  
   \bibitem[BBD]{BBD} 
    J. Baik, R.  Buckingham, J.  DiFranco,  {\it Asymptotics of Tracy-Widom distributions and the total
  integral of a Painlev\'e II function},   Comm.  Math. Phys.  {\bf 280},
  (2008), 463--497; arXiv:0704.3636 [math.FA].
  
   \bibitem[BBDI]{BBDI} J. Baik, R. Buckingham, J. DiFranco, A. Its, \textit{Total integrals of global solutions to Painlev\'{e} II}, Nonlinearity~\textbf{22}, (2009), 1021--1061; arXiv:0810.2586 [math.CA].
  
   \bibitem[BT]{BT} E. L. Basor, C. A. Tracy, {\it Some problems associated with the asymptotics
    of $\tau$-functions},   Surikagaku (Mathematical Sciences) {\bf 30}, no. 3,  (1992), 71--76.  
 
 \bibitem[BS]{BS} M. Bershtein, A. Shchechkin, {\it Bilinear equations on Painlev\'e tau functions from CFT}, Comm. Math. Phys.~\textbf{339}, (2015), 1021--1061; arXiv:1406.3008v5 [math-ph].  
 
 \bibitem[Ber]{Bertola} M. Bertola, \textit{The dependence on the 
 monodromy data of the
 isomonodromic tau function}, Comm. Math. Phys.~\textbf{294}, (2010),  539--579; arXiv: 0902.4716 [nlin.SI].
 
 \bibitem[Ber1]{Bertola1} M. Bertola, \textit{Corrigendum: The dependence on the monodromy data of the isomonodromic tau function}, 	arXiv:1601.04790 [math-ph].
 
 
  \bibitem[BIK]{BIK} A. A. Bolibruch, A. R. Its, A. A. Kapaev, 
  \textit{On the Riemann-Hilbert-Birkhoff inverse monodromy problem
 and the Painleve equations}, Algebra and Analysis
 {\bf 16 }, No.~1, (2004), 121--162.
 
 \bibitem[BGT]{BGT}
 G. Bonelli, A. Grassi, A. Tanzini, {\it Seiberg-Witten theory as a Fermi gas}, Lett. Math. Phys. {\bf 107}, (2017), 1--30; 	arXiv:1603.01174 [hep-th].
 
  	\bibitem[BoI]{BoI} T. Bothner, A. Its, \textit{Asymptotics of a cubic sine kernel determinant},
  	St. Petersburg Math. J. 26:4, (2015), 515--565; arXiv:1303.1871v1.
 
 \bibitem[BK]{BK}
 E. Br\'ezin, V. A. Kazakov, {\it Exactly solvable field theories of
 closed strings}, Phys. Letts. {\bf B236}, (1990), 144--150.
 
 \bibitem[BB] {BB}    A. M. Budylin, V. S. Buslaev,  {\it Quasiclassical asymptotics of the resolvent of 
  an integral convolution operator with a sine kernel on a finite interval},
  (Russian)   Algebra i Analiz  {\bf 7},  no. 6, (1995), 79--103;  
  translation in  St. Petersburg Math. J.  {\bf 7},  no. 6 (1996), 925--942.
  
  \bibitem[BFT]{BFT} V. S. Buslaev, L. D. Faddeev, L. A. Takhtajan, {\it Scattering theory for Korteweg-de-Vries (KdV)
  equation and its Hamiltonian interpretation}, Physica {\bf 18D}. (1986), 255--266.  
  
 \bibitem[CCR]{CCR}
 B. Carneiro da Cunha, M. Carvalho de Almeida, A. Rabelo de Queiroz,
 {\it On the existence of monodromies for the Rabi model}, J. Phys. A: Math. Theor. \textbf{49}, (2016), 194002; arXiv:1508.01342 [math-ph]. 
  
    \bibitem[DIK]{DIK} 
    P. Deift, A. Its, I. Krasovsky, {\it Asymptotics of the Airy-kernel determinant},
  Comm. Math. Phys. {\bf 278}, no. 3, (2008), 643--678;  arXiv:math/0609451 [math.FA].
  
   \bibitem[DIKZ]{DIKZ} 
    P. Deift, A. Its, I. Krasovsky, X. Zhou, {\it The Widom-Dyson constant
    for the gap probability in random matrix theory},   J. Comput. Appl. Math. {\bf 202}, no. 1, (2007),  26--47; arXiv:math/0601535 [math.FA].
    
 \bibitem[DKV]{DKV}
 P. Deift, I. Krasovsky, J. Vasilevska, {\it Asymptotics for a determinant with a confluent hypergeometric kernel},
 Int. Math. Res. Not.,  (2010), rnq150;  	arXiv:1005.4226 [math-ph].     
 
 \bibitem[DZ1]{DZ0}
 P. Deift, X. Zhou, {\it A steepest descent method for oscillatory Riemann-Hilbert problems. Asymptotics for the MKdV equation,}
 Ann. of Math. {\bf 137}, (1993), 296--368.
 
  \bibitem[DZ2]{DZ1}
 P.A. Deift, X. Zhou, {\it Asymptotics for the Painlev\'e II equation,}
  Comm.\ Pure Appl.\ Math. {\bf 48}, no.~3, (1995),  277--337.
  
  \bibitem[DZ3]{DZ2}
 P.A. Deift, X. Zhou, {\it Long-time asymptotics for integrable
 systems. Higher order theory}, Comm.\ Math.\ Phys., {\bf 165}, (1995), 175--191.   
  
 \bibitem[DS]{DS}
 M. Douglas, S. Shenker, {\it Strings in less than one dimension},
  Nucl. Phys. {\bf B335}, (1990), 635--654.   
  
   \bibitem[Ehr]{E} 
  T. Ehrhardt, {\it Dyson's constant in the asymptotics of the
  Fredholm determinant of the sine kernel},  Comm. Math. Phys. {\bf 262}, (2006), 317--341; arXiv:math/0401205 [math.FA].

 
 
 \bibitem[FIKN]{FIKN} A. S. Fokas, A. R. Its, A. A. Kapaev, V. Yu. Novokshenov, \textit{Painlev\'e transcendents:
  the Riemann-Hilbert approach}, Mathematical Surveys and Monographs~\textbf{128}, AMS, Providence,
  RI, (2006).
 
 \bibitem[FN1]{FN1}
 H.~Flaschka, A. C.~Newell,  {\it Monodromy- and spectrum-preserving
 deformations I},  Comm.\ Math.\ Phys. {\bf 76}, (1980) 65--116.
 
 \bibitem[FN2]{FN2}
 H.Flaschka, A. C. Newell,  {\it The inverse monodromy transform is a canonical transformation},
 Math.Studies {\bf  61}, North Holland, (1982), 65--91.
 
 \bibitem[GIL12]{GIL12}
  O. Gamayun, N. Iorgov, O. Lisovyy,  \textit{Conformal field theory of Painlev\'e~VI},
  JHEP \textbf{10}, (2012), 038; arXiv:1207.0787 [hep-th].
  
 \bibitem[GIL13]{GIL13}
 O. Gamayun, N. Iorgov, O. Lisovyy,  \textit{How instanton combinatorics solves Painlev\'e~VI, V and III's},
  J.~Phys.~\textbf{A46}, (2013), 335203;
   {arXiv:1302.1832 [hep-th]}.
  
 \bibitem[Gav]{Gav}
 P.~Gavrylenko, \textit{Isomonodromic $\tau$-functions and $W_N$
 conformal blocks}, JHEP \textbf{09}, (2015), 167;  arXiv:1505.00259v1 [hep-th].
 
 \bibitem[GL]{GL}  P. Gavrylenko, O. Lisovyy, \textit{Fredholm determinant and Nekrasov sum representations of isomonodromic tau functions}, \textbf{363}, (2018), 1--58; arXiv:1608.00958 [math-ph].
 
 \bibitem[GM]{GM}
 P. Gavrylenko, A. Marshakov, \textit{Exact conformal blocks for the W-algebras, twist fields and isomonodromic deformations}, JHEP~\textbf{02}, (2016), 181;  	arXiv:1507.08794 [hep-th].
 
 \bibitem[Gol]{Gol} W. M. Goldman, \textit{
 Invariant functions on Lie groups and Hamiltonian flows of surface group representations}, Invent. Math. \textbf{85}, (1986), 263--302.
 
 
 \bibitem[GrM]{GrM} D. Gross, A. Migdal, {\it A nonperturbative treatment of
 two-dimensional quantum gravity}, Nucl. Phys. \textbf{B340}, (1990), 333--365.
 
 \bibitem[HXZ]{HXZ}
  M. Huang, S. Xu, L. Zhang, \textit{Location of poles for the Hastings-McLeod solution to the second Painlevé Equation}, Constr. Approx.~\textbf{43}, (2016), 463; arXiv:1410.3338v2 [math.CA].
 
 \bibitem[ILST]{ILST}
  N. Iorgov, O. Lisovyy, A. Shchechkin, Yu. Tykhyy, \textit{Painlev\'e functions and conformal blocks},
  Constr. Approx.~\textbf{39}, (2014), 255--272.
 
 \bibitem[ILT13]{ILT13} N. Iorgov, O. Lisovyy, Yu. Tykhyy, \textit{Painlev\'e VI connection problem and monodromy of
 $c = 1$ conformal blocks}, JHEP~\textbf{12}, (2013), 029;
 arXiv:1308.4092v1 [hep-th].
 
 \bibitem[ILTe]{ILTe}
  N. Iorgov, O. Lisovyy, J. Teschner,  \textit{Isomonodromic tau-functions from Liouville conformal blocks}, Comm. Math. Phys. \textbf{336}, (2015),  671--694; arXiv:1401.6104 [hep-th].
 
 \bibitem[ILT14]{ILT14} 
 A. Its, O. Lisovyy, Yu. Tykhyy, \textit{Connection problem for the sine-Gordon/Painlev\'e III tau function
  and irregular conformal blocks}, Int. Math.
  Res. Not. \textbf{2015}, no. 18, (2015), 8903--8924; arXiv: 1403.1235 [math-ph].
  
  \bibitem[IN]{IN}
 A. R. Its, V. Yu. Novokshenov, {\it The Isomonodromy Deformation
 Method in the Theory of Painlev\'e Equations},
 Lect.\ Notes in Math. {\bf 1191}, Springer-Verlag, (1986). 
  
 \bibitem[IP]{IP} A. Its, A. Prokhorov, \textit{Connection problem for the tau-function of the sine-Gordon reduction of Painlev\'e-III equation via the Riemann-Hilbert approach}, Int. Math. Res. Not. \textbf{2016}, No. 22, (2016), 6856--6883; arXiv:1506.07485v2 [math-ph].
 
  

\bibitem[Jim]{Jimbo} M. Jimbo,
 {\it Monodromy problem and the boundary condition for some Painlev\'e
 equations}, Publ. Res. Inst. Math. Sci.~\textbf{18}, (1982),  1137--1161.
 
 \bibitem[JMMS]{JMMS}
  M. Jimbo, T. Miwa, Y. M\^ori, M. Sato, \textit{Density matrix of an impenetrable Bose gas
  and the fifth Painlev\'e transcendent}, Physica~\textbf{1D}, (1980), 80--158.
 
 \bibitem[JMU]{JMU} M. Jimbo, T. Miwa, K. Ueno,  {\it Monodromy preserving deformation of linear ordinary differential equations with rational coefficients. I},
  Physica {\bf D2}, (1981), 306--352. 
  
\bibitem[Kap]{Kap}  A. A. Kapaev, {\it Global asymptotics of the second Painlev\'e
transcendent},  Phys. Lett. {\bf A167}, (1992), 356--362.

  \bibitem[KK]{KK}
  A. V. Kitaev, D. A. Korotkin, \textit{On solutions of the Schlesinger equations
  in terms of $\Theta$-functions}, Int. Math. Res. Notices~\textbf{17}, (1998), 877--905; arXiv:math-ph/9810007.

  \bibitem[Kra]{K} 
I. Krasovsky, {\it Gap probability in the spectrum of random matrices and asymptotics
of polynomials orthogonal on an arc of the unit circle},  {Int. Math. Res. Not.} {\bf 2004}, (2004), 1249--1272;  	arXiv:math/0401258 [math.FA]

\bibitem[Kri]{Kri} I. Krichever, {\it Isomonodromy equations on algebraic curves, cannonical transformations
and Whitham equations}, Moscow Math. J.
{\bf 2}, (2002), 717 -- 752; arXiv:hep-th/0112096.  	

   \bibitem[L09]{Liso11}
 O. Lisovyy, {\it Dyson's constant for the hypergeometric kernel}, in
  ``New trends in quantum integrable systems'' (eds. B. Feigin, M. Jimbo, M. Okado),
  World Scientific, (2011), 243--267; arXiv:0910.1914 [math-ph]. 
  
 \bibitem[LT]{LT}
 O. Lisovyy, Yu. Tykhyy, {\it Algebraic solutions of the sixth Painlev\'e equation}, J. Geom. Phys.~\textbf{85}, (2014), 124--163; arXiv:0809.4873 [math.CA].  
  

  
 \bibitem[Mal]{Malgrange}
 B. Malgrange, \textit{Sur les d\'eformations isomonodromiques, I. Singularit\'es r\'eguli\`eres}, in
 ``Mathematics and Physics'', (Paris, 1979/1982); Prog. Math.~\textbf{37}, Birkh\"auser, Boston, MA, (1983), 401--426.
 
 \bibitem[Miw]{Miwa} T. Miwa, {\it Painlev\'e property of monodromy preserving deformation
equations and the analyticity of $\tau $-functions}, 
Publ. Res. Inst. Math. Sci. {\bf 17}, (1981), 703--712.
  
 \bibitem[Nag]{Nagoya}
 H. Nagoya, \textit{Irregular conformal blocks, with an application to the fifth and fourth Painlev\'e equations}, J. Math. Phys.~\textbf{56}, (2015), 123505; arXiv:1505.02398v3 [math-ph].
 
 \bibitem[NC]{NC}
 F. Novaes, B. Carneiro da Cunha, \textit{Isomonodromy, Painlev\'e transcendents and
 scattering off of black holes}, JHEP \textbf{2014}:132, (2014); arXiv:1404.5188v1 [hep-th].
 
\bibitem[Pal]{Palmer} 
J. Palmer,  {\it Zeros of the Jimbo, Miwa, Ueno tau function}, J. Math. Phys., {\bf 40}, (1999), 6638--6681; arXiv:solv-int/9810004.
 
 \bibitem[Tra]{T}  
C. A. Tracy, {\it Asymptotics of the $\tau$-function arising in the two-dimensional Ising model},
Comm. Math. Phys. {\bf 142}, (1991), 297--311.

   \bibitem[TW1]{TW1}
 C. A. Tracy, H. Widom, \textit{Level-spacing distributions and the Airy kernel},
 Comm. Math. Phys.~\textbf{159}, (1994), 151--174;
 arXiv:hep-th/9211141.
 \bibitem[TW2]{TW2}
 C. A. Tracy, H. Widom, \textit{Fredholm determinants, differential equations and matrix models},
 Comm. Math. Phys.~\textbf{163}, (1994), 33--72 ; 
 arXiv:hep-th/9306042.

 \bibitem[Was]{Was} W. Wasow, {\it Asymptotic expansions for ordinary differential equations}, Dover, New York, (2002).

  \bibitem[WMTB]{WMTB}
  T. T. Wu, B. M. McCoy, C. A. Tracy, E. Barouch, \textit{Spin-spin correlation functions for
  the two-dimensional Ising model: exact theory in the scaling region}, Phys. Rev.~\textbf{B13},
  (1976), 316--374.
  
 \end{thebibliography}
\end{document}